\documentclass{article}                    
\pdfoutput=1

\usepackage[pdftex]{graphicx}
\usepackage{amssymb}
\usepackage{amsmath}
\usepackage{authblk}
\usepackage{multirow}
\usepackage{bstnotationsi}
\usepackage{natbib}
\usepackage{xcolor}
\usepackage[margin=30mm]{geometry}

\graphicspath{{./}}

\begin{document}

\title{Characterization of random stress fields obtained from
  polycrystalline aggregate calculations using multi-scale stochastic
  finite elements}


\author{Bruno Sudret$^1$, Hung Xuan Dang$^2$, Marc
  Berveiller$^3$, Asmahana Zeghadi$^3$, Thierry Yalamas$^2$ \\

  $^1$ ETH Zurich, Chair of Risk, Safety and Uncertainty Quantification 
  Stefano-Franscini-Platz 5, CH-8093 Zurich, Switzerland 
  \and
  $^2$ Phimeca Engineering S.A., Centre d'affaires du
  Z\'enith, 34 rue de Sarli\`eve, F-63800 Cournon, France
  \and
  $^3$  EDF R\&D, Dept. of Materials and Mechanics
  of Components, Site des Renardi\`eres, 77250 Moret-sur-Loing Cedex,
  France}

\date{Submitted: October 20th, 2014 / Revised: January 12th, 2015}

\maketitle

\begin{abstract}
  The spatial variability of stress fields resulting from
  polycrystalline aggregate calculations involving random grain geometry
  and crystal orientations is investigated. A periodogram-based method
  is proposed to identify the properties of homogeneous Gaussian random
  fields (power spectral density and related covariance structure).
  Based on a set of finite element polycrystalline aggregate
  calculations the properties of the maximal principal stress field are
  identified. Two cases are considered, using either a fixed or random
  grain geometry. The stability of the method w.r.t the number of
  samples and the load level (up to 3.5 \% macroscopic deformation) is
  investigated.

\end{abstract}

\section{Introduction}
\label{sec:1}

In pressurized water reactors of nuclear plants, the pressure vessel
constitutes one element of the second safety barrier between the
radioactive fuel rods and the external environment. It is made of 16MND5
(A508) steel which is forged and welded. In case of operating accidents
such as LOCA (\emph{loss of coolant accident}), the pressure vessel is
subjected to a pressurized thermal shock due to fast injection of cold
water into the primary circuit. If some defects (\emph{e.g} cracks) were
present in the vessel wall this may lead to crack initiation and
propagation and the brittle fracture of the vessel. The detailed study
of the embrittlement of 16MND5 steel under irradiation is thus a great
concern for electrical companies such as EDF.

The brittle fracture behavior of the 16MND5 steel has been thoroughly
studied in the last decade using the local approach of fracture theory
\citep{Tanguy:2001} and the so-called
Beremin model \citep{Beremin:1983}, which assumes that cleavage is
controlled by the propagation of the weakest link between a population
of pre-existing micro-defects in the material.  This approach has been
recently coupled with polycrystalline aggregates simulations
\citep{Mathieu:2006b}, \citep{Mathieu:2010}.

The main idea is to model a material representative volume element (RVE)
as a polycrystalline synthetic aggregate and compute the stress field
under given load conditions. As a post-processing a statistical
distribution of defects (carbides) is sampled over the volume. In each
Gauss point of the finite element mesh the cleavage criterion is
attained somewhere along the load path if a) the equivalent plastic
strain has attained some threshold (cleavage initiation) and b) a
Griffith-like criterion applied to the size of the carbide in this Gauss
point is reached (cleavage propagation). Within the weakest link theory
the failure of a single critical carbide induces the failure of the RVE.

From a single RVE simulation (i.e. a single stress field) various
distributions of carbides are drawn, each realization leading to a
maximal principal stress associated to failure. Then the distribution of
these quantities is fitted using a Weibull law \citep{Mathieu:2010}. In
such an approach, the current practice of computational micromechanics
assumes that the RVE is large enough to represent the behavior of the
material so that a single polycrystalline analysis is carried out (the
large CPU required by polycrystalline simulations also favours the use
of a single simulation). However it is believed that numerous parameters
such as grain geometry and orientation may influence the stress field
and thus the final result.

The connection between micromechanics and stochastic methods has been
given much attention in the past few years, as shown in
\citet{GrahamBrady2006, Stefanou2009,Xu2009}. Many papers are devoted to
developing probabilistic models for reproducing a random microstructure,
\eg \citet{Graham2001, Arwade2007, Gutierrez2005, Rahman2008}. The
specific representation of polycrystalline microstructures has been
addressed in \citet{Arwade2004, Grigoriu2010, Zabaras2010a,
  Zabaras2010b} among others. The propagation of the uncertainty on the
microstructure through a micromechanical model in order to study the
variability of the resulting strain and stresses has not been addressed
much though (see \eg \citet{Zabaras2009}).

In this paper it is proposed to identify the properties of a stress
random field resulting from the progressive loading of a polycrystalline
aggregate. More precisely, assuming that the stress random field is
Gaussian, a procedure called {\em periodogram method} is devised, which
allows one to identify the correlation structure of the resulting stress
field.

The paper is organized as follows: in Section~2 basics of Gaussian
random fields are recalled and the periodogram method is presented
\citep{SudretDangIcasp2011}. The polycrystalline aggregate computational model is
detailed in Section~3. The methodology for identifying the correlation
structure of the resulting stress field is presented in Section~4. Two
application cases are then investigated, namely an aggregate with fixed
grain boundaries and random crystallographic orientations (Section~5)
and an aggregate with both random geometry and orientations (Section~6).
The variance of the resulting stress field as well as the spatial
covariance function and its correlation lengths is investigated in
details. The properties of the identified random fields will be used in a
forthcoming study in the context of the local approach to fracture, as
explained above.

\section{Inference of the properties of a Gaussian random field}
\label{sec:2}
In this section an identification method called \emph{periodogram} is
presented, which uses an estimator of the \emph{Power Spectral Density}
(PSD) in order to identify the correlation structure of a \emph{Gaussian
  homogeneous random field}. Based on original developments by
\citet{Stoica:1997} and \citet{Li:2005} for unidimensional fields, it
has been extended to two-dimensional cases by \citet{SudretDangIcasp2011}.
As it relies upon the use of the Fast Fourier Transform
(FFT) its computational efficiency is remarkable. 

\subsection{Definitions}
\label{section-2.1}

A Gaussian random field $Z(\ve{x}, \omega)$ is completely defined by its
mean value $\mu(\ve{x} )$, its standard deviation $\sigma(x)$ and its
auto-covariance function $C(\ve{x} ,\ve{x} ')$. It is said
\emph{homogeneous} if the mean value $\mu(\ve{x} )$ and the standard
deviation $\sigma(\ve{x} )$ are constant in the domain of definition of
$\ve{x} $ and the auto-covariance function $C(\ve{x} ,\ve{x} ')$ only
depends on the shift $\ve{h} = \ve{x} -\ve{x}'$. Let us introduce the
{\em $n-$th statistical moment} $m^n_Z$ and the spatial average
$m^n_{\ve{V}}$:
\begin{eqnarray}
  & m^n_Z & = \Esp{Z^n(\ve{x}_0,\omega)} = \int\limits_{-\infty}^{\infty} z^n(\ve{x}_0,\omega)f_Z(z,\ve{x}_0)dz
  \label{eq-II.1}\\
  & m^n_{\ve{V}} & = \lim\limits_{\ve{V}\rightarrow \infty} \frac{1}{\ve{V}} \int\limits_{V}
  z^n(\ve{x},\omega_0)d\ve{x} \label{eq-II.2}  
\end{eqnarray}

\begin{figure}[!ht]
    \begin{center}
      \includegraphics[width = 0.49\textwidth]{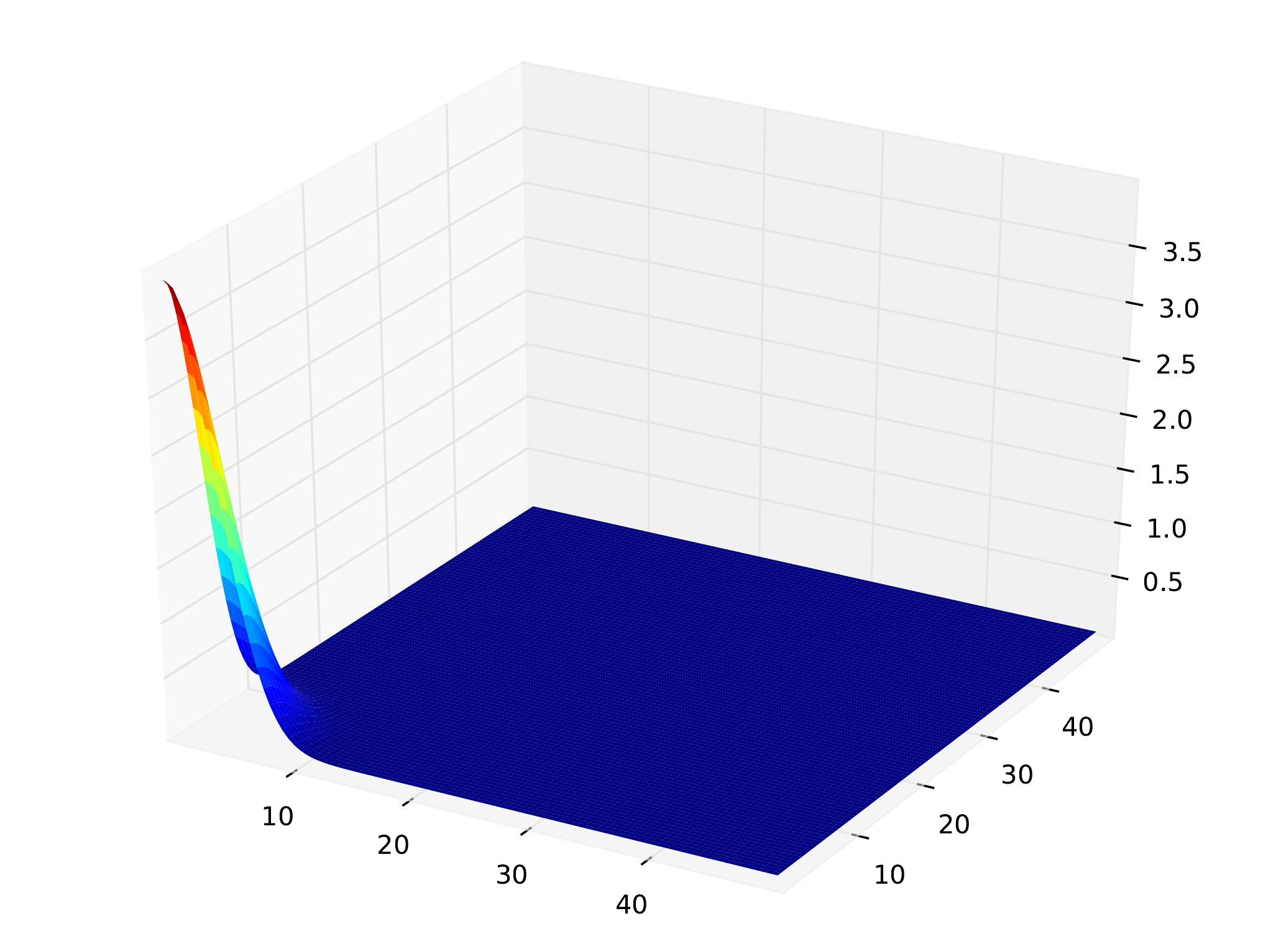}
      \includegraphics[width = 0.49\textwidth]{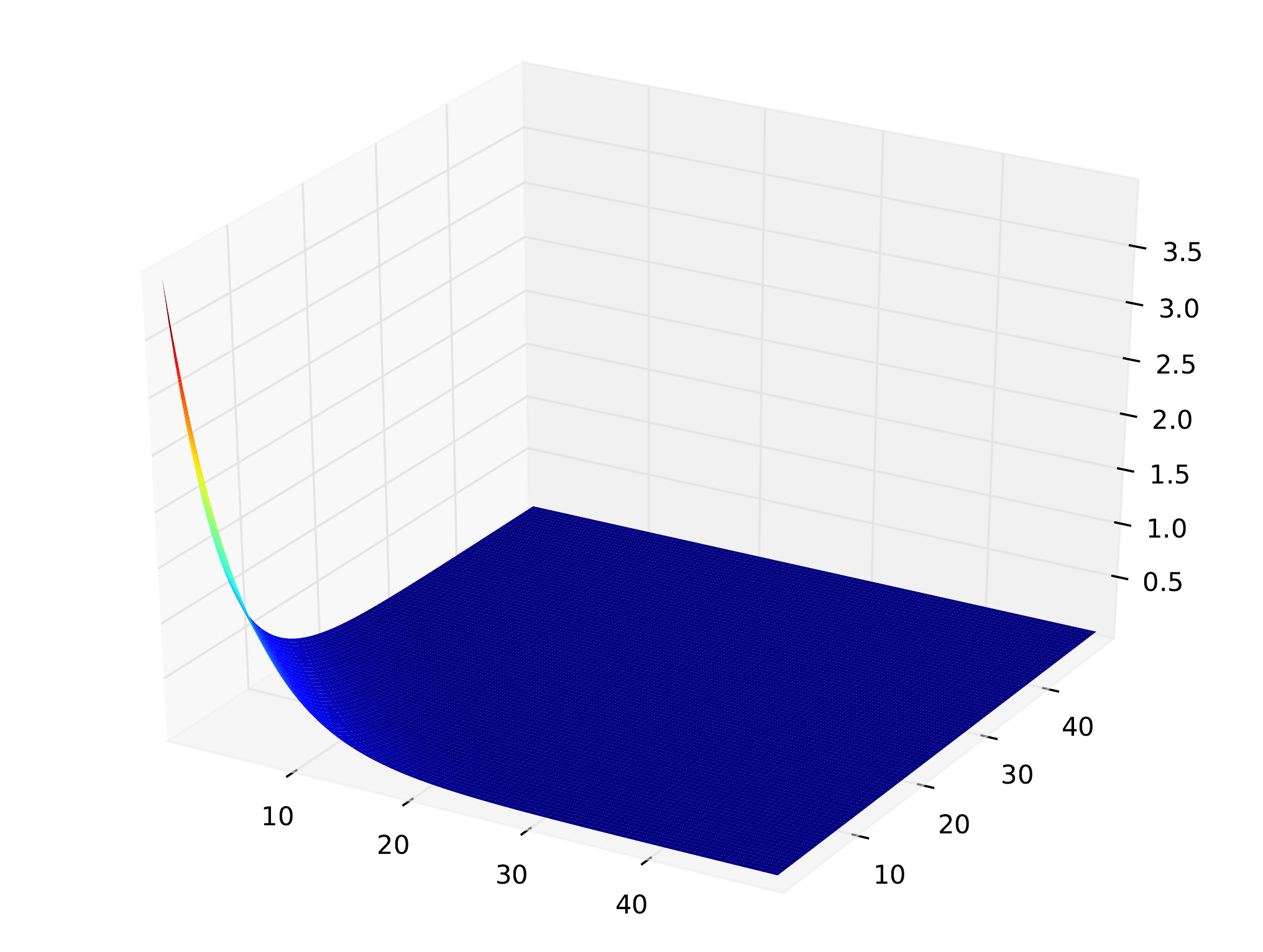}
      \caption{ Gaussian covariance model (left) and Exponential
        covariance model (right) : $\sigma = 2$, $l_1 = l_2 = 5$}
      \label{fig-II.1}
    \end{center}
\end{figure}

The field is said \emph{ergodic} if its ensemble statistics is equal to
the spatial average, \ie $m^n_Z = m^n_{\ve{V}}$ \citep{Cramer1967}.
Several popular covariance models for two-dimensional homogeneous random
fields are presented in Table~\ref{tab:01}. In this table, $\sigma$ is
the constant standard deviation of the field, $h_1, h_2$ are the
components of the shift $\ve{h}$ in the two directions, $l_1, l_2$ are
the correlation lengths in the two directions. Gaussian and exponential
models are plotted in Figure~\ref{fig-II.1} for the sake of
illustration. Note that we call {\em correlation length} the
  parameters that appear in the definition of the covariance functions. 
  This is not to be confused with the {\em scale of fluctuation}
  \citep{Vanmarcke1983}, which combine both the shape of the covariance
  function and the lengths $l_1, l_2$. In one dimension, denoting by
  $\rho(x;\, l)$ the autocorrelation function, the scale of fluctuation may
  be defined by:
  \begin{equation*}
    \label{eq:1001}
    2\, l_c = \int_{-\infty}^\infty \rho(x; \,l)\, dx
  \end{equation*}
which reduces to $l_c = l$ for the exponential correlation function
and $l_c = \sqrt{\pi}/2 \approx 0.886 \, l$ for the Gaussian
case. Similar expression are available in two and three dimensions, see
\eg \citet{Xu2009}.

\begin{table}[!ht] 
  \caption{Covariance functions and associated power spectral densities
    for homogeneous twodimensional random fields}
  \label{tab:01} 
  \begin{tabular}{lll}
    \hline\noalign{\smallskip}
    Model & Covariance function & Power spectral density \\
    \noalign{\smallskip}\hline\noalign{\smallskip}
    Exponential & $\sigma^2 \mbox{exp}
    \left[-(\frac{|h_1|}{l_1}+\frac{|h_2|}{l_2}) \right]$
    & $\sigma^2 \;\frac{2l_1}{1 + 4\pi^2 l_1^2 f_1^2} \; \frac{2l_2}{1 + 4\pi^2 l_2^2 f_2^2}$\\ 
    Gaussian    & $\sigma^2 \mbox{exp} \left[-(\frac{h_1^2}{l_1^2} +
      \frac{h_2^2}{l_1^2})\right]$ 
    & $\sigma^2 \;\pi l_1 \mbox{exp} \left(\pi^2 l_1^2 f_1^2 \right) \; \pi l_2 \mbox{exp} \left(\pi^2 l_2^2 f_2^2
\right)$\\
    Wave        & $\sigma^2  \mbox{sinc} (\frac{|h_1|}{l_1})
    $sinc$(\frac{|h_2|}{l_2})$
    & $\sigma^2 \;\pi l_1 \mbox{rect}_1(\pi l_1 f_1)\; \pi
    l_2\mbox{rect}_1(\pi l_2 f_2)$ \\
    Triangle    & $\sigma^2  \mbox{tri} (\frac{|h_1|}{l_1})
    $tri$(\frac{|h_2|}{l_2})$
    & $\sigma^2 \;l_1\mbox{sinc}^2(\pi f_1l_1)\; l_2\mbox{sinc}^2(\pi
    f_2l_2)$ \\
    \noalign{\smallskip}\hline
  \end{tabular}\\
  $ \text{sinc}(x) = \sin x / x $ \\$ \mbox{tri(x)}  =
  1-|x|$ if $|x|\le 1$ and 0 otherwise \\
  $ \mbox{rect}_{\tau}(f)  = 1$  if $|f| \leq \frac{\tau}{2}$ and 0 otherwise
\end{table}

The \emph{power spectral density} (PSD) of the random field is the
Fourier transform of its covariance function as a result of the
Wiener-Khintchine relationship \citep{Preumont1990}. The following
relationships hold:
\begin{eqnarray}
   S(f_1,f_2)  =& \int\limits_{-\infty}^{\infty} \int\limits_{-\infty}^{\infty} C(h_1,h_2) e^{-i 2\pi f_1 h_1} e^{-i 2 \pi
    f_2 h_2} dh_1 dh_2  \label{eq-II.3}\\
   C(h_1,h_2)  =& \int\limits_{-\infty}^{\infty} \int\limits_{-\infty}^{\infty} S(f_1,f_2) e^{i 2\pi f_1 h_1 } e^{i 2 \pi
    f_2 h_2} df_1 df_2  \label{eq-II.4}
\end{eqnarray}
The PSD of the Gaussian and exponential covariance models are presented
in Table~\ref{tab:01}.

\subsection{Identification of a PSD}
\label{section-2.2}
\subsubsection{Empirical periodogram} 
\label{section-2.2.1}

One considers an ergodic homogeneous random field $Z(x_1,x_2)$, $(x_1,
x_2) \in \cd_1 \times \cd_2 \subset \Rr^2$ for which a {\em single}
realization $z(x_1,x_2)$ is available. If the random field was defined
over an {\em infinite} domain, the classical estimation of the covariance
function would be:
\begin{equation} 
  \label{eq-II.5}
  \hat{C}(h_1,h_2) = \frac{1}{4MN}
  \sum\limits_{n = -N}^{N-1} \sum\limits_{m = -M}^{M-1} Z(x_{1n} +
  h_{1},x_{2m} + h_{2}) Z(x_{1n},x_{2m})
\end{equation} 
By definition, the Fourier transform of the covariance
estimation is an estimation of the PSD.
\begin{equation} 
  \label{eq-II.6} 
  \hat{S}(f_1,f_2) = \frac{1}{4MN}
  \tilde{Z}(f_1,f_2) \tilde{Z}^*(f_1,f_2) = \frac{1}{4MN}
  |\tilde{Z}(f_1,f_2)|^2
\end{equation} 
where $|.|$ denotes the modulus operator.

In practice, the problem is to estimate the periodogram from a
\emph{limited amount} of data gathered on $N \times M$ grid
$\{z(x_{1i},x_{2j}), \, i = 1 \enu N; j = 1 \enu M \}$. Due to symmetry,
the covariance estimation in Eq.(\ref{eq-II.5}) is recast as follows:
\begin{equation} 
  \label{eq-II.7}
  \hat{C}(h_{1k},h_{2l}) = \frac{1}{NM}
  \sum\limits_{n = 1}^{N-k} \sum\limits_{m = 1}^{M-l} Z(x_{1n} +
  h_{1k},x_{2m} + h_{2l})\, Z(x_{1n},x_{2m})
\end{equation} 
By taking the expectation of the above equation one gets:
\begin{equation} 
  \label{eq-II.8} 
  \begin{split}
    \mathbb{E} \left[ \hat{C}(h_{1k},h_{2l}) \right] &= \frac{N-k}{N}
    \frac{M-l}{M} \mathbb{E} \left[ Z(x_{1n} + h_{1k},x_{2m} + h_{2l})
      Z(x_{1n},x_{2m}) \right] \\ &= \frac{N-k}{N} \frac{M-l}{M}
    C(h_1,h_2)
  \end{split}
\end{equation} 
The latter equation exhibits some bias term between the expectation
  of the estimator and the covariance function $ C(h_1,h_2)$. Using the
symmetry of the covariance function, one can write:
\begin{equation}
  \label{eq-II.9} 
  \mathbb{E} \left[
    \hat{C}(h_{1k},h_{2l}) \right] = w_B(k,l) C(h_1,h_2)
\end{equation}
where $w_B(k,l)$ is the triangle window, also known as the {\em Bartlett
  window} (Figure~\ref{fig-II.3}):
\begin{equation}
  \label{eq-II.10} w_B(k,l) = \left\{
    \begin{array}{l l} \frac{N - |k|}{N} \frac{M-|l|}{M} &
      \mbox{if $|k| \leq N$; $|l| \leq M$}\\ 0 & \mbox{otherwise} \\
    \end{array} \right.
\end{equation} 
Consequently the expectation of the periodogram estimation becomes:
\begin{equation} 
  \label{eq-II.11} 
  \begin{split}
    \mathbb{E} \left[ \hat{S}(f_{1k},f_{2l}) \right] &= \mathcal{F}
    \left\{ \mathbb{E} \left[ \hat{C}(h_{1k},h_{2l}) \right] \right\} =
    \mathcal{F} \{ w_B(k,l) C(h_1,h_2)\} \\ &= W_B(f_1,f_2) * S(f_1,f_2)
  \end{split}\end{equation} 
where $\mathcal{F}$ and $W_B(f_1,f_2)$ respectively denote the 2D
Fourier transform operator and the Fourier transform of the Bartlett
window and $*$ denotes the convolution product. This window tends to a
Dirac pulse when $N, M$ tend to infinity and $w_B$ tends to a unit
constant. Thus the periodogram estimation is asymptotically unbiased.
However it is not consistent since its variance does not tend to zero
\citep{Preumont1990}.  Furthermore using this window leads to a
convolution product which introduces additional computational burden.
Hence in practice, the \emph{modified periodogram} presented in the next
section is used to estimate the PSD of the random field.

\subsubsection{Modified periodogram}
\label{section-2.2.2}
\begin{figure}[!ht]
    \begin{center}
      \includegraphics[width = 0.48\textwidth]{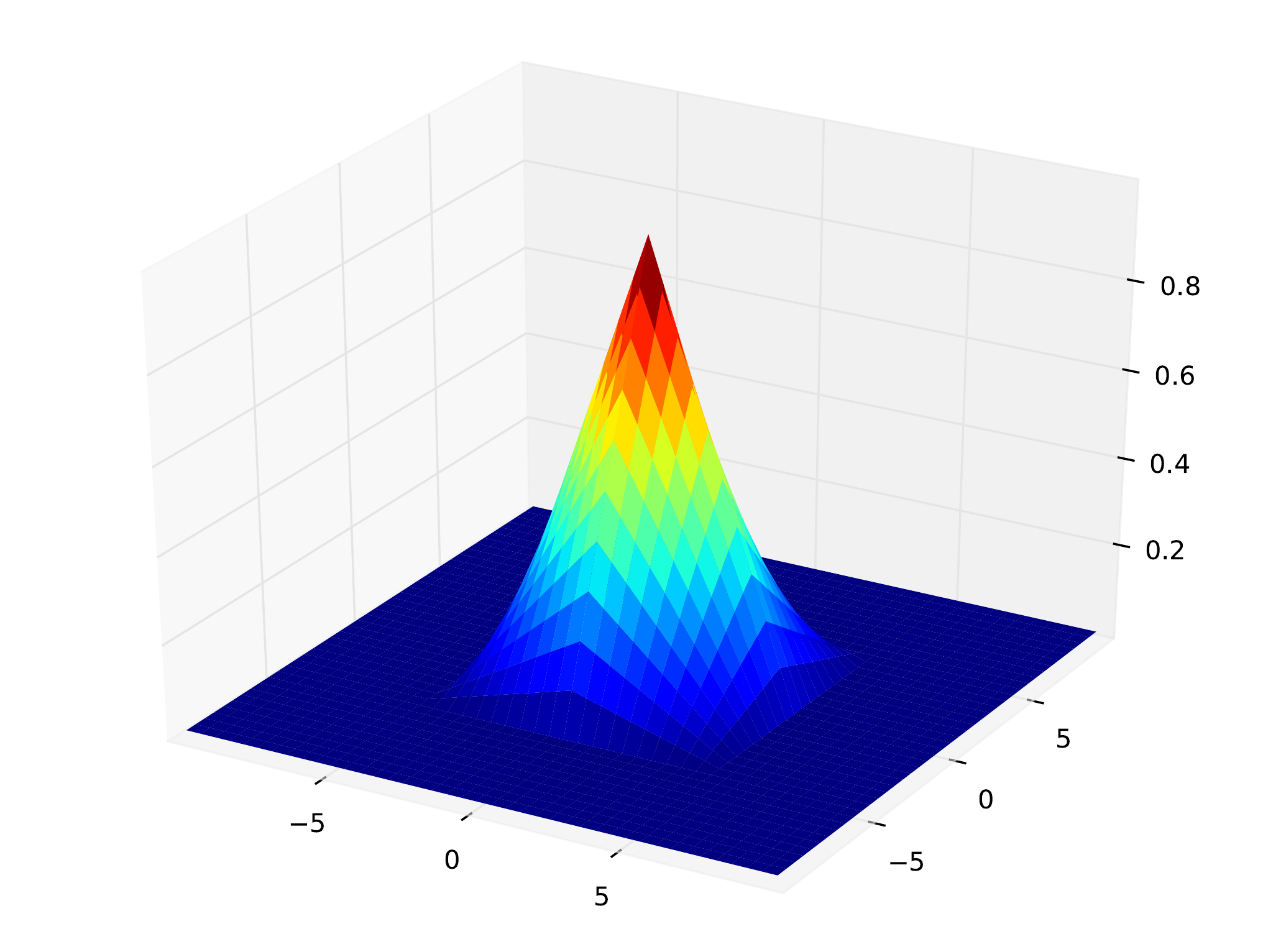}
      \includegraphics[width = 0.48\textwidth]{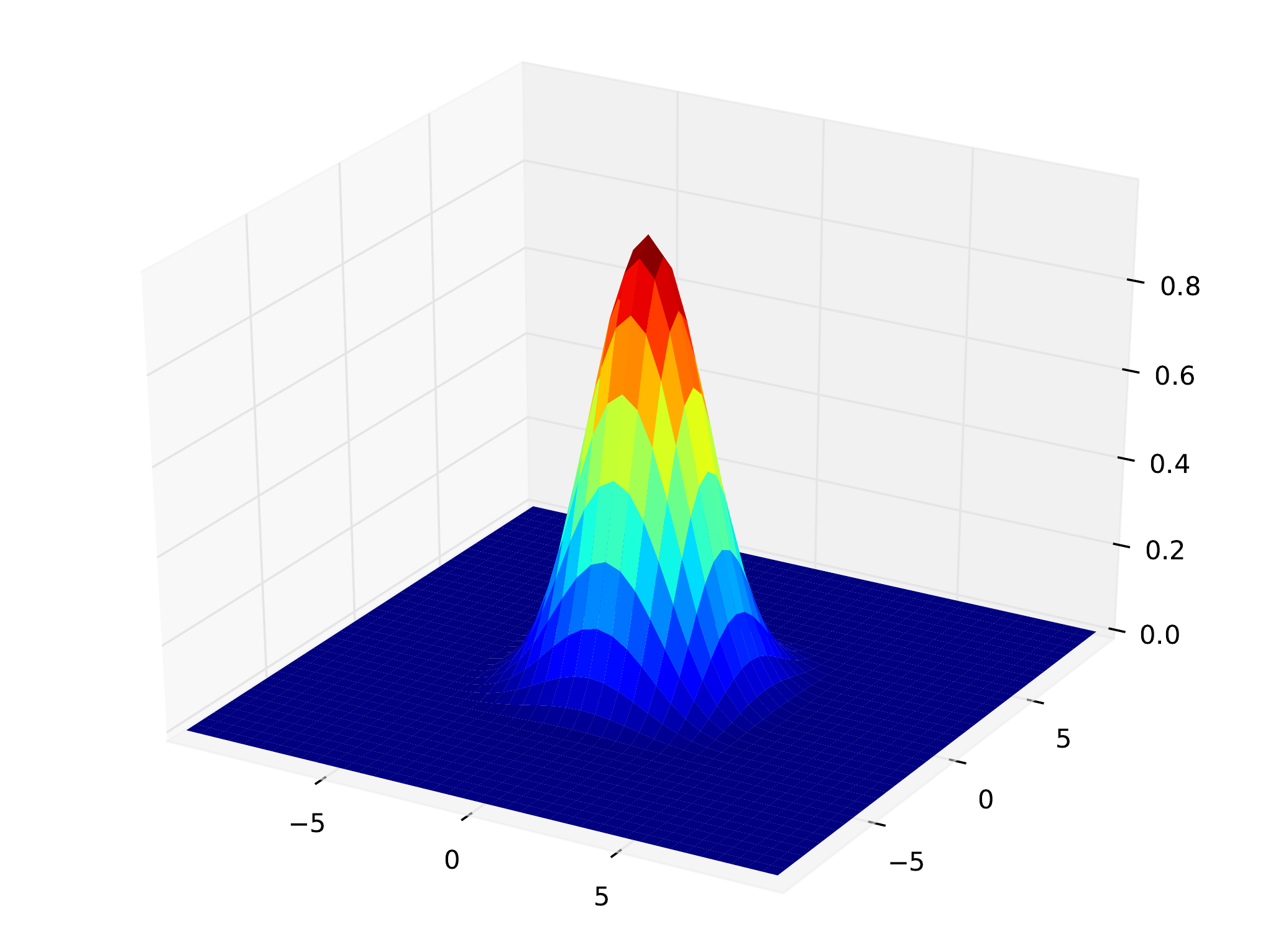}
      \caption{(a) Bartlett window ; (b) Blackman window } 
      \label{fig-II.3} 
    \end{center}
\end{figure}

The modified periodogram is built up in order to avoid the convolution
product with the transformed window $W_B(f_1,f_2)$ in
Eq.(\ref{eq-II.11}).  In this approach, the data is multiplied directly
with the window $w(x,y)$ {\em before} the Fourier transform is carried
out. It aims at filtering the data to limit the influence of long
distance terms and to focus on the information given by the short
distance terms. This leads to the following estimate:
\begin{equation} 
  \label{eq-II.12} \hat{S}(f_1,f_2) = \frac{1}{NMU} |
  \mathcal{F} \left\{ z(x_1,x_2).w(x_1,x_2) \right\} |^2
\end{equation} 
where $U$ is the energy of the window calculated by:
\begin{equation} 
  \label{eq-II.13}
  U = \frac{1}{D_1D_2} \sum\limits_{i =
    1}^{N} \sum\limits_{j = 1}^{M} w^2(x_{1i},x_{2j})
\end{equation} 
and $D_1,D_2$ denote the size of the two-dimensional domain $\cd_1
\times \cd_2$. Various window functions are proposed in
\citet{Preumont1990}, see Table~\ref{tab:03}. In this paper we will use
mainly the Blackman window (Figure~\ref{fig-II.3}).

\begin{table} [!ht]
  \caption{Window functions used in the modified periodogram approach}
  \label{tab:03} 
  \begin{tabular}{ll}
    \hline\noalign{\smallskip}
    Model & Window equation\\
    \noalign{\smallskip}\hline\noalign{\smallskip}
    Bartlett    & $\left\{
      \begin{array}{l l}
        \frac{N - |k|}{N} \frac{M-|l|}{M}		& 	\mbox{if $|k| \leq N$; $|l| \leq M$}\\
        0 						& 	\mbox{otherwise} \\
      \end{array}
    \right. $ \\
    \\
    Hann        & $\left\{
      \begin{array}{l l}
        \left[ 0.5 + 0.5 \mbox{cos}(\frac{\pi k}{N}) \right] \left[ 0.5 + 0.5 \mbox{cos}(\frac{\pi l}{M})\right]
        & \mbox{if $|k| \leq N$; $|l| \leq M$}\\
        0 				& 	\mbox{otherwise} \\
      \end{array}
    \right. $ \\
    \\ 
    Hamming    & $\left\{
      \begin{array}{l l}
        \left[ 0.54 + 0.46 \mbox{cos}(\frac{\pi k}{N}) \right] \left[ 0.54 + 0.46 \mbox{cos}(\frac{\pi l}{M})\right]
        & \mbox{if $|k| \leq N$; $|l| \leq M$}\\
        0 				& 	\mbox{otherwise} \\
      \end{array}
    \right. $ \\
    \\
    Blackman    & $\left\{
      \begin{array}{l l}
        \left[ 0.42 + 0.5 \mbox{cos}(\frac{\pi k}{N}) + 0.08 \mbox{cos}(\frac{2\pi k}{N})\right] 
        \left[ 0.42 + 0.5 \mbox{cos}(\frac{\pi l}{M}) + 0.08 \mbox{cos}(\frac{2\pi l}{M})\right]
        & \mbox{if $|k| \leq N$; $|l| \leq M$}\\
        0 				& 	\mbox{otherwise} \\
      \end{array}
    \right. $ \\
    \noalign{\smallskip}\hline
  \end{tabular}
\end{table}

\subsubsection{Average modified periodogram}
 \label{section-2.2.3}

 As shown in Section~\ref{section-2.2.1}, the estimation of the
 periodogram is asymptotically unbiased, however not consistent since
 its variance does not tend to zero when $N, M$ tend to infinity. The
 averaging of the modified periodogram will solve this problem. Assume
 that $L$ realizations of the random field are available. For each
 realization $ z^l(x_1,x_2)$, one calculates the periodogram as in
 Eq.(\ref{eq-II.12}):
 \begin{equation} 
   \label{eq-II.14} 
   \hat{S}^l(f_1,f_2) = \frac{1}{NMU} |
   \mathcal{F}\left\{ z^l(x_1,x_2).w(x_1,x_2) \right\} |^2
 \end{equation} 
 with $1 \leq l \leq L$. Then one calculates the average periodogram by:
 \begin{equation} 
   \label{eq-II.15} 
   \overline{S}(f_1,f_2) = \frac{1}{L}
   \sum\limits_{l=1}^{L} \hat{S}^l(f_1,f_2)
 \end{equation} 
 
 Therefore the variance of the average periodogram is:
 \begin{equation} 
   \label{eq-II.16} 
   \Var{ \overline{S}(f_1,f_2)} =
   \frac{1}{L} \Var{ \hat{S}(f_1,f_2) }
 \end{equation} 
 It is then obvious that this variance tends to zero when $L$ tends to
 infinity, making the ``average modified periodogram'' approach more
 robust.
 
 \subsubsection{Final algorithm for PSD estimation}
\label{section-2.2.4}
As a summary, the algorithm to estimate the PSD of a random field from
$L$ realizations may be decomposed into the four following steps:
\begin{enumerate}
  \item multiplication of each realization by a selected window, \emph{e.g.}
the Blackman window (see Table~\ref{tab:03});
  \item computation of 2D Fourier transform of the product of the
    current realization by the filtering window;
  \item computation of the modulus of the result to obtain the PSD
estimation of each realization;
  \item averaging of the $L$ PSD estimations.
\end{enumerate} 

Once the empirical periodogram has been computed, a {\em theoretical}
periodogram is selected (\eg Gaussian, exponential, etc., see
Table~\ref{tab:01}) and fitted
using a least-square procedure \citep{Marquardt1963}. In case of
multiple potential forms for the theoretical periodogram the best
fitting is selected according to the smallest residual.

\section{Modeling polycrystalline aggregates}
\label{section-3}

In this section the computational mechanical model used in this study is
presented. It simulates a tensile test on a bidimensional
polycrystalline aggregate under plane strain conditions. The various
ingredients are discussed, namely:
\begin{itemize}
 \item the microstructure of the material and its synthetic
representation;
 \item the material constitutive law;
 \item the boundary conditions applied onto the aggregate;
 \item the mesh used in the finite element simulation.
\end{itemize} 

\subsection{Material characterization} \label{section-3.1}

The material is a 16MND5 ferritic steel with a granular microstructure.
The ferrite has a body centered cubic (BCC) structure.  Three families
of slip system should be taken into account, namely $\{110\}\langle 111
\rangle$, $\{112\}\langle 111 \rangle$, $\{123\}\langle 111 \rangle$.
However, following \citet{Franciosi:1985} it is assumed that the glides
on the plane {123} are a succession of micro-glides on the planes {110},
{112}. This leads to consider only the two first families, which yields
24~slip systems by symmetry.

\subsection{Crystal plasticity} 
\label{section-3.2}
The model for crystal plasticity chosen in this work has been originally
formulated in \citet{Meric:1991} within the small strain framework. The
total strain rate $\dot{\varepsilon}_{ij}$ is classically decomposed as
the sum of the elastic strain rate $\dot{\varepsilon}^e_{ij}$ and
plastic strain rate $\dot{\varepsilon}^p_{ij}$.
\begin{equation}
  \label{eq-III.17} 
  \dot{\varepsilon}_{ij} =
  \dot{\varepsilon}^e_{ij} + \dot{\varepsilon}^p_{ij}
\end{equation} 
The elastic part follows the Hooke's law and the plastic part is
calculated from the shear strain rates of the $24$ active slip systems.
\begin{equation}
  \label{eq-III.18} 
  \dot{\varepsilon}^p_{ij} =
  \sum\limits_{g = 1}^{24} \dot{\gamma}^g R^g_{ij}
\end{equation}
where $\dot{\gamma}^g$ is the shear strain rate of the slip system $g$
and $R^g_{ij}$ is the Schmid factor which presents the geometrical
projection tensor. The latter is calculated from the normal vector to the
gliding plane $\textbf{n}$
and the direction of gliding $\textbf{m}$.
\begin{equation} 
  \label{eq-II.19}
  R^g_{ij} = \frac{1}{2}(m_i n_j + m_j n_i)
\end{equation}
The Resolved Shear Stress (RSS) $\tau^g$ of the slip system $g$ is the
projection of the stress tensor via the Schmid factor.
\begin{equation} \label{eq-III.20}
   \tau^g = R^g_{ij} \sigma_{ij}
\end{equation}
The shear strain rates $\dot{\gamma}^g$ of each slip system $g$ are the
internal variable that describes plasticity. The evolution of these
variables depends on the difference between the RSS $\tau^g$ and the
actual {\em critical RSS} $\tau^g_c$ in an elastoviscoplastic setting:
\begin{equation} 
  \label{eq-III.21}
  \dot{\gamma}^g = \left( \frac{\tau^g - \tau^g_c}{K} \right)^n \mbox{sign}(\tau^g)
\end{equation}
where $K$ and $n$ are material constants, and $\mbox{sign} (a) = a/|a|$
if $a \neq 0 $ and 0 otherwise. Note that this formula corresponds to an
elastoviscoplastic constitutive law but the viscous effect will be
negligible if sufficiently large values of $K$ and $n$ are selected.
Its power form allows one to automatically detect the active slip
systems. All the systems are considered active but the slip rate is
significant only if the RSS $\tau^g$ is much higher than the critical
RSS $\tau^g_c$.  This procedure allows one to numerically smooth the
elastoplastic constitutive law.

The critical RSS $\tau^g_c$ evolves according to the following isotropic
hardening law:
\begin{equation} 
  \label{eq-III.22}
  \tau^g_c =  \tau^g_{c0} + Q^g \sum\limits_{s=1}^{24} h^{gs} (1-e^{-b^g \gamma^s_{cum}})
\end{equation}
where $\gamma^s_{cum} = \int\limits_{t_0}^{t} |\dot{\gamma^s}|dt$. The
exponential term presents the hardening saturation in the material when
the accumulated slip is high. $\tau^g_{c0}$ is the {\em initial critical
  RSS} on the considered system $g$. $Q^g$ and $b^g$ are parameters
which depend on the material. $h^{gs}$ is the hardening matrix of size
$24 \times 24$ whose component $h^{gs}$ presents the hardening effect of
the system $g$ on the system $s$. In the present work, one considers
only two families of slip systems named ${110}\langle 111 \rangle$,
${112}\langle 111 \rangle$. Thus the hardening matrix $h^{gs}$ is
completely defined by four coefficients $h_1, h_2, h_3, h_4$ only. The
values of these coefficients and this matrix are presented in
\citet{Mathieu:2006a}.  All the parameters describing crystal plasticity
for 16MND5 steel are gathered in Table \ref{tab:04}.

\begin{table}[!ht] 
  \caption{Parameters of the crystal plasticity constitutive law for the
    16MND5 steel \citep{Mathieu:2006a}} 
  \label{tab:04} 
  \begin{tabular}{c  c  c  c  c  c  c  c  c  c  c}
    \hline\noalign{\smallskip}
    \multicolumn{2}{c}{Isotropic elasticity} & \multicolumn{2}{c}{Viscoplasticity} & \multicolumn{7}{c}{Isotropic
      hardening}\\ \noalign{\smallskip}\hline
    $E$     &   $\nu$     &$K$             &   $n$   &   $\tau_{c_0}$   &   $Q$   &   $b$  &  $h_1$  &  $h_2$  &  $h_3$  & $h_4$\\
    $(MPa)$ &	 	  &$(MPa.s^{1/n})$ &         &   $(MPa)$        & $(MPa)$ &        &         &         &         &   \\
   \hline
    $210,000$&   $0.3$     & $15$           &   $12$  &    $175$         &   $20$  &  $30$  &   $1$   &   $1$   &   $1$   &  $1$ \\
    \noalign{\smallskip}	\hline
  \end{tabular}
\end{table}

\subsection{Microstructure and boundary conditions} 
\label{section-3.3}

\begin{figure}[!ht]
    \begin{center}
      \includegraphics[width = 0.45\textwidth]{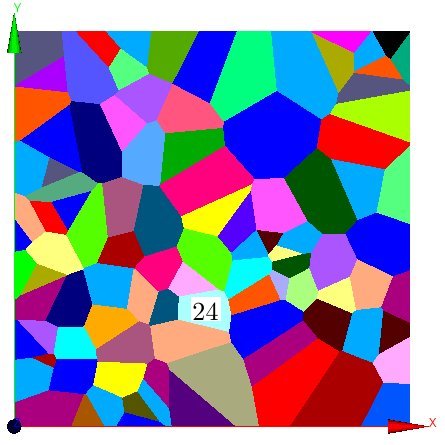}
      \includegraphics[width = 0.45\textwidth]{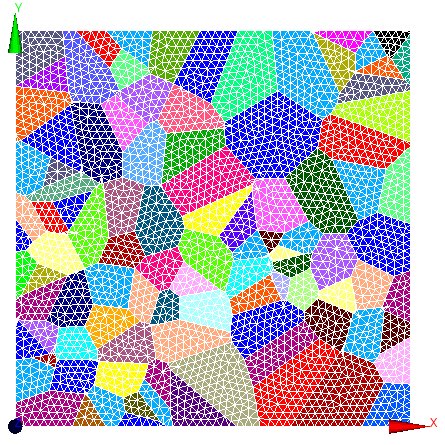}
      \caption{\label{fig-III.4} (a) Two-dimensional polycrystalline
        aggregate modelling a volume of 16MND5 steel (100 grains)  (b) Mesh of the
        specimen (11,295 nodes)}
    \end{center}
\end{figure}

The construction of the aggregate is based on the Voronoi polyhedra
model \citep{Gilbert:1962}, generated in this work with the Quickhull
algorithm \citep{Barber:1996}. The geometry of the resulting synthetic
aggregate, which is a simplified representation of the real
microstructure of the 16MND5 steel, is shown in Figure \ref{fig-III.4}.
It corresponds to a square of size 1,000 (this is a relative length
which shall be scaled with a real length depending on the grain size).
Grain boundaries are considered as perfect interfaces. Note that
  more detailed models of grains have been proposed recently using
  so-called Laguerre tessellations \citep{Lautensack2008} in order to
  better fit the observed distributions of grain size, see \eg
  \citet{ZhangBalint2011, Leonardia2012}.

The same crystallographic orientation, defined by the three Euler angles
$\varphi_1$, $\phi$, $\varphi_2$, is randomly assigned to all
integration points inside each individual grain using a uniform
distribution. In Figure \ref{fig-III.4}-a, the color of each grain
corresponds to a given crystallographic orientation.  The mesh is
generated by the $BLSURF$ algorithm \citep{Laug:1999} of the Salome
software (http://www.salome-platform.org). The mesh of the generated
specimen is presented in Figure \ref{fig-III.4}-b. The finite elements
are quadratic 6-node triangles with $3$ integration points.

\begin{figure}[!ht]
    \begin{center}
      \includegraphics[width = 0.5\textwidth]{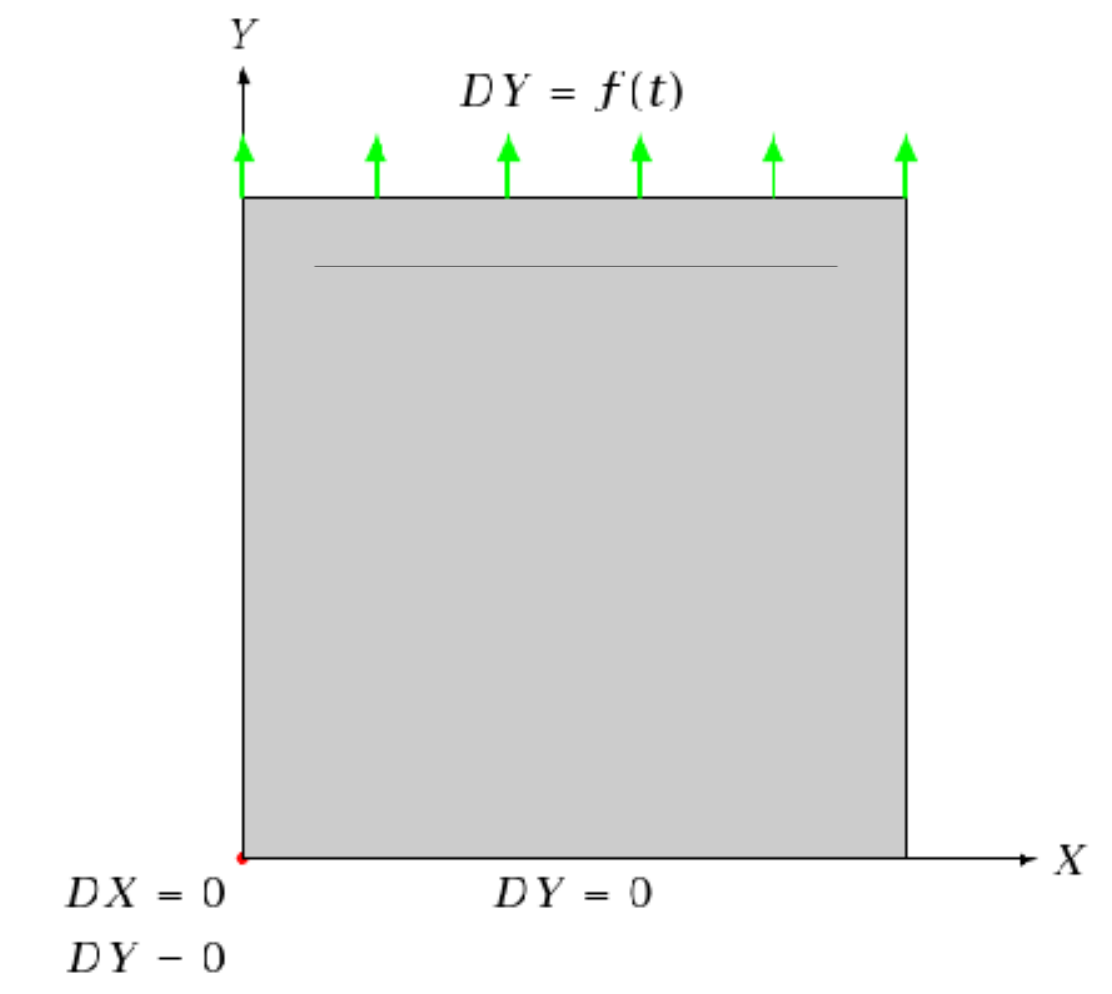}
      \caption{\label{fig-III.5} Boundary conditions used for simulating
        the tensile test}
    \end{center}
\end{figure}

The boundary conditions applied onto the aggregate are sketched in
Figure~\ref{fig-III.5}. The lower surface is blocked along the $Y$
direction. The displacements $DX=DY=0$ are blocked at the origin of the
coordinate system (lower left corner). On the upper surface, an
homogeneous displacement is applied by steps in the $Y$ direction up to
a macroscopic strain equal to $3.5\%$. The computation is carried out
using the open source finite element software Code\_Aster
(http://www.code-aster.org).

The computational cost for such a non linear analysis is high. The
number of degrees of freedom of the finite element model is $33,885$. A
parallel computing method based on sub-domain decomposition is used. One
simulation of a full tensile test up to 3.5\% strain requires about $2$
hours computation time when distributed over $4$ processors.

\subsection{Results} \label{section-3.4} In this section, we present the
result of the simulation of a tensile test on the 2D aggregate at
different scales. We define the mean stress and strain tensor calculated
in a volume $V$ by:
\begin{eqnarray}
  \ma{\Sigma} = \frac{1}{V} \int\limits_{V} \ma{\sigma} dV  \label{eq-III.23}\\
  \ma{E} = \frac{1}{V} \int\limits_{V} \ma{\varepsilon} dV     \label{eq-III.24}
\end{eqnarray}

\begin{figure}[!ht]
    \begin{center}
      \includegraphics[width = 0.85\textwidth]{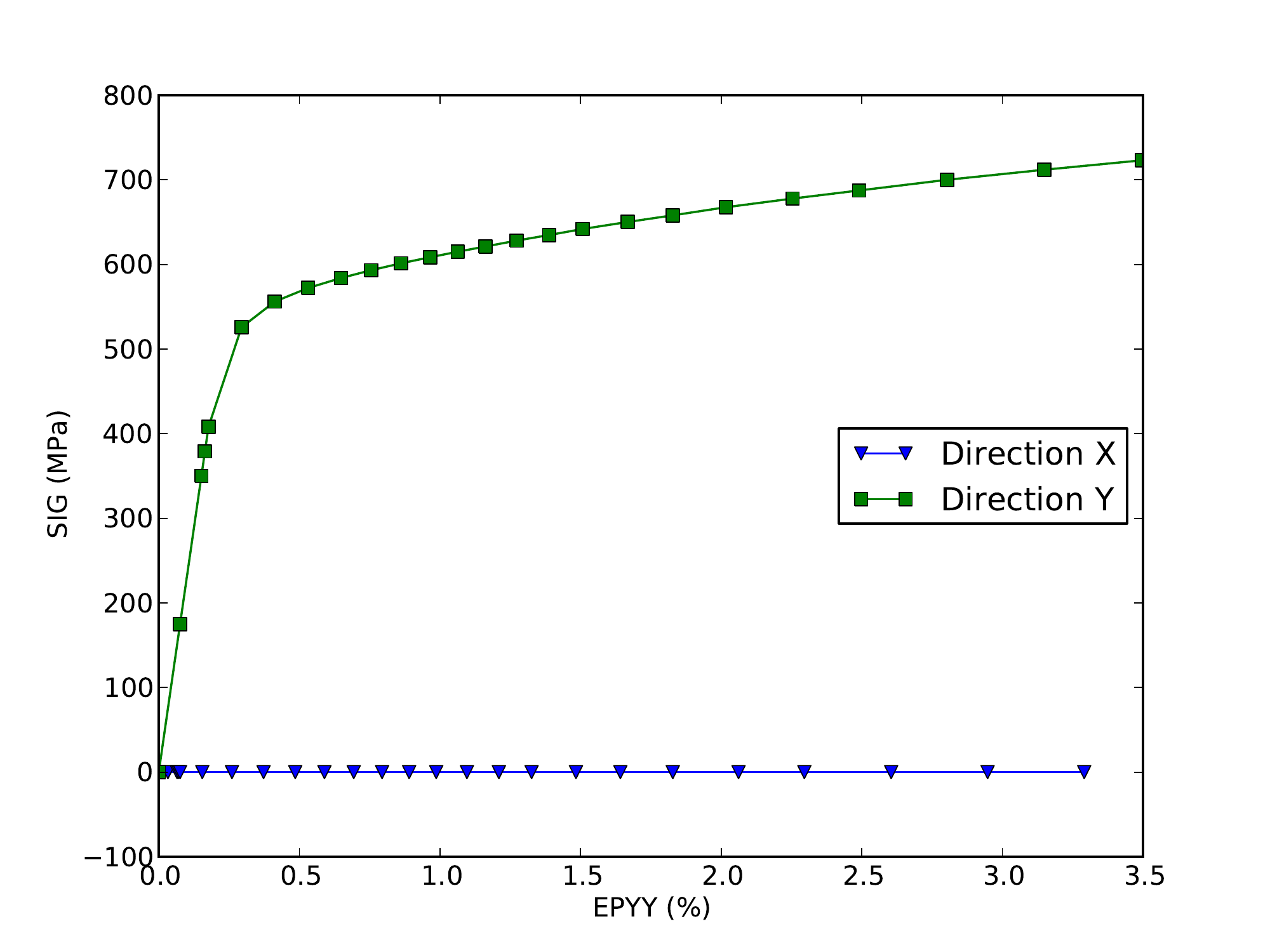}
      \caption{\label{fig-III.6} Macroscopic strain-stress relationship
        in the $X$ and $Y$ directions}
    \end{center}
\end{figure}

Figure~\ref{fig-III.6} shows the macroscopic strain/stress curve. It is
observed that $\Sigma_{XX}=0$ as expected whereas the uniaxial
behaviour shows a global elastoplastic behaviour.

\begin{figure}[!ht]
  \begin{center}
    \includegraphics[width = 0.48\textwidth]{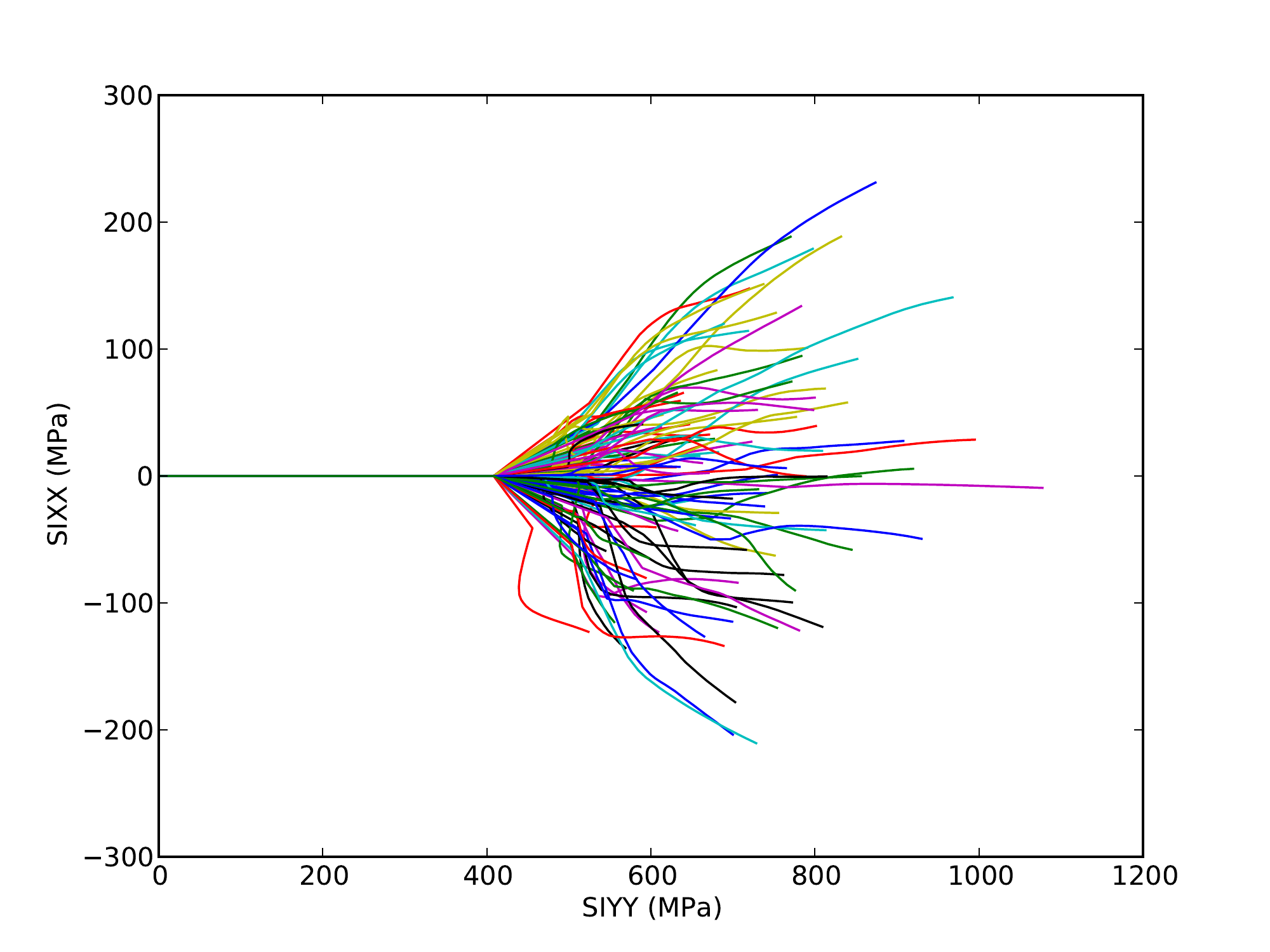}
    \includegraphics[width = 0.48\textwidth]{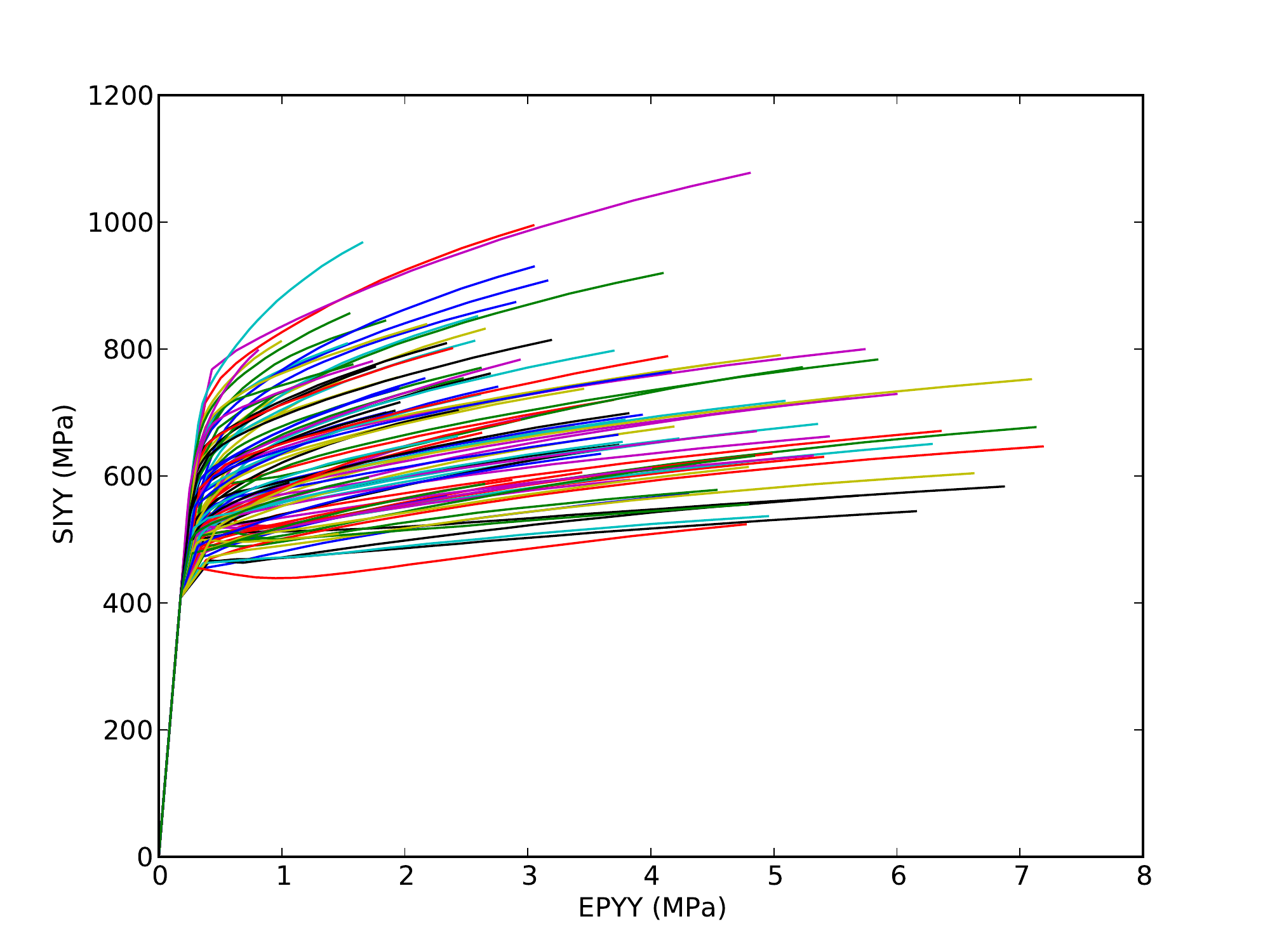}
    \caption{\label{fig-III.7} Mesoscopic behavior in each grain in
      the $X$ (left) and $Y$ (right) directions}
  \end{center}
\end{figure}

At the mesoscopic scale one can observe the mean strain-stress
relationship in each grain as shown in Figure \ref{fig-III.7}. Because
of the different crystallographic orientations in each grain, the mean
elastoplastic behaviour is different from grain to grain. Furthermore,
whereas the mean stress $\Sigma_{XX}$ calculated in all the specimen is
zero, the mean values calculated in each single grain are scattered
around zero. This observation shows the first scale of heterogeneity
of the material.

\begin{figure}[!ht]
  \begin{center}
    \includegraphics[width = 0.85\textwidth]{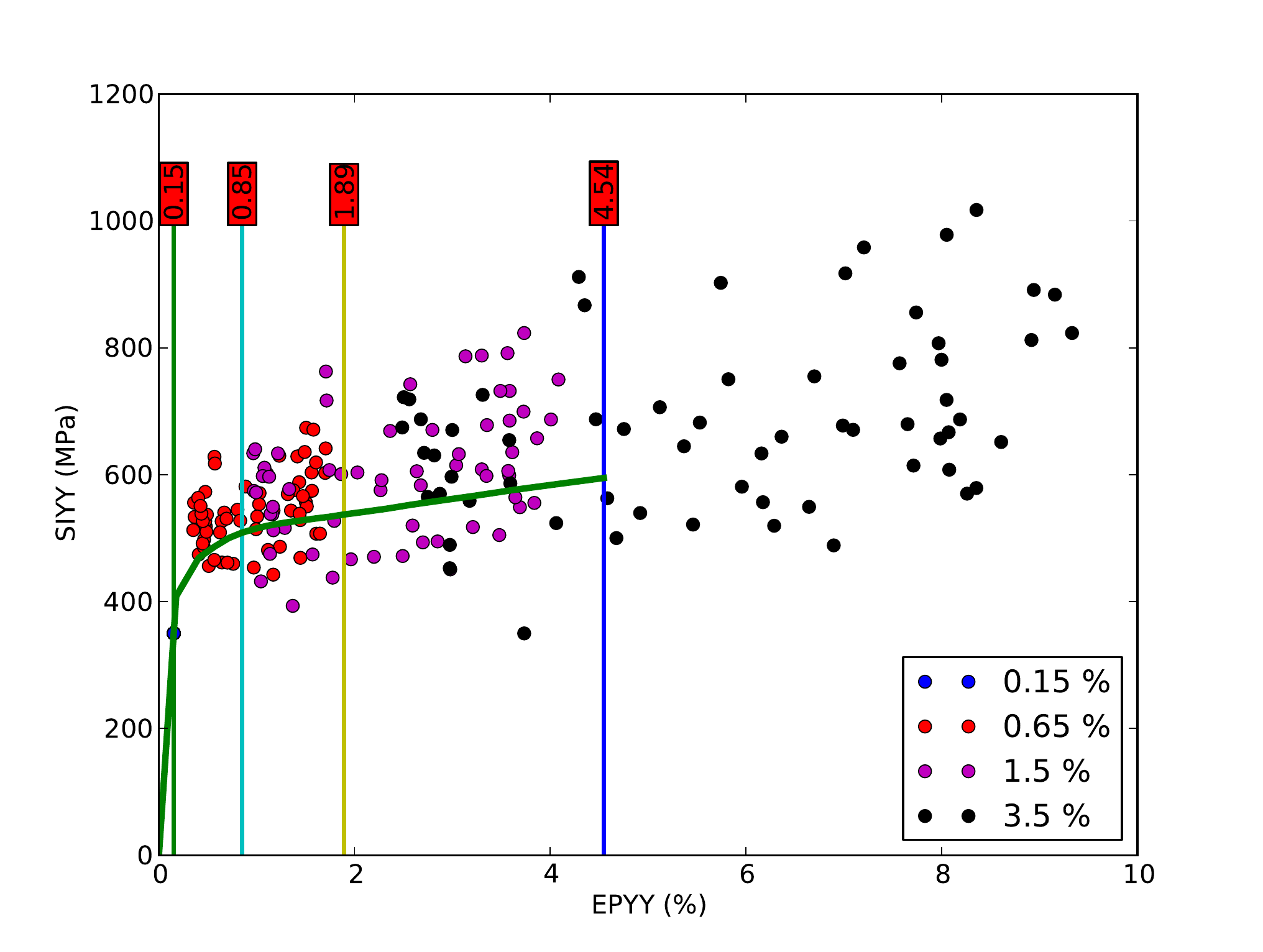}
    \caption{\label{fig-III.8} Microscopic strain-stress relationship
      for various nodes within Grain \#24 and mean tensile curve}
  \end{center}
\end{figure}

The microscopic behaviour of a single grain (Grain \#24, see tag in
Figure~\ref{fig-III.4}) is finally studied. The mean behavior and the
strain-stress relationship at each node of this grain are plotted in
Figure~\ref{fig-III.8} for four levels of macroscopic strain, namely
$E_{YY} = 0.15\%, 0.65\%, 1.5\%, 3.5\%$.

In this figure the blue point represents the stress field within the
grain for a macroscopic strain level $E_{YY} = 0.15\%$. This single
point shows that the stress field is homogeneous within the grain in the
elastic domain. The red points represent the strees values $\sigma_{YY}$
in each node of the grain at $E_{YY}=0.65\%$ macroscopic strain. One
observes that the mean strain calculated for this single grain is
$0.85\%$ and the maximal strain value $\varepsilon_{YY}$ in a specific
node may attain about $2\%$. Similar effects are observed at other
levels of macroscopic strain, which show the heterogeneity of the strain
and stress fields at the microscopic scale. It is observed that the
scattering around the mean curve increases with the macroscopic strain.
Indeed, for the final loading step corresponding to $E_{YY}=3.5\%$ the
mean strain in the grain is about 4.54\%, while the local strain varies
form 2.4 to 9\%.

\section{Identification of the maximal principal stress field} 
\label{section-4}

In this section the method developed in Section~2 is applied to the
identification of the properties of the random stress field in
polycristalline aggregate calculations.  More specifically the {\em
  maximal principal stress} field $\sigma_I$ that is computed from
repeated polycrystalline simulations is considered. Throughout the
  paper this stress field is considered {\em Gaussian}. This is a strong
  assumption which shall be considered as a first approximation.  Indeed
  the maximal principal stress is positive in nature under the uniaxial
  loading that is considered and a Gaussian assumption cannot totally
  fit this feature. Yet it is believed that the results obtained in
  terms of the description of the spatial variability (covariance
  functions), which is the main outcome of the paper, will not be
  strongly influenced by this assumption. Note that methods for
  identifying the properties of non Gaussian random fields have been
  recently developed, see \eg \citet{PerrinSoize2014}.

\subsection{Finite element calculations and projection}
\label{section-4.1}
The maximal principal stress field is assumed to be Gaussian and
homogeneous (the latter assumption will be empirically checked as shown
in the sequel). The periodogram method is applied using $K=35$
realizations of stress fields, \ie 35~full elastoplastic analysis of
aggregates up to a macroscopic strain of 3.5 \%. The identification is
carried out successively at various levels of the macroscopic strain.
Two cases are considered:
\begin{itemize}
\item Case \#1: the grains geometry is the same for all the finite
  element calculations. Only the crystallographic orientations are
  varying from one calculation to the other.
\item Case \#2: both the grains geometry and the crystallographic
  orientations vary.
\end{itemize}

\begin{figure}[!ht]
    \begin{center}
      \includegraphics[width = 0.55\textwidth]{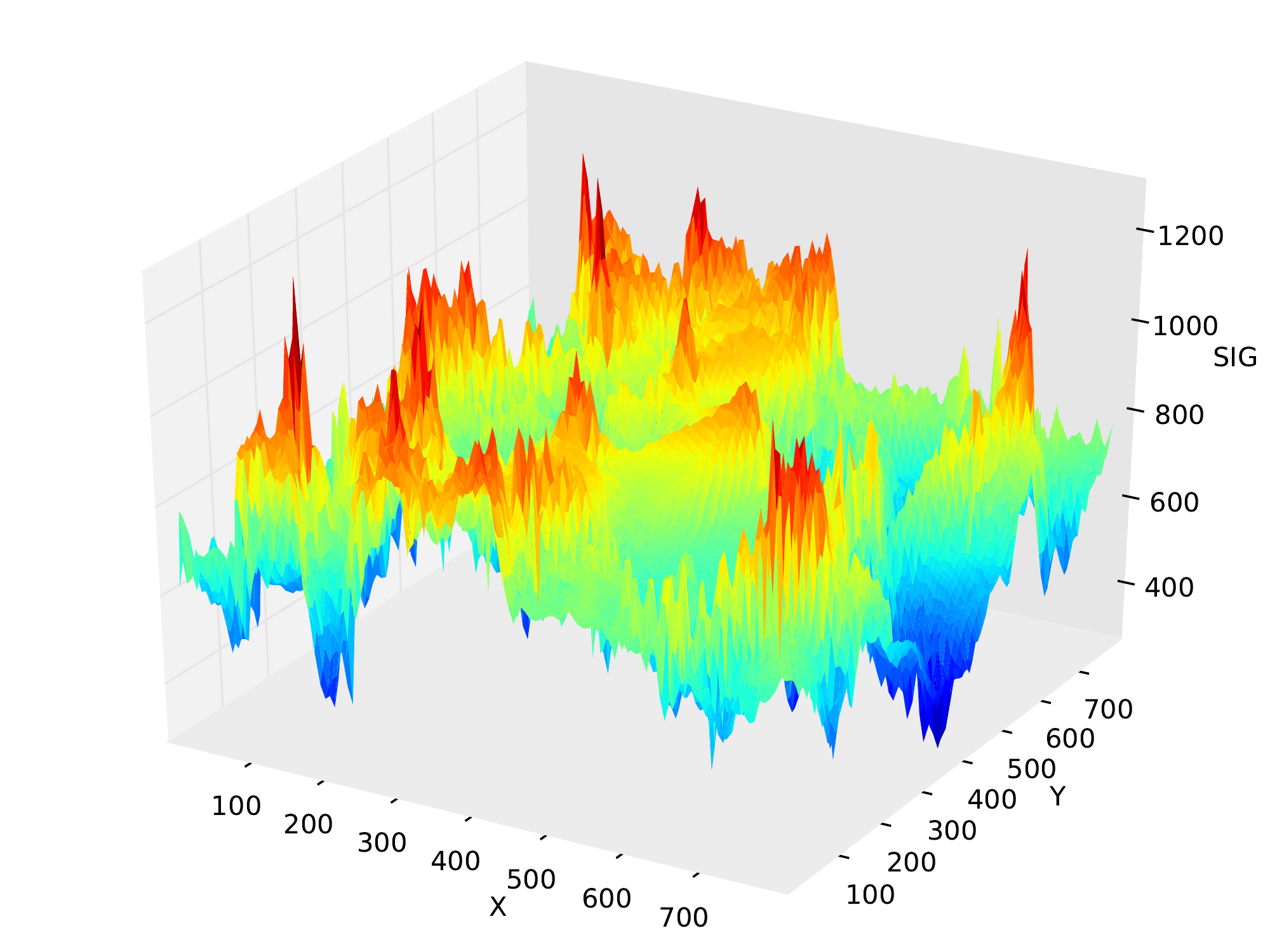}
      \includegraphics[width = 0.37\textwidth]{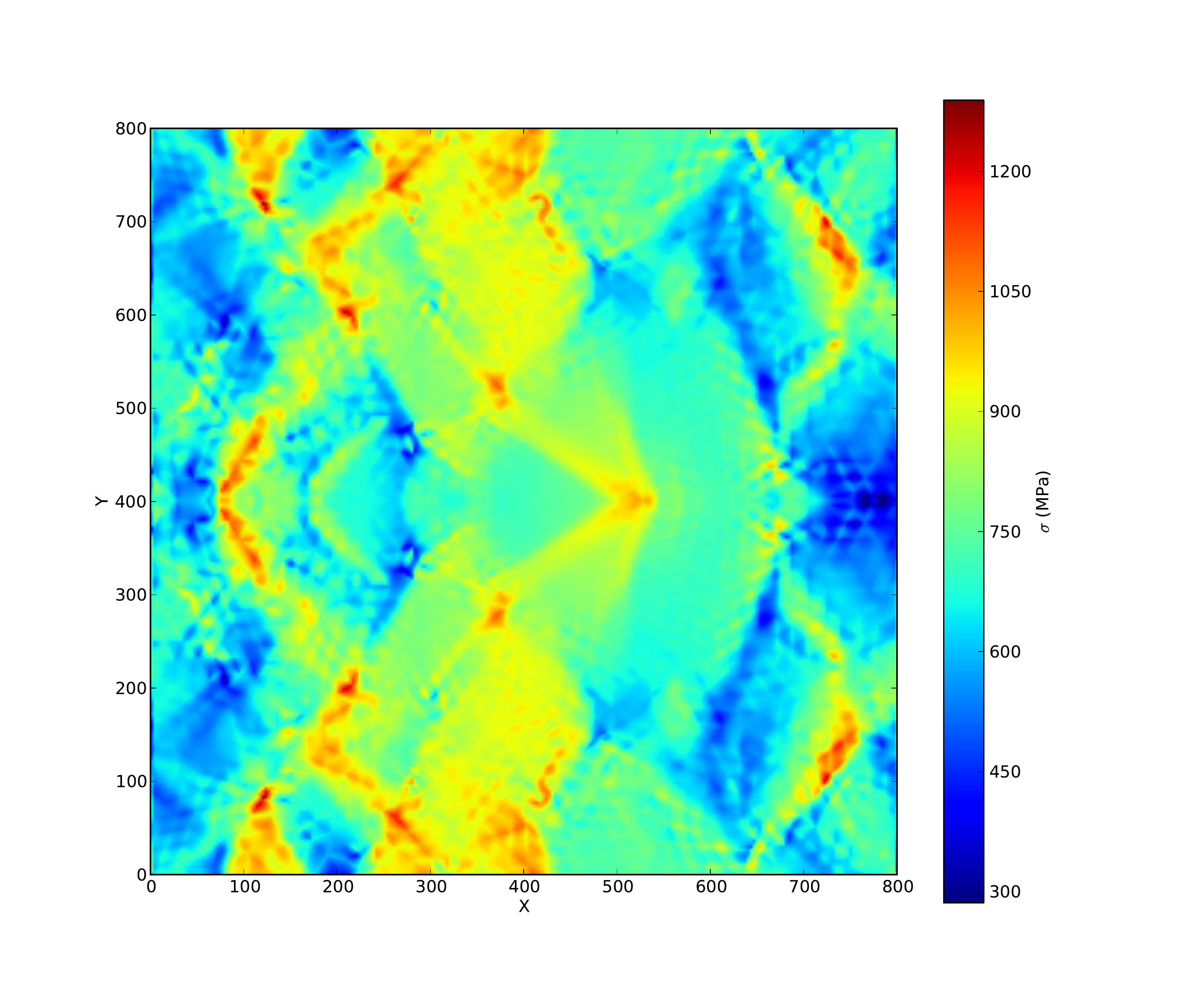}
      \caption{\label{fig-IV.10} A realization of the maximal principal
        stress field $\sigma_I$}
    \end{center}
\end{figure}

The input data of the identification problem is the maximal principal
stress field $\sigma_I$ obtained from the finite element calculations.
As the periodogram method is based on a regular sampling of the random
field under consideration, the brute result (\ie the maximal principal
stress at the nodes of the mesh) has to be projected onto a regular
grid. This operation is carried out using internal routines of
Code\_Aster. Note that a slice of width $100$ (\ie $10\%$ of total size)
is discarded along the edges of the aggregate in order to avoid the
effect of boundary conditions on the computed stress field, as suggested
in \citet{Mathieu:2006a}. A typical maximal principal stress field is
shown in Figure~\ref{fig-IV.10}.

\subsection{Check of the homogeneity of the field}
\label{section-4.2}
As it was described in Section~2 the periodogram method assumes that the
random field under consideration is {\em homogeneous}. From the
available realizations $SIG_i(x,y), i=1 \enu 35$ one first checks
empirically this assumption using the following approach:
\begin{itemize}
\item The {\em ensemble mean and variance} of the field is computed
  point-by-point throughout the grid for an increasing number of
  realizations $K=2\enu 35$:
\begin{eqnarray}
   \mu_K(x,y) & =& \frac{1}{K} \sum\limits_{i=1}^{K} SIG_i(x,y)  \label{eq-IV.25}\\
   \sigma^2_K(x,y)  & = &\frac{1}{K-1} \sum\limits_{i=1}^{K} \prt{
     SIG_i(x,y) - \mu(x,y) }^2 \label{eq-IV.26} 
\end{eqnarray}
If the field is homogeneous these quantities should tend to constant
values that are independent from the position $(x,y)$ when $K$ tends to
infinity.
\item In order to measure the magnitude of the spatial fluctuation of
  the latter, the {\em spatial average} and {\em spatial variance} of a
  realization of a field $Z(x,y)$ sampled onto a $N \times M$ grid is
  defined by:
\begin{eqnarray}
  \bar{\mu}_Z & = &\frac{1}{NM} \sum\limits_{i=1}^{N} \sum\limits_{j=1}^{M} Z(x_i,y_j)  \label{eq-IV.27}\\
   \bar{\sigma}^2_Z & =& \frac{1}{NM} \sum\limits_{i=1}^{N}
   \sum\limits_{j=1}^{M} \prt{ Z(x_i,y_j) - \bar\mu_Z}^2      \label{eq-IV.28}
 \end{eqnarray}
 whereas the associated ``spatial'' coefficient of variation is defined by:
\begin{equation}
  \label{eq-IV.29}
  CV_Z = \frac{\bar{\sigma}_Z}{\bar{\mu}_Z}
\end{equation}
\item The spatial coefficient of variation of the ensemble mean and
  variance (Eqs.(\ref{eq-IV.25})-(\ref{eq-IV.26})) are computed and
  plotted as a function of $K$. If the underlying random field is
  homogeneous it is expected that the curves of $CV_{\mu_K}$ and
  $CV_{\sigma_K^2}$ converge to zero.
\end{itemize}

\subsection{Choice of theoretical periodograms and fitting}
\label{section-4.3}
From a visual inspection of the obtained empirical periodograms it
appears that a Gaussian or an exponential model of periodogram such as
those presented in Table~\ref{tab:01} may be consistent with the data.
However it appeared in the various analyses that the peak of the
periodogram is not always in zero. An {\em initial frequency} is thus
introduced which shifts the theoretical periodogram. Finally, due to
lack of fitting of the single-type periodogram (\eg Gaussian and
exponential), a combination thereof is also fitted. The most general
model finally reads:
 \begin{equation} \label{eq-IV.31}
  \begin{split}
    S(f_x, f_y) = & \sigma_1^2 \pi l_{x1} \mbox{exp} \left[ \pi^2
      l_{x1}^2 (f_x - f^{(1)}_{x0})^2 \right] l_{y1} \mbox{exp}
    \left[\pi^2 l_{y1}^2 (f_y - f^{(1)}_{y0})^2 \right] \\
    & + \sigma_2^2 \frac{2l_{x2}}{1 + 4\pi^2 l_{x2}^2 (f_x -
      f^{(2)}_{x0})^2} \frac{2l_{y2}}{1 + 4\pi^2 l_{y2}^2 (f_y -
      f^{(2)}_{y0})^2}
  \end{split}
\end{equation}
where $l_{x1}, l_{y1}, l_{x2}, l_{y2}$ are correlation lengths in each
direction $X$ and $Y$ (aniso-tropic field) for each component (1)
(Gaussian part) and (2) (exponential part). Similarly $f^{(1)}_{x0}, f^{(1)}_{y0},
f^{(2)}_{x0}, f^{(2)}_{y0}$ are initial shift frequencies.

Note that Eq.(\ref{eq-IV.31}) corresponds only to positive values of
$f_x,f_y$. The periodogram is then extended by symmetry for negative
frequencies. In terms of associated covariance models, the linear
combination of periodograms leads to a linear combination of covariance
models. The initial frequency shift in the periodogram leads to
oscillatory cosine terms in the covariance by inverse Fourier
transform:
\begin{equation} \label{eq-IV.32}
  \begin{split}
    C(h_x,h_y) = & \sigma_1^2\mbox{exp} \left[-(\frac{h_x^2}{l_{x1}^2} +
      \frac{h_y^2}{l_{y1}^2})\right] \mbox{cos}(2\pi
    f^{(1)}_{x0}h_x) \mbox{cos}(2\pi f^{(1)}_{y0}h_y)\\
    & + \sigma_2^2
    \mbox{exp}\left[-(\frac{|h_x|}{l_{x2}}+\frac{|h_y|}{l_{y2}}) \right]
    \mbox{cos}(2\pi f^{(2)}_{x0}h_x) \mbox{cos}(2\pi f^{(2)}_{y0}h_y)
  \end{split}
\end{equation}

In order to compare the various fittings the least-square residual
between the empirical periodogram $\bar{S}(f_x,f_y)$
(Eq.(\ref{eq-II.15})) and the fitted periodogram $S^{theor}(f_x,f_y)$ is
finally computed. The following non dimensional error estimate is used:
\begin{equation} 
  \label{eq-IV.33}
  \epsilon = \sqrt{\frac{1}{NM} \sum\limits_{i=1}^{N}
    \sum\limits_{j=1}^{M} \left[\bar{S}(f_{xi}, f_{yj}) -
      S^{theor}(f_{xi}, 
      f_{yj})\right]^2} / \max_{(f_x,f_y)}\bar{S}(f_x,f_y)
\end{equation}

\section{Results -- Case \#1: fixed grain geometry} 
\label{section-5}

\subsection{Check of the homogeneity}
\label{sec:5-1}

\begin{figure}[!ht]
    \begin{center}
      \includegraphics[clip, trim = 50mm 152mm 60mm 130 mm, width =0.48\textwidth]{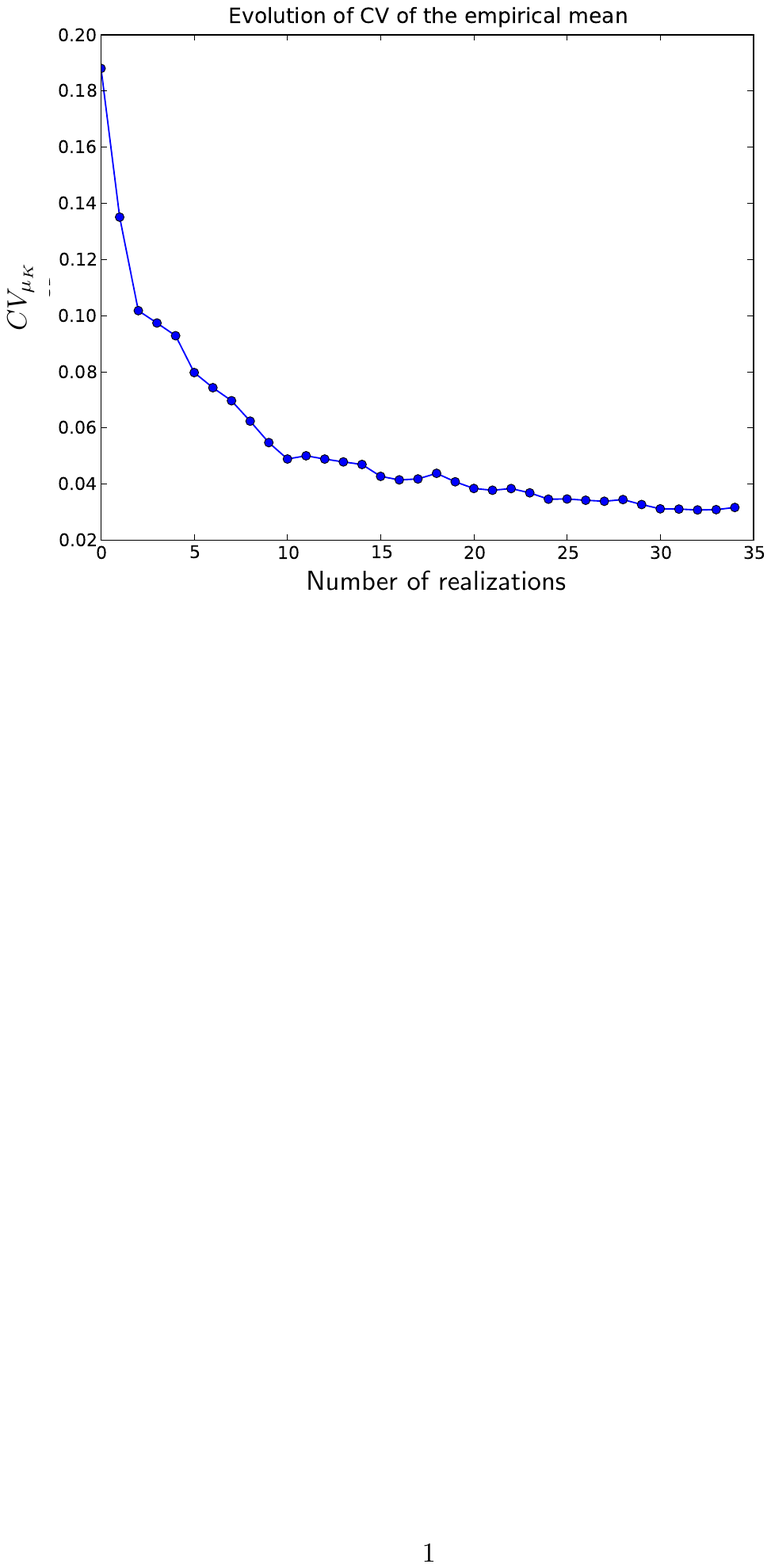}
      \includegraphics[clip, trim = 50mm 152mm 60mm 130 mm,width = 0.48\textwidth]{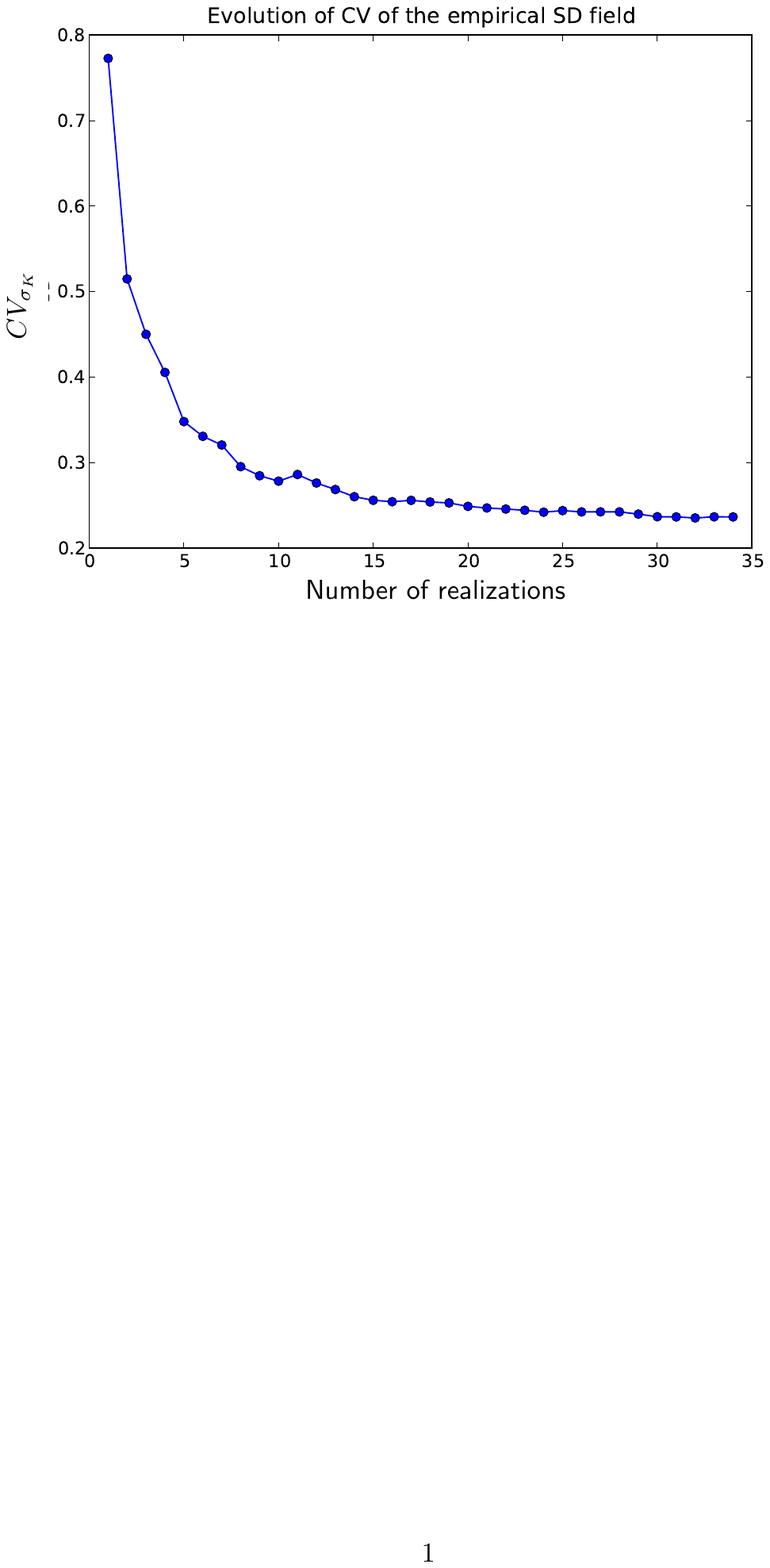}
      \caption{Case \#1: Evolution of $CV_{\mu_K}$ and $CV_{\sigma}^K$
        with respect to the number of realizations} \label{fig-IV.12}
    \end{center} 
\end{figure}

First the homogeneity of the maximal principal stress field is checked
using the methodology proposed in Section~\ref{section-4.2}.
Figure~\ref{fig-IV.12} shows the evolution of $CV_{\mu_K}$ and
$CV_{\sigma_K^2}$. These quantities regularly decrease and it is seen
that they would tend to zero if a larger number of realizations was
available. This leads to accepting the assumption
that the random field is homogeneous since the fluctuations around the
constant spatial average tend to zero when $K$ increases.

\subsection{Identification of periodograms at 3.5\% macroscopic
strain}
\label{sec:5-2}
\begin{figure}[!ht]
    \begin{center}
      \includegraphics[width=0.48\textwidth]{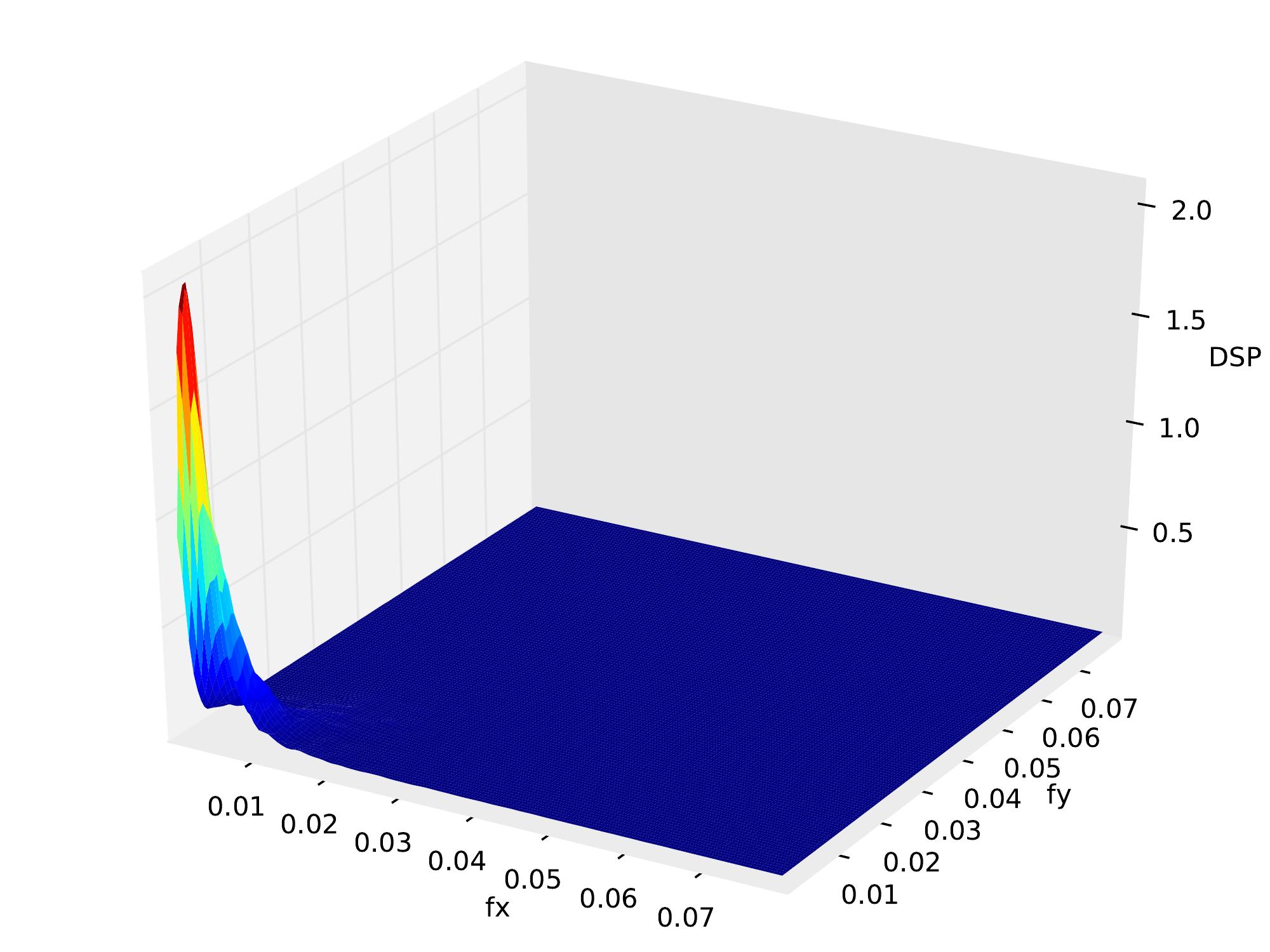}
      \includegraphics[width=0.48\textwidth]{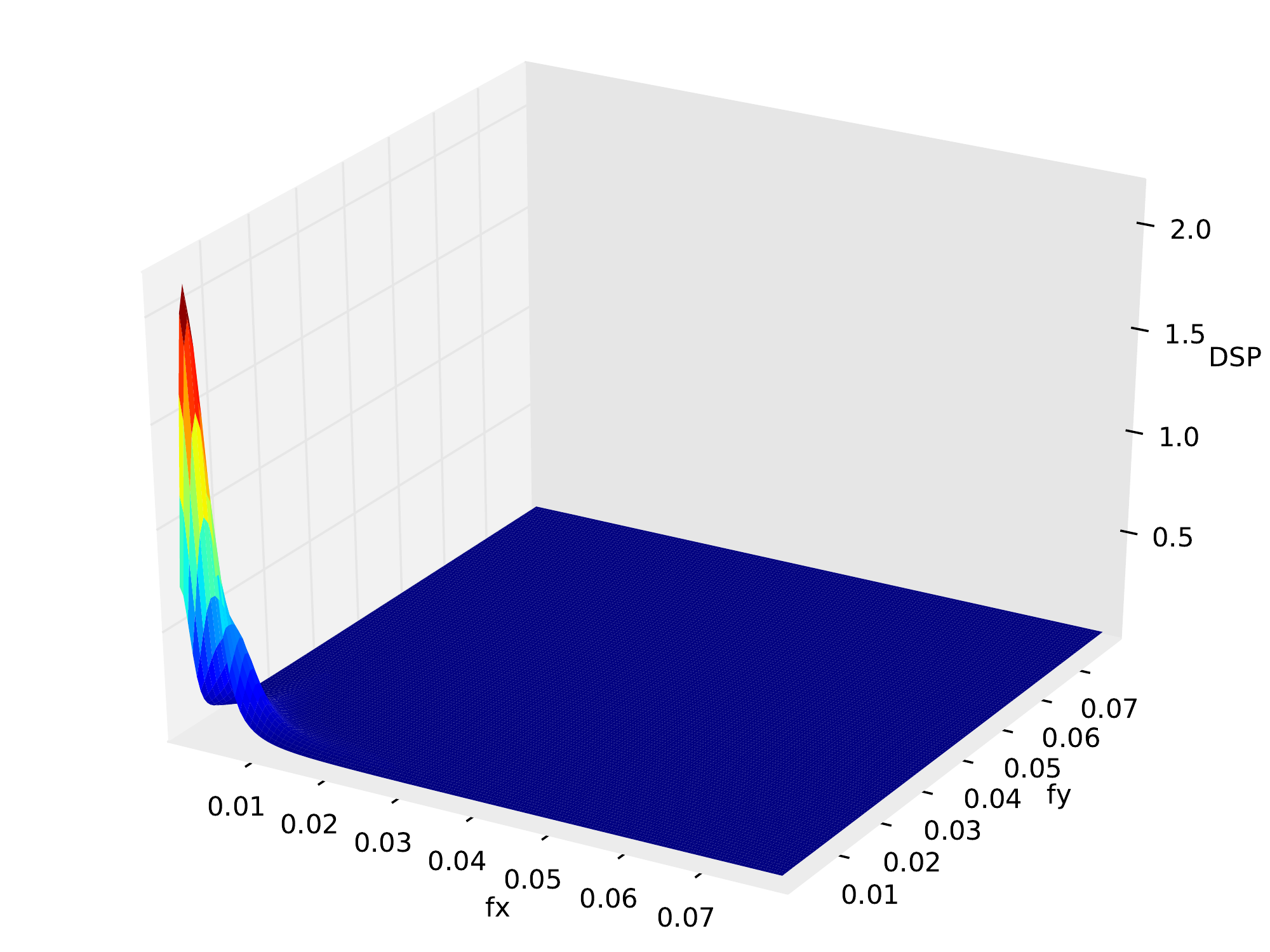}
      \caption{ Case \#1: (a) Average empirical periodogram of the
        stress field at $3.5\%$ macroscopic strain -- (b) best fitted
        periodogram}\label{fig-IV.13}
    \end{center}
\end{figure}

The average empirical periodogram obtained from $L=35$ realizations of
the maximal principal stress field $\sigma_I$ at 3.5\% of macroscopic
strain is plotted in Figure~\ref{fig-IV.13}-a.

Table \ref{tab:05} presents the results of the fitting of the average
empirical periodogram calculated from 35~realizations of the field using
three models, namely Gaussian, exponential and a mixed ``Gaussian +
exponential'' as in Eq.(\ref{eq-IV.31}).

\begin{table}[!ht] 
  \caption{Fitted parameters and error estimates  for the three fitted models:
    Gaussian, exponential and mixed ``Gaussian + exponential''} \label{tab:05} 
  \begin{tabular}{cccccccccccc}
     \hline\noalign{\smallskip}
    \multirow{2}{*}{Model} & $\epsilon$ & \multicolumn{5}{c}{Gaussian} & \multicolumn{5}{c}{Exponential} \\
    \cline{3-12}
    & (Eq.(\ref{eq-IV.33})) & $\sigma_1$ & $l_{x1}$ & $l_{y1}$ & $f^{(1)}_{x0}$ & $f^{(1)}_{y0}$ & $\sigma_2$ & $l_{x2}$ & $l_{y2}$ &
    $f^{(2)}_{x0}$  &   $f^{(2)}_{y0}$ \\
    \hline
    Gaussian & $0.0043$ & $69.4$ & $104.6$ & $102.9$ & $0.00287$ & $0$ & $\_$ & $\_$ & $\_$ & $\_$ & $\_$\\
    Exponential & $0.0039$ & $\_$ & $\_$ & $\_$ & $\_$ & $\_$ & $84.2$ & $73.8$ & $87.5$ & $0.00275$ & $0$\\
    Mixed & $0.0017$ & $54.7$ & $138.4$ & $159.1$ & $0.00244$ & $0$ & $57.6$ & $57.5$ & $63.5$& $0.00562$ & $0.0028$\\
    \noalign{\smallskip}\hline
  \end{tabular}
\end{table}

From the results in Table~\ref{tab:05} it appears that the mixed model
provides a significantly smaller least-square error than that obtained
from the Gaussian and exponential models respectively. The corresponding
fitted periodogram is plotted in Figure~\ref{fig-IV.13}-b. 

\begin{figure}[!ht]
    \begin{center}
      \includegraphics[width=0.48\textwidth]{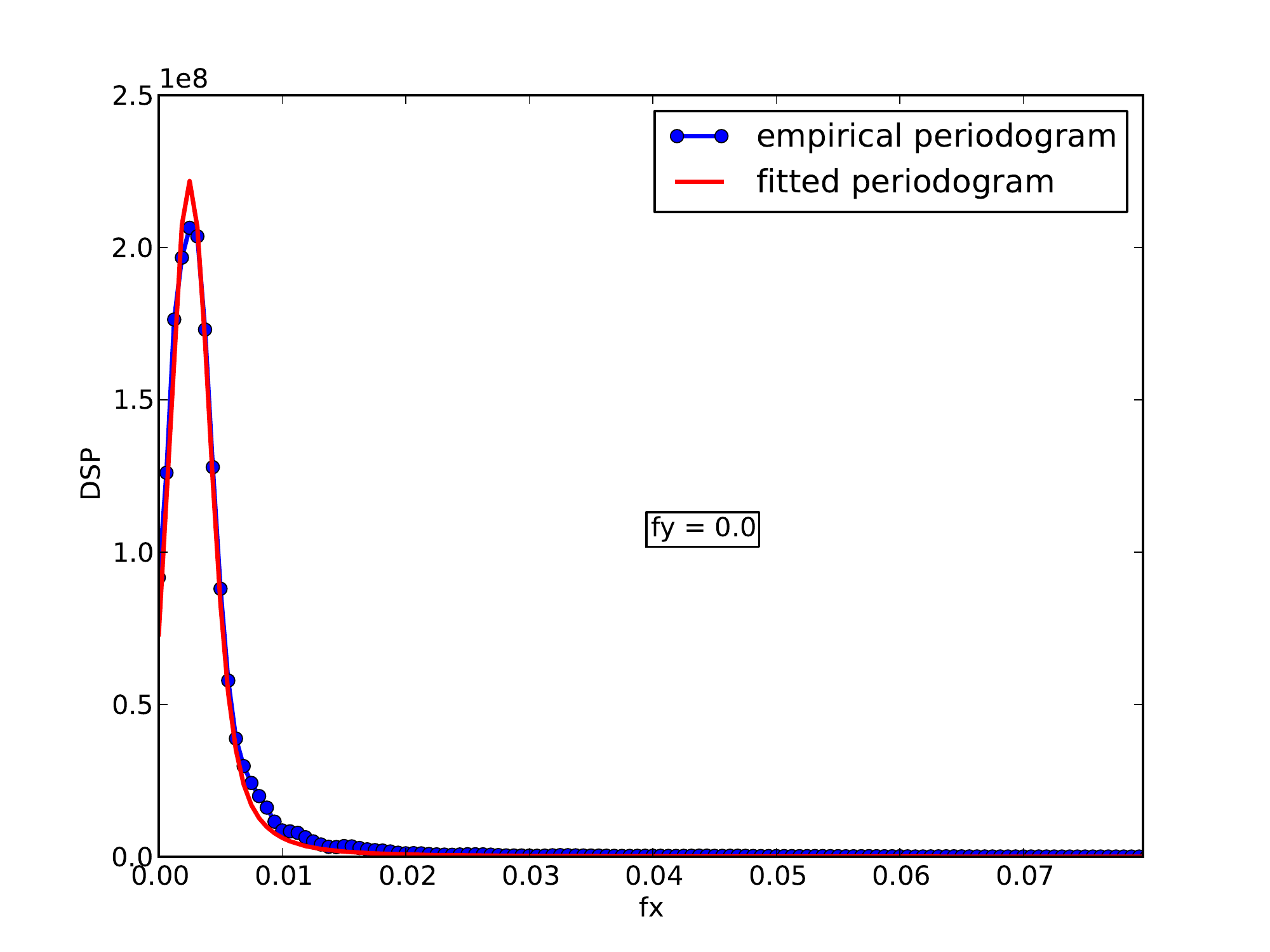}
      \includegraphics[width=0.48\textwidth]{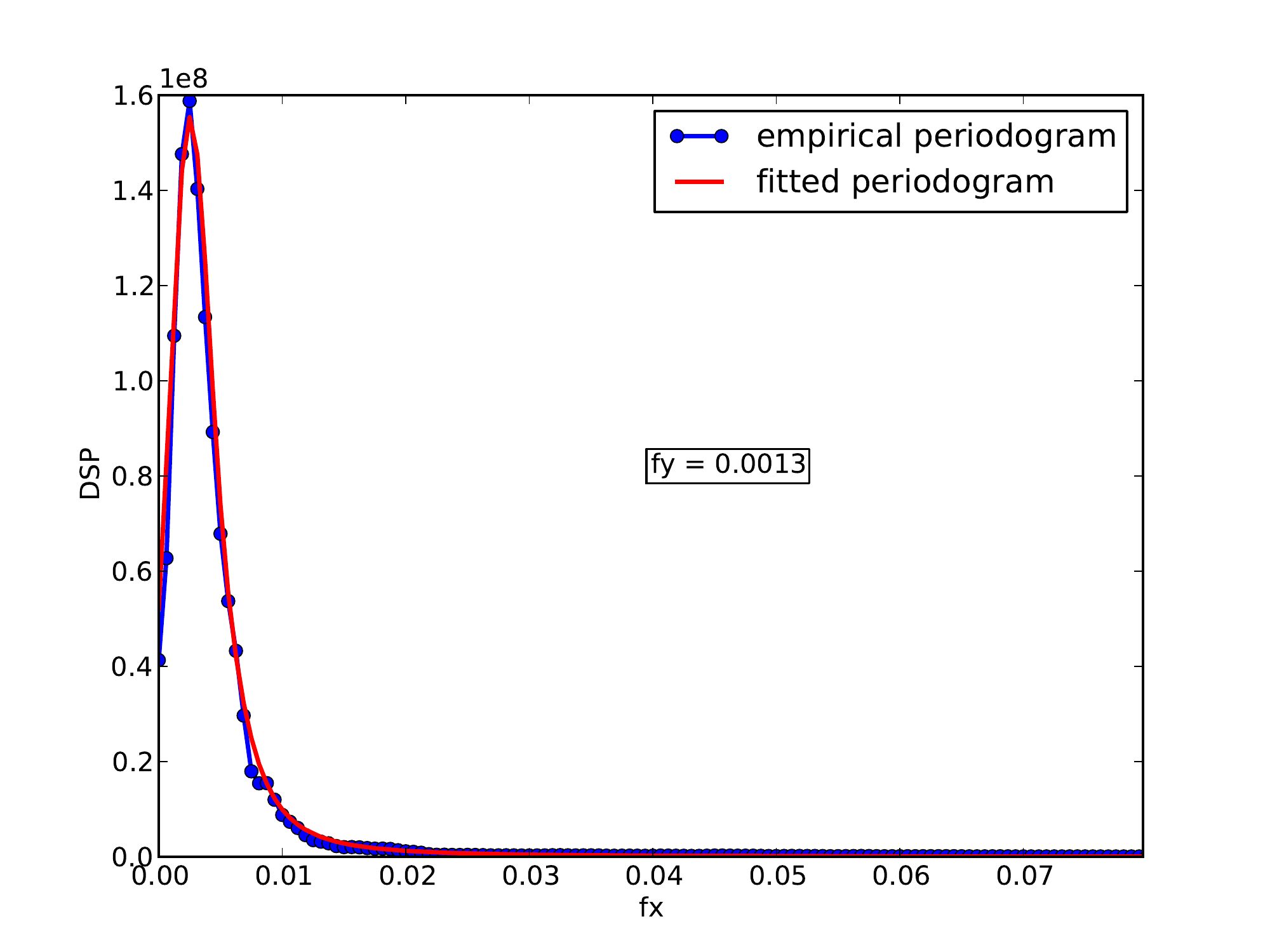}
      \caption{Case \#1: Cut of the periodograms in the $X$ direction}\label{fig-IV.16}
    \end{center}
\end{figure}

\begin{figure}[!ht]
    \begin{center}
      \includegraphics[width=0.48\textwidth]{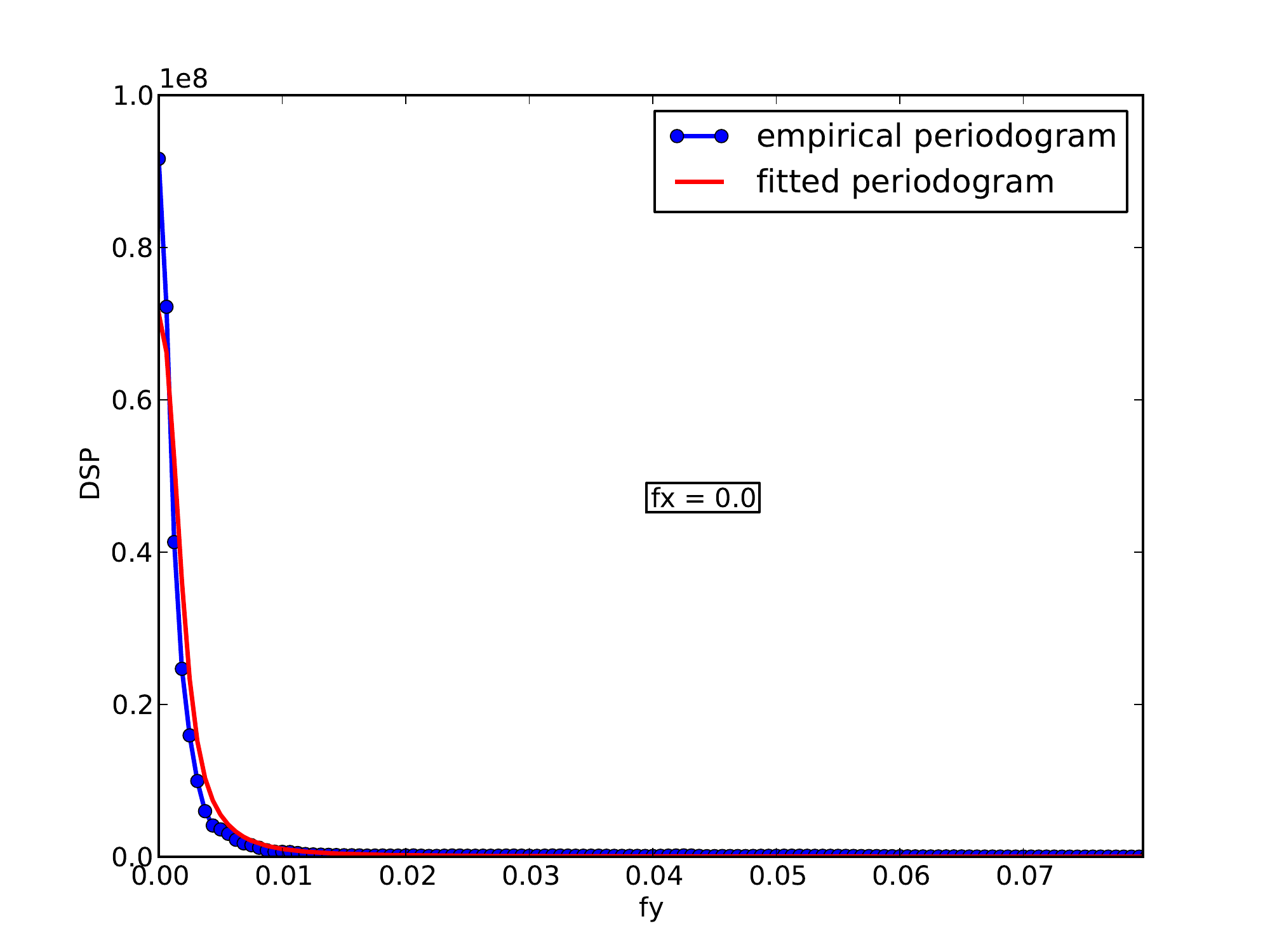}
      \includegraphics[width=0.48\textwidth]{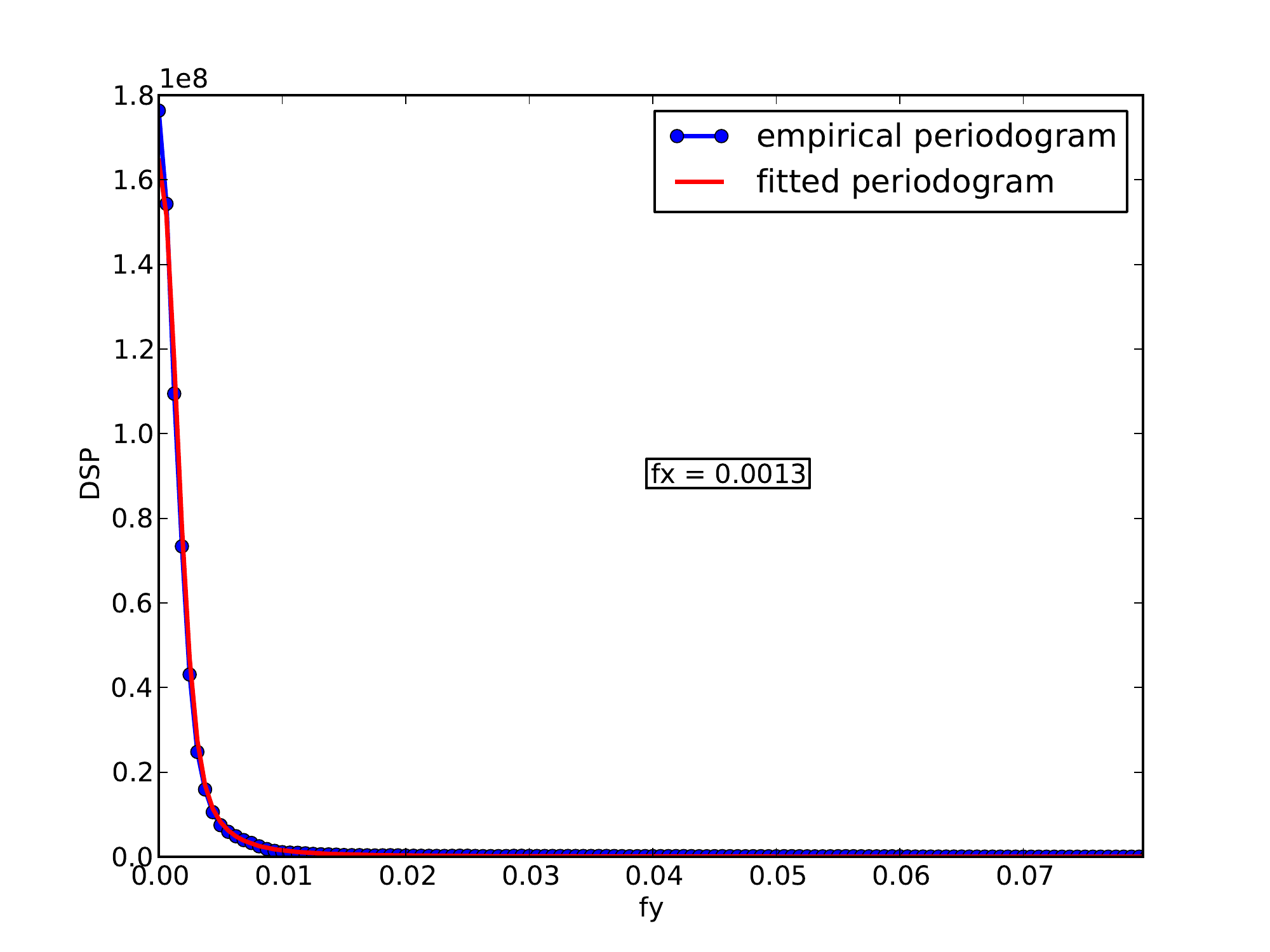}
      \caption{Case \#1: Cut of the periodograms in the $Y$ direction}
      \label{fig-IV.17}
    \end{center}
\end{figure}

\begin{figure}[!ht]
    \begin{center}
      \includegraphics[width=0.48\textwidth]{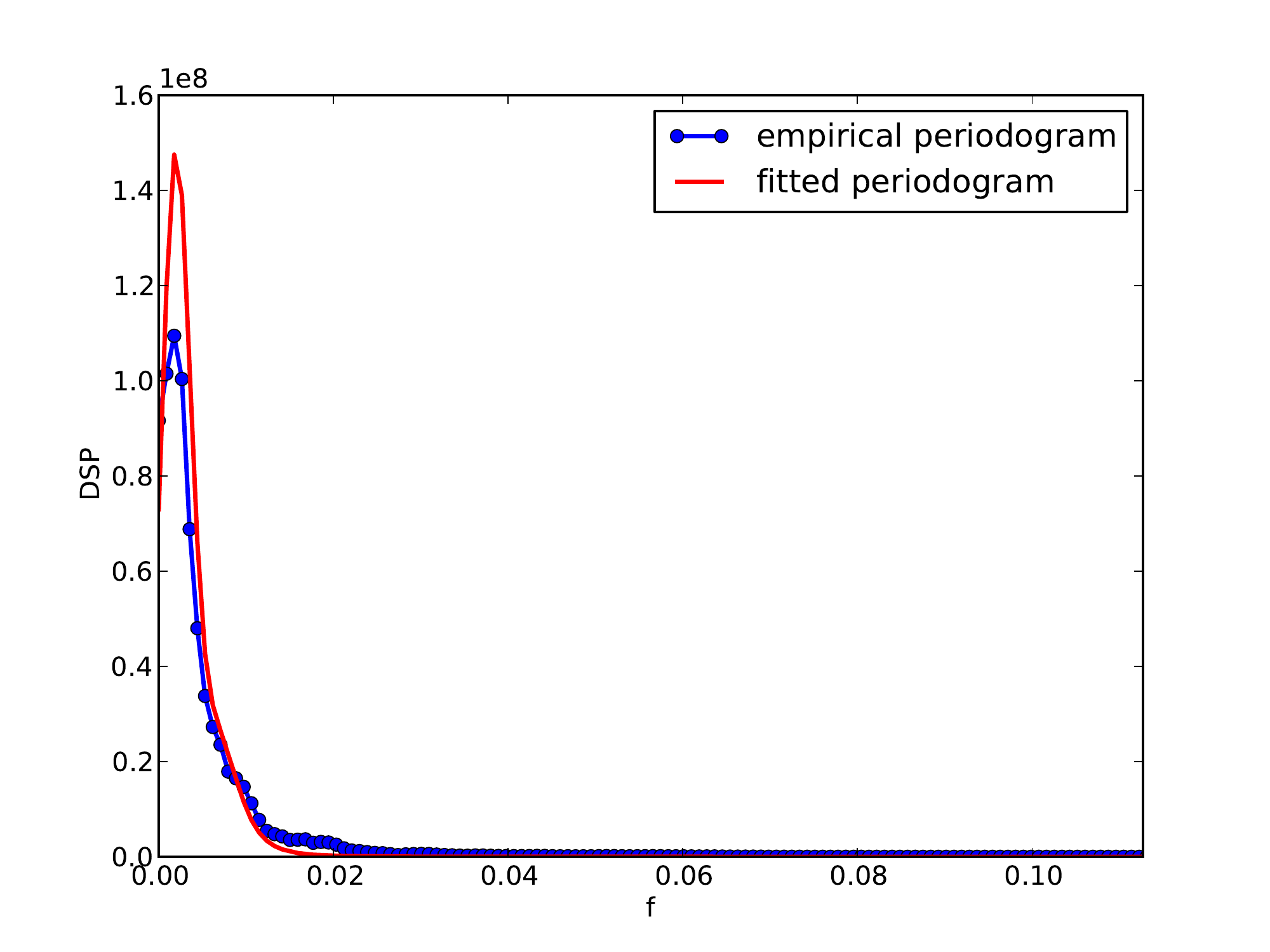}
      \caption{ Case \#1: Cut of the periodograms along
        the diagonal $f_x = f_y$} \label{fig-IV.15}
    \end{center}
\end{figure}

In order to better appreciate the quality of the fitting,
two-dimensional cuts of the empirical (resp. fitted) periodogram are
given in Figures~\ref{fig-IV.16}--\ref{fig-IV.15}. Figure~\ref{fig-IV.16} corresponds
to a cut along the $X$ direction for two values of $f_y =0 \,;\, 0.0013$.
Figure~\ref{fig-IV.17} corresponds to a cut along the $Y$ direction for two
values of $f_x =0 \,;\, 0.0013$. Finally Figure~\ref{fig-IV.15} corresponds to a
cut along the diagonal $f_x=f_y$.

From the above figures it appears that the fitting of the empirical
periodogram by a mixed model is remarkably accurate. It is interesting
to interpret the fitted parameters reported in Table~\ref{tab:05}. First
it is observed that the amplitude of each component of the mixed
periodogram is similar since $\sigma_1 \approx \sigma_2$. The variance
of the field is equal to $\sigma_1^2 + \sigma_2^2 \approx$
6,309~MPa${}^2$. The associated standard deviation is 79.4~MPa. As the
mean principal stress is 720~MPa at 3.5\% macroscopic strain, the
coefficient of variation of the field is about 11\%.

In order to interpret the correlation length parameters let us define
the mean size of a grain $S_g$ such a two-dimensional aggregate. As the
volume of edge length equal to 1,000 corresponds to 100~grains, the
equivalent diameter of a single grain reads:
\begin{equation} 
  \label{eq-IV.34}
  D_g = \sqrt{\frac{4}{\pi}S_{g}} = \sqrt{\frac{4}{\pi} \frac{ 1,000 \times 1,000}{100}} = 112.8
\end{equation}
Thus the correlation lengths obtained from the fitting vary from 0.55 to
1.3$D_g$. This shows that the characteristic dimension of the underlying
microstructure (\ie $D_g$) is of the same order of magnitude as these
parameters. In other words the scale of local fluctuation of the stress
field is related to the grain size, as heuristically expected. Moreover,
it appears that the lengths in the $X$ and $Y$ directions are almost
identical. The stress field does not show any significant anisotropy in
this case.

\subsection{Influence of the number of realizations}
\label{sec:5-3}

\begin{figure}[!ht]
    \begin{center}
      \includegraphics[clip, trim = 50mm 152mm 60mm 130 mm,width = 0.48\textwidth]{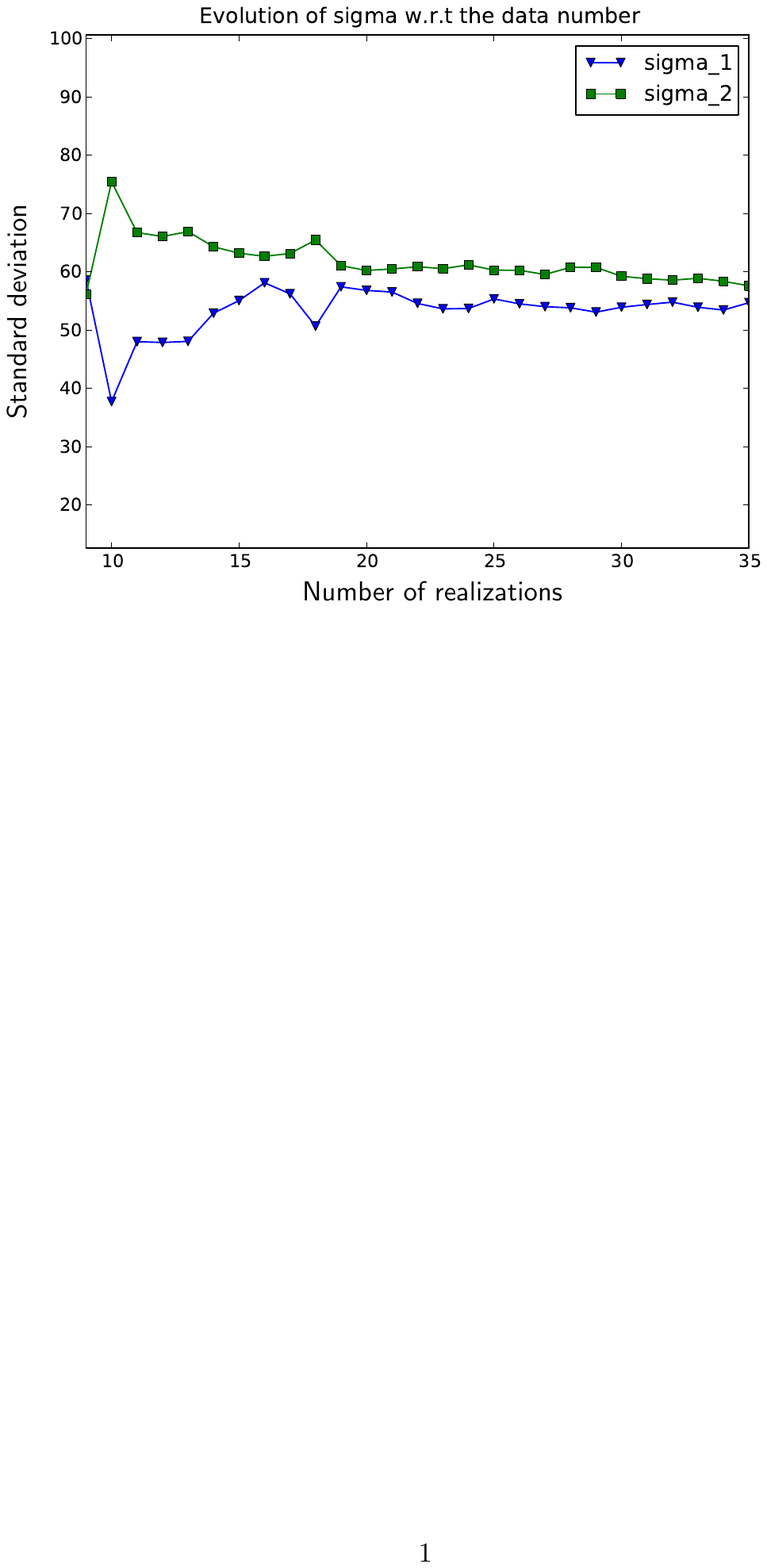}
      \caption{Case \#1: Evolution of the fitted standard deviations
        with respect to the number of realizations $K=8 \enu 35$}
      \label{fig-IV.18}
    \end{center}
\end{figure}

\begin{figure}[!ht]
    \begin{center}
      \includegraphics[clip, trim = 50mm 152mm 60mm 130 mm,width = 0.48\textwidth]{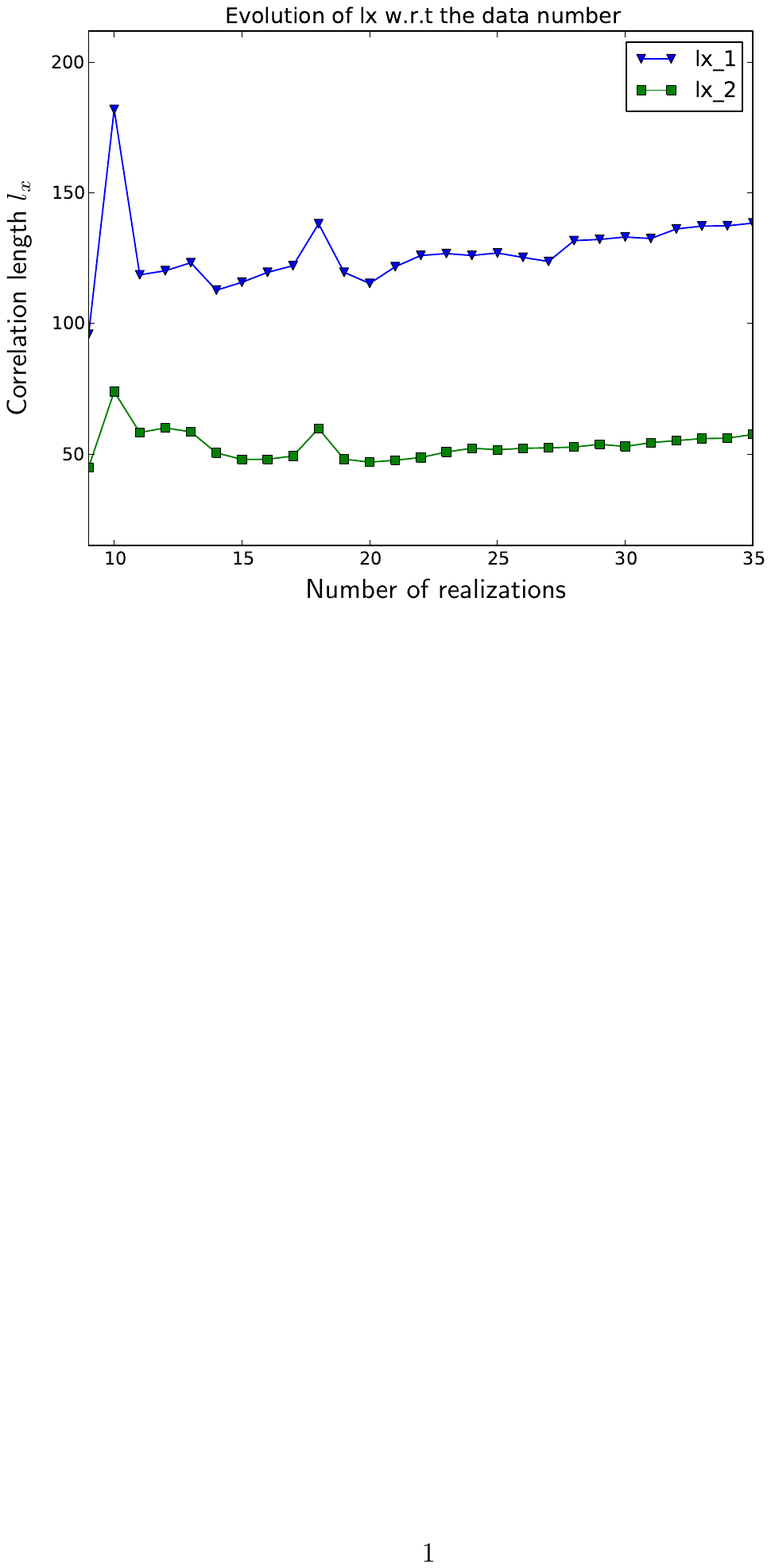}
      \includegraphics[clip, trim = 50mm 152mm 60mm 130 mm,width = 0.48\textwidth]{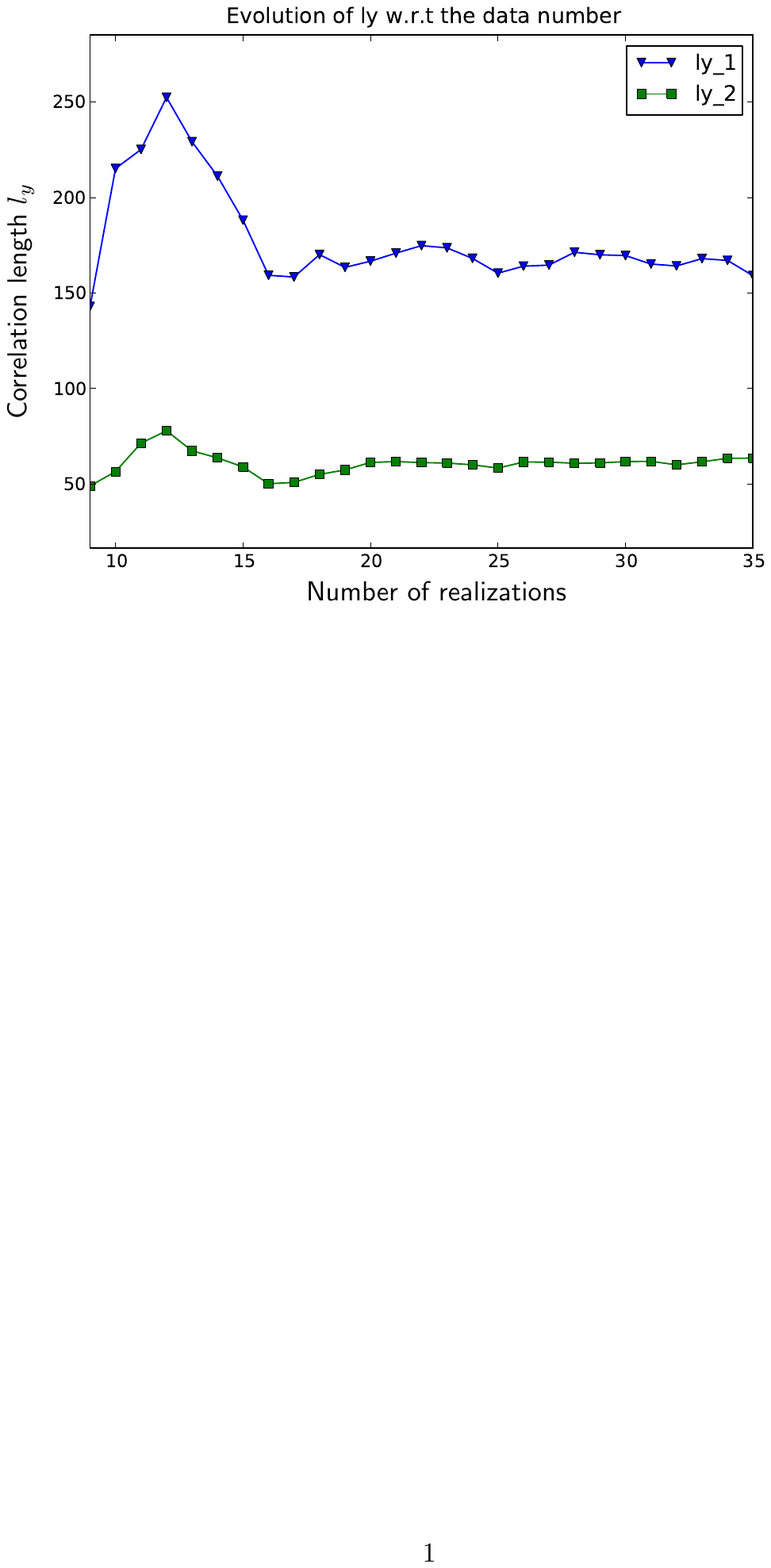}
      \caption{Case \#1: Evolution of the fitted correlation lengths in
        the $X, Y$ directions with respect to the number of realizations
        $K=8 \enu 35$} 
      \label{fig-IV.19}  
    \end{center}
\end{figure}

\begin{figure}[!ht]
    \begin{center}
      \includegraphics[clip, trim = 50mm 152mm 60mm 130 mm,width = 0.48\textwidth]{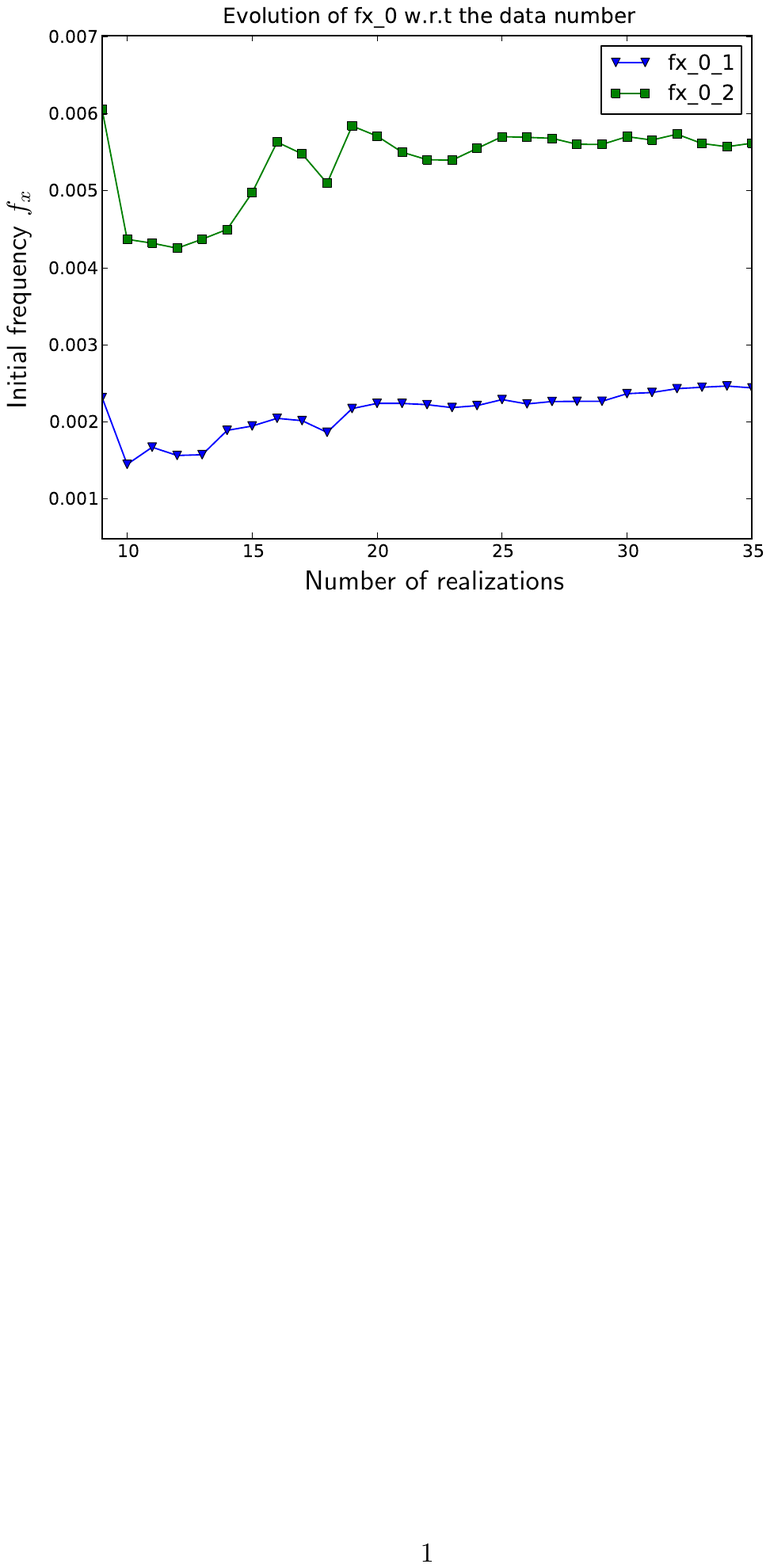}
      \includegraphics[clip, trim = 50mm 152mm 60mm 130 mm,width = 0.48\textwidth]{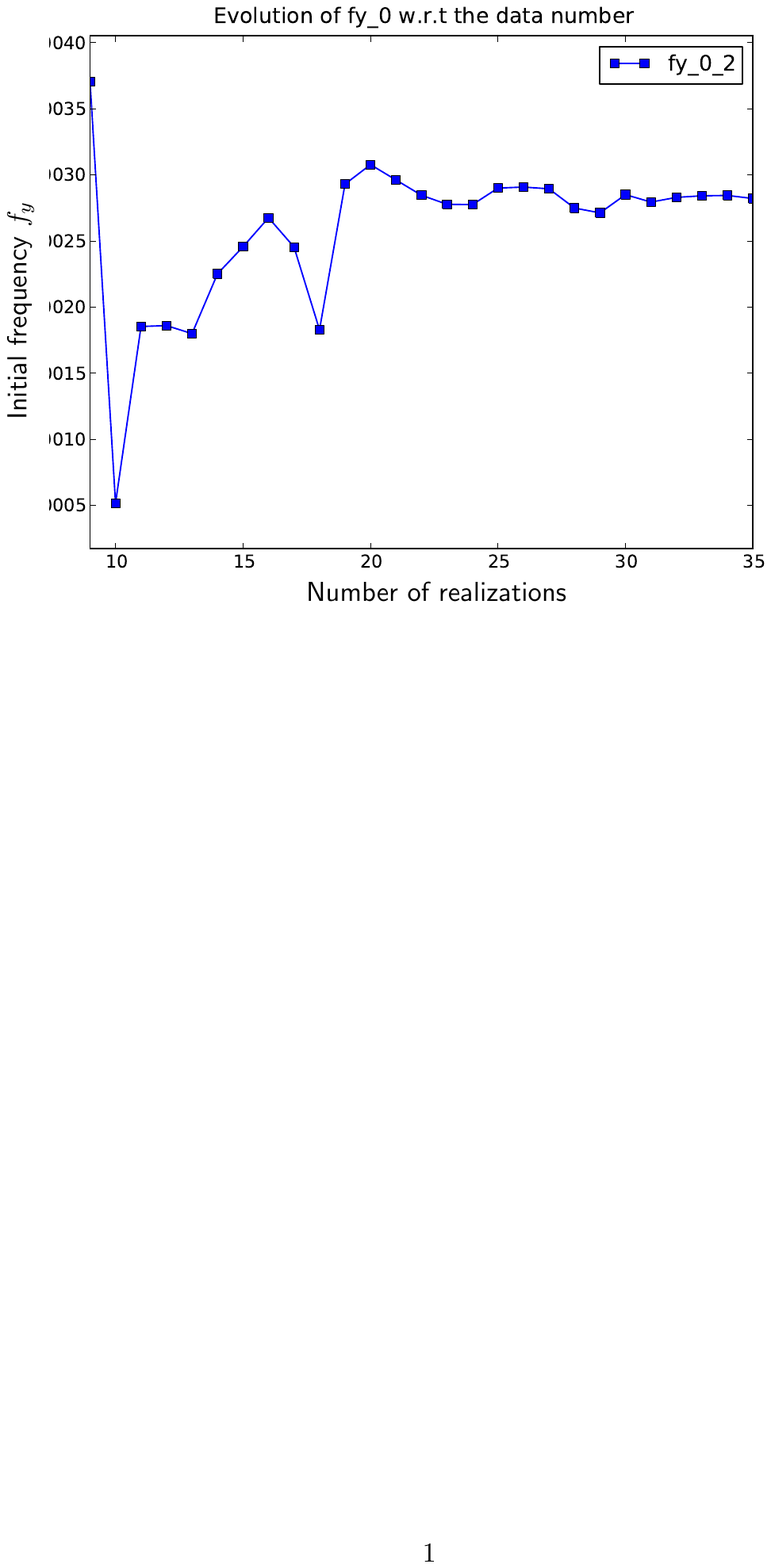}
      \caption{Case \#1: Evolution of the fitted initial frequency in
        the $X, Y$ directions with respect to the number of realizations
        $K=8 \enu 35$} \label{fig-IV.20}
    \end{center}
\end{figure}

In this section the stability of the fitted parameters as a function of
the number of available realizations $K$ used in the average periodogram
method is considered. In practice the procedure applied in the previous
paragraph is run using $K=8,9 \enu 35$ realizations of the stress field.
The evolution of the standard deviations $(\sigma_1, \sigma_2)$ is shown
in Figure~\ref{fig-IV.18}. The evolution of the correlation lengths
$l_{(x,y)(1,2)}$ is shown in Figure~\ref{fig-IV.19}.  The evolution of
the initial frequencies $f_{(x,y)_0}^{(1,2)}$ is shown in
Figure~\ref{fig-IV.20}.

From these figures it appears that the fitted parameters tend to a
converged value when at least 20~realizations of the stress field are
used for their estimation. 

\subsection{Influence of the macroscopic strain level}
\label{sec:5-4}
In this section the evolution of the parameters of the fitted
periodograms as a function of the macroscopic strain is investigated.
For this purpose the methodology presented in Section~\ref{sec:5-2} is
applied using the realizations of the maximal principal stress fields
corresponding to various levels of the loading curve, \ie various values
of the equivalent macroscopic strain $E_{YY} = 0. \enu 3.5\%$.

\begin{figure}[!ht]
    \begin{center}
     \includegraphics[clip, trim = 50mm 152mm 60mm 130 mm,width = 0.48\textwidth]{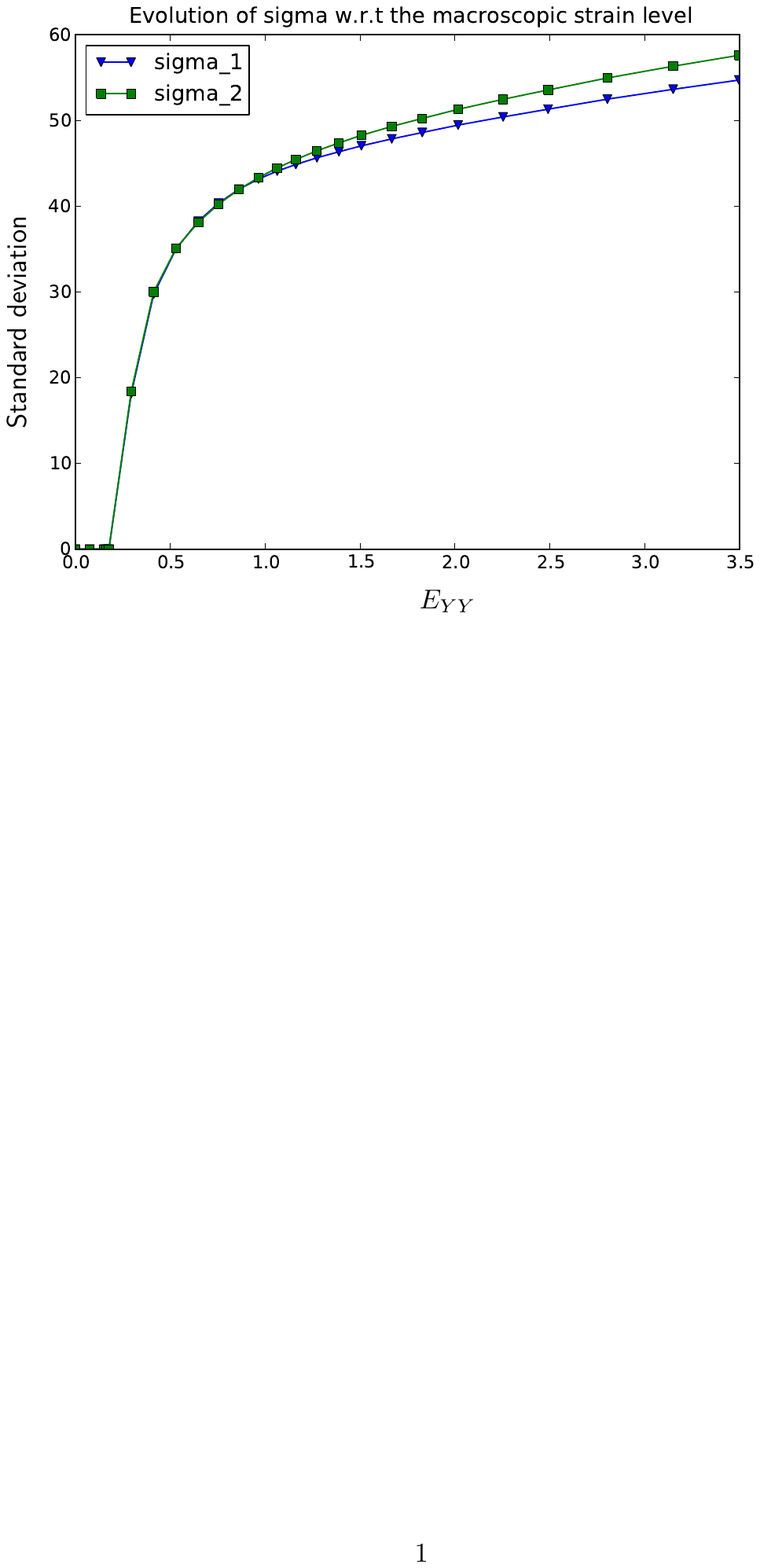}
     \caption{Case \#1: Evolution of the fitted standard deviations
        with respect to the load level (macroscopic strain $E_{YY} =
        0. \enu 3.5\%$)}
      \label{fig-IV.18a}
    \end{center}
\end{figure}

The evolution of the standard deviations $(\sigma_1, \sigma_2)$ is shown
in Figure~\ref{fig-IV.18a}. The two components of the periodogram (\eg
Gaussian and exponential) contribute for approximately the same
proportion to the total variance of the field since the curves are
almost superimposed. Note that these standard deviations increase with
the applied load in the same way as the mean load curve
(Figure~\ref{fig-III.6}).

\begin{figure}[!ht]
    \begin{center}
      \includegraphics[clip, trim = 50mm 152mm 60mm 130 mm,width = 0.48\textwidth]{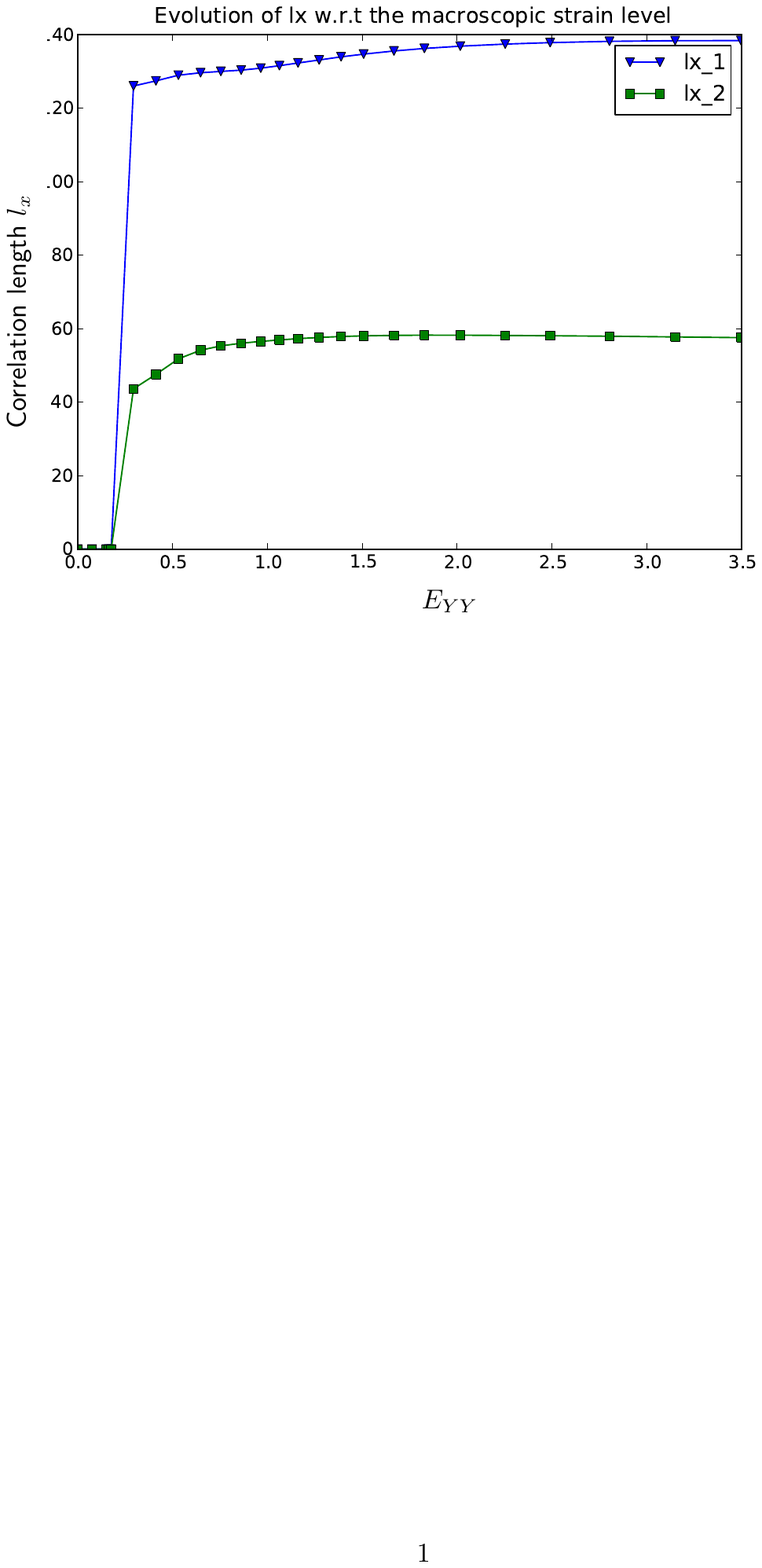}
      \includegraphics[clip, trim = 50mm 152mm 60mm 130 mm,width = 0.48\textwidth]{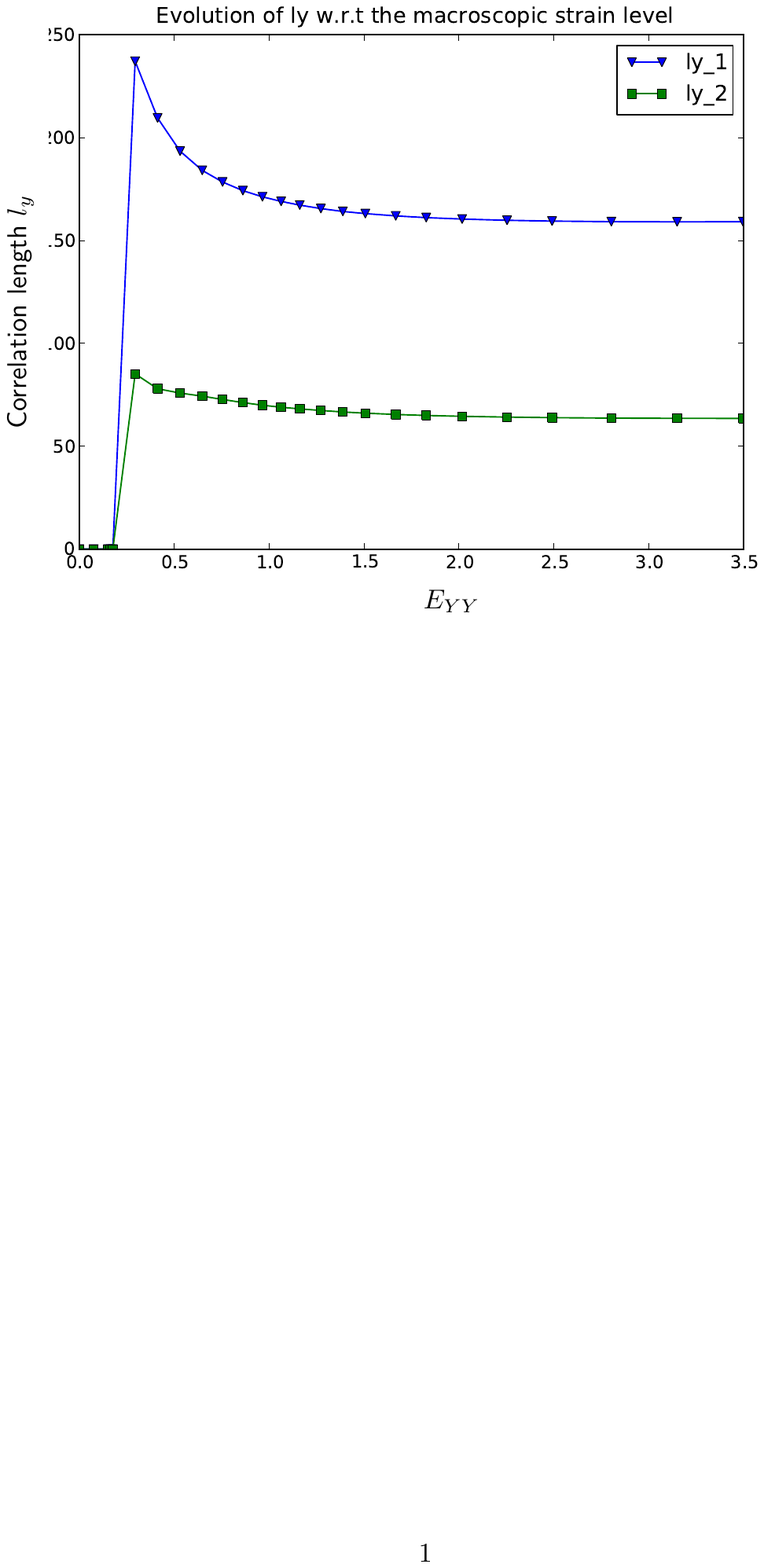}
      \caption{Case \#1: Evolution of the fitted correlation lengths in
        the $X, Y$ directions with respect to the load level
        (macroscopic strain $E_{YY} = 0. \enu 3.5\%$)}
      \label{fig-IV.19a}  
    \end{center}
\end{figure}

\begin{figure}[!ht]
    \begin{center}
      \includegraphics[clip, trim = 50mm 152mm 60mm 130 mm,width = 0.48\textwidth]{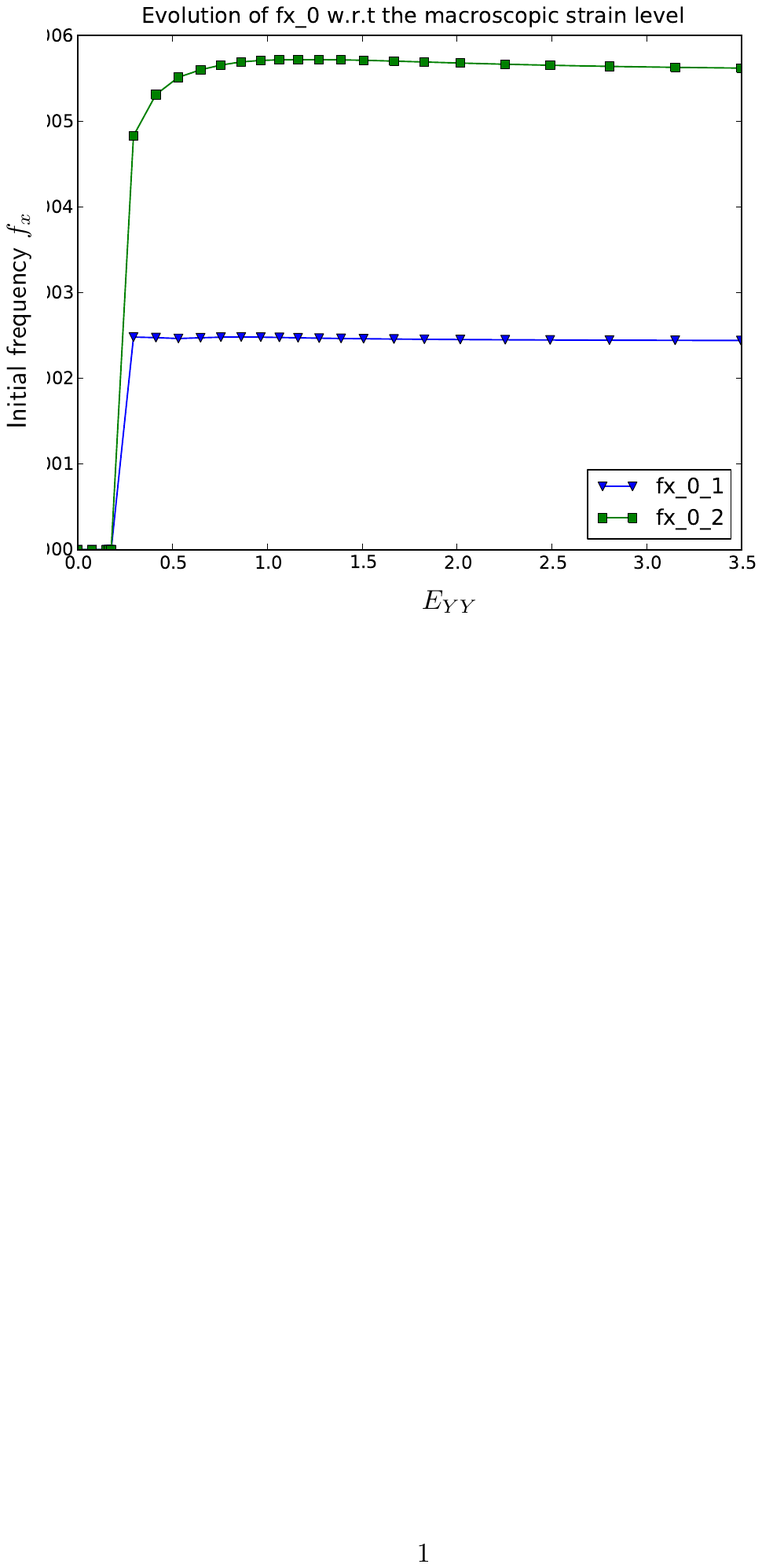}
      \includegraphics[clip, trim = 50mm 152mm 60mm 130 mm,width = 0.48\textwidth]{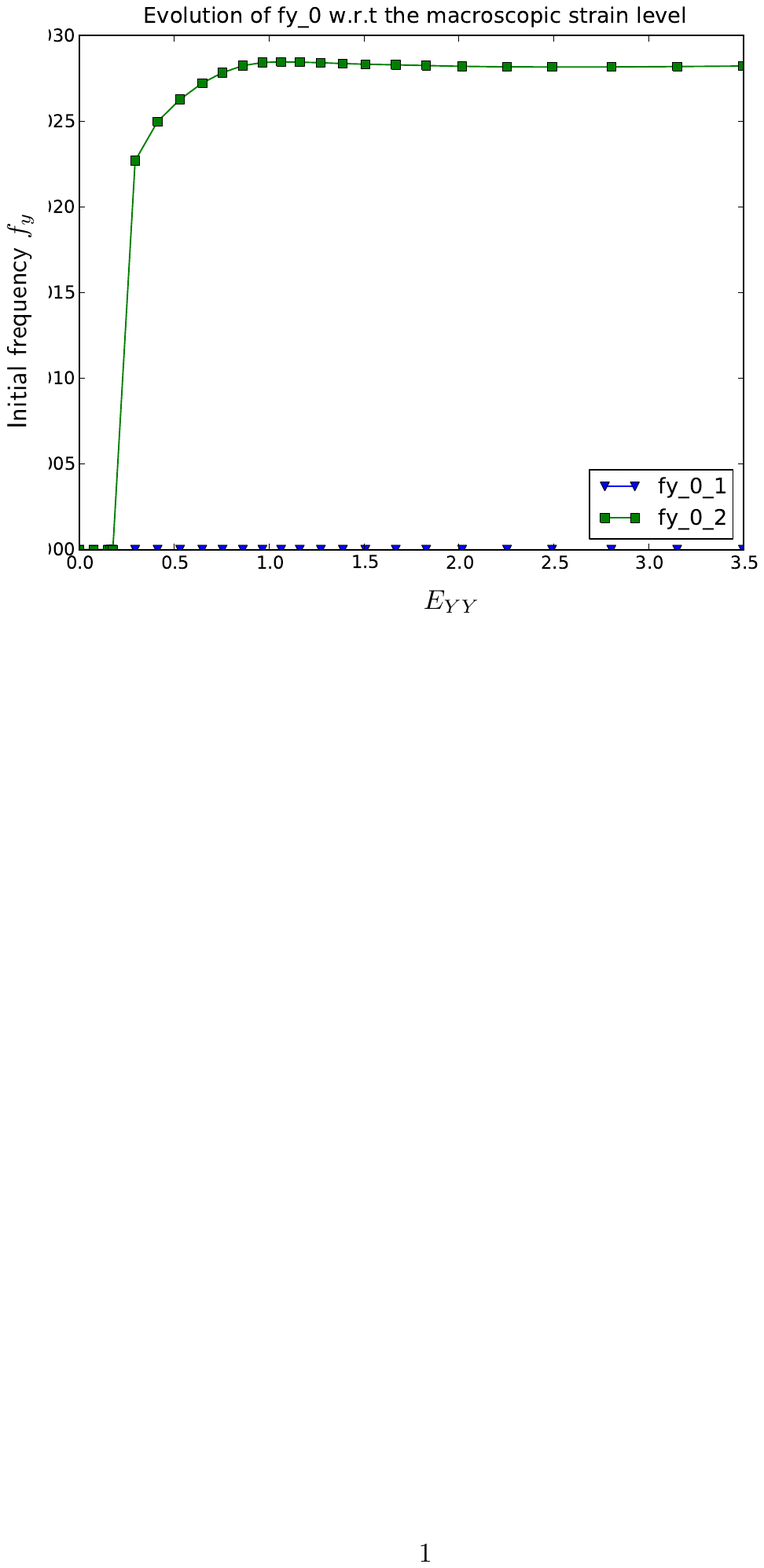}
      \caption{Case \#1: Evolution of the fitted initial frequency in
        the $X, Y$ directions with respect to the load level
        (macroscopic strain $E_{YY} = 0. \enu 3.5\%$)}
      \label{fig-IV.20a}
    \end{center}
\end{figure}

The evolution of the correlation lengths $l_{(x,y)(1,2)}$ is shown in
Figure~\ref{fig-IV.19a}.  The evolution of the initial frequencies
$f_{(x,y)_0}^{(1,2)}$ is shown in Figure~\ref{fig-IV.20a}.  It is
observed that once plasticity is settled (\ie once the macroscopic
strain $E_{YY}$ is greater than $\sim 0.5\%$) the parameters describing
the fluctuations of the maximal principal stress field are almost
constant. This conclusion is valid for both the correlation lengths and
the initial frequencies. Note that the convergence is faster for the
parameters related to the $X$ direction, \ie the direction that is
transverse to the one-dimensional loading. Finally it is also observed
that $ f^{(1)}_{y0}$ is almost equal to zero whatever the load level,
thus the zero value in Table~\ref{tab:05}.

\clearpage

\section{Results -- Case \#2: random grain geometry} 
\label{section-6}
In this section both the randomness in the grain geometry and in the
crystallographic orientations are taken into account. A total number of
35~finite element models are run. In each case, the grain geometry is
obtained from a uniform sampling of points from which a Vorono\"i
tessellation is built.

\subsection{Check of the homogeneity}
\label{sec:6-1}

\begin{figure}[!ht]
    \begin{center}
      \includegraphics[clip, trim = 50mm 152mm 60mm 130 mm,width = 0.48\textwidth]{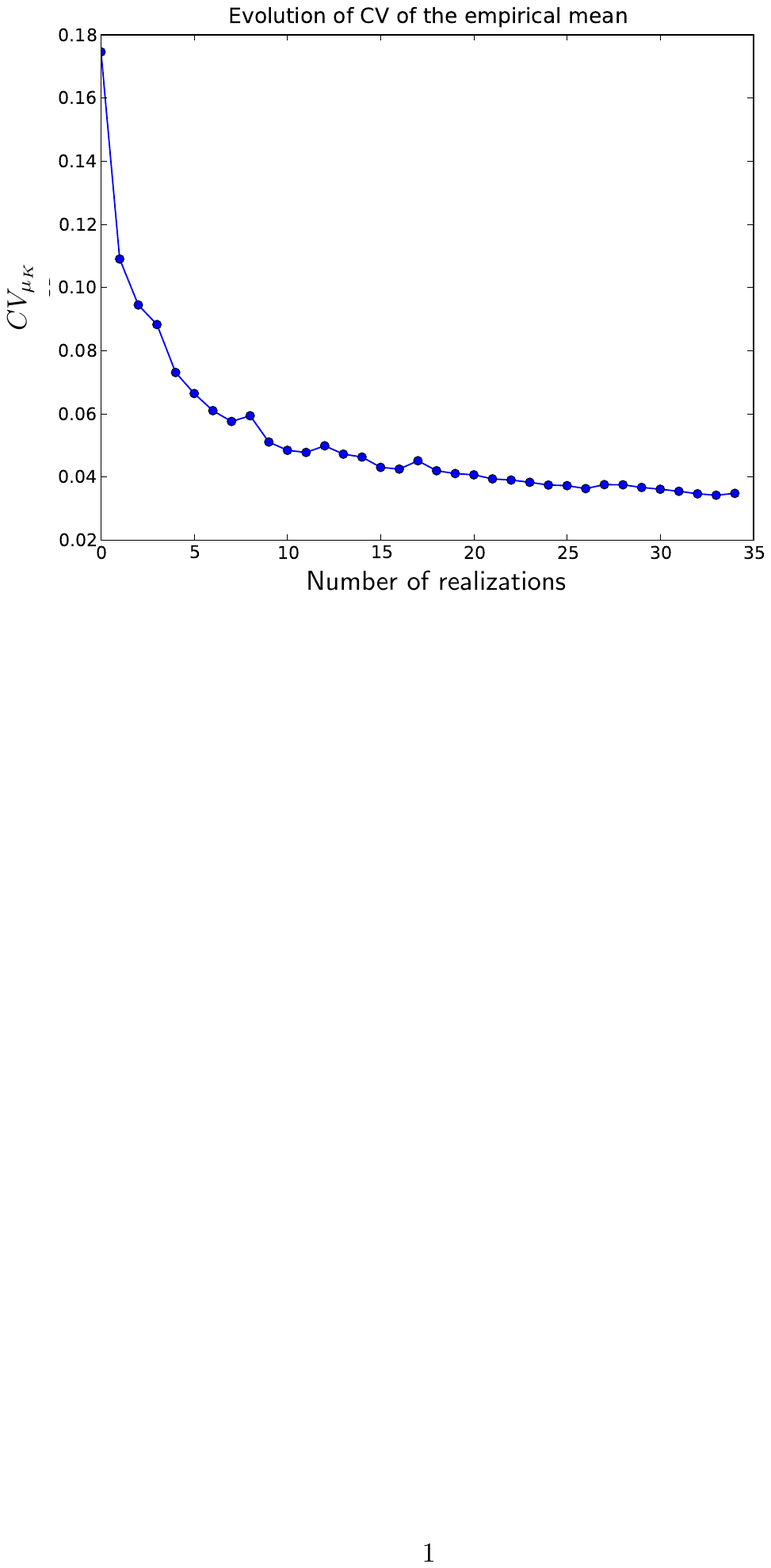}
      \includegraphics[clip, trim = 50mm 152mm 60mm 130 mm,width = 0.48\textwidth]{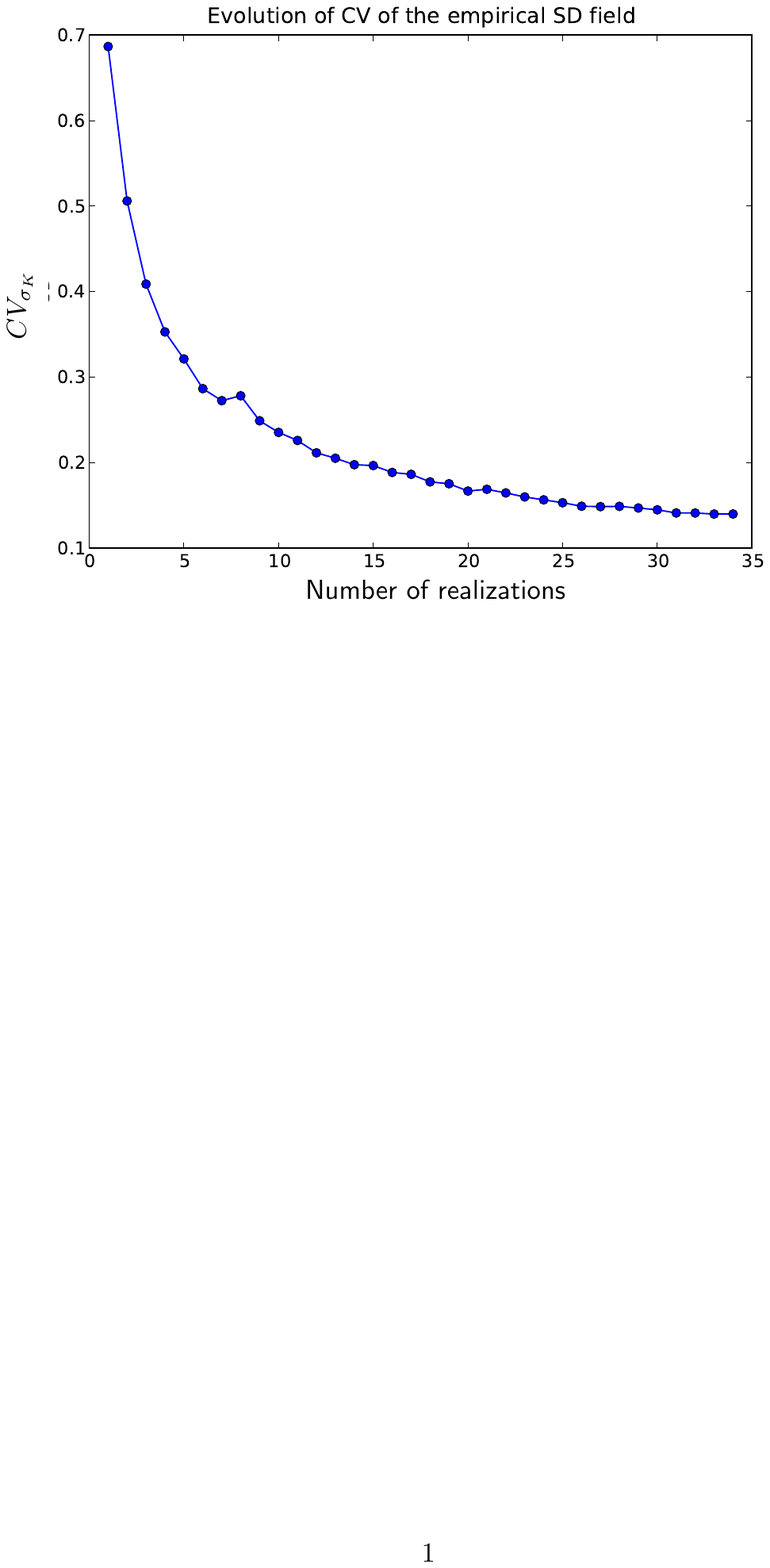}
      \caption{Case \#2: Evolution of $CV_{\mu_K}$ and $CV_{\sigma}^K$
        with respect to the number of realizations} \label{fig-IV.22}
    \end{center} 
\end{figure}

As in Section~\ref{section-5} the homogeneity of the maximal principal
stress field is checked using the methodology proposed in
Section~\ref{section-4.2}.  Figure~\ref{fig-IV.22} shows the evolution
of $CV_{\mu_K}$ and $CV_{\sigma_K^2}$. These quantities regularly
decrease and it is seen that they would tend to zero if a larger number
of realizations was available. This leads to accepting the assumption
that the random field is homogeneous.

\subsection{Identification of periodograms at 3.5\% macroscopic
strain}
\label{sec:6-2}
\begin{figure}[!ht]
    \begin{center}
      \includegraphics[width=0.48\textwidth]{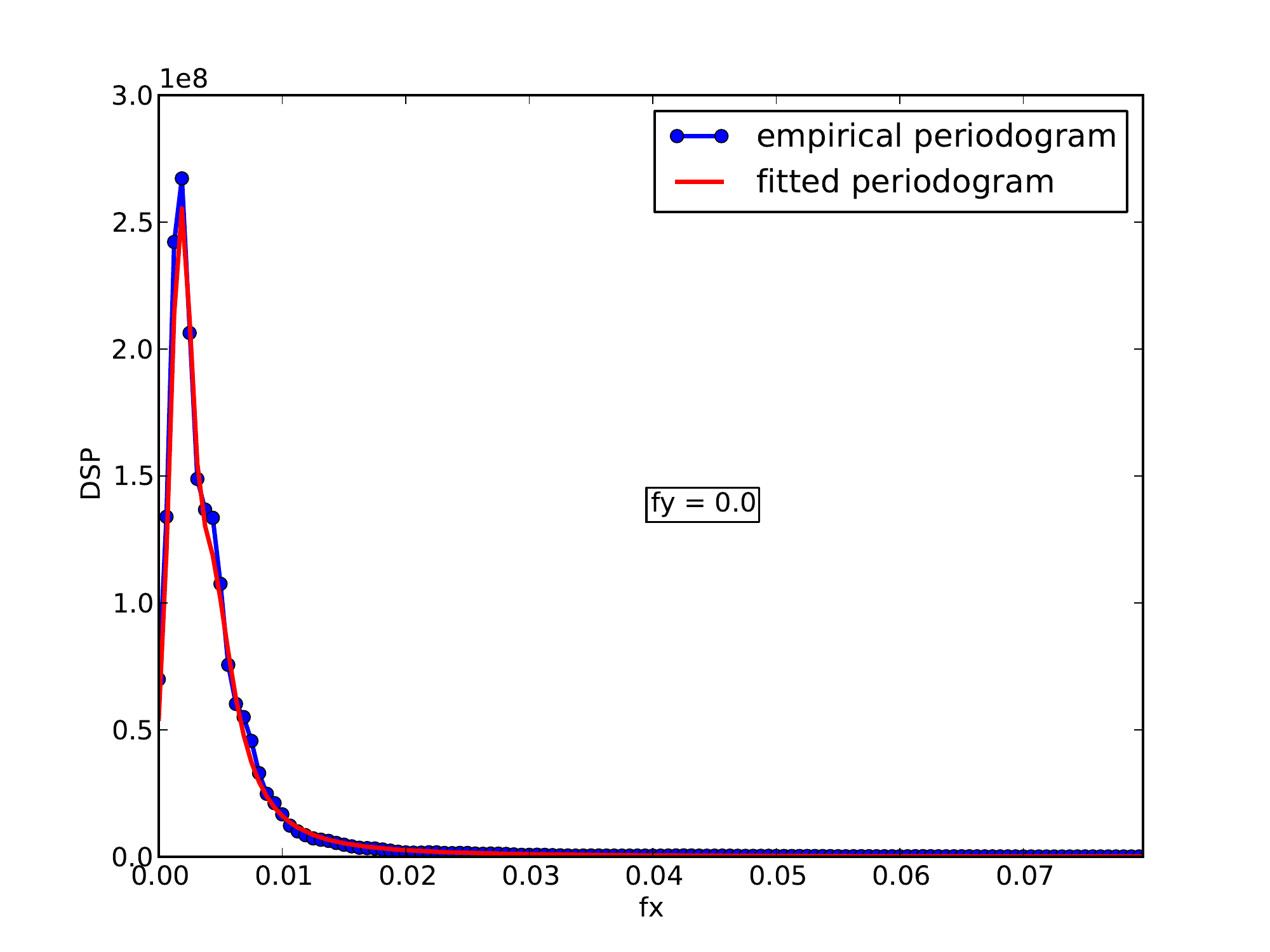}
      \includegraphics[width=0.48\textwidth]{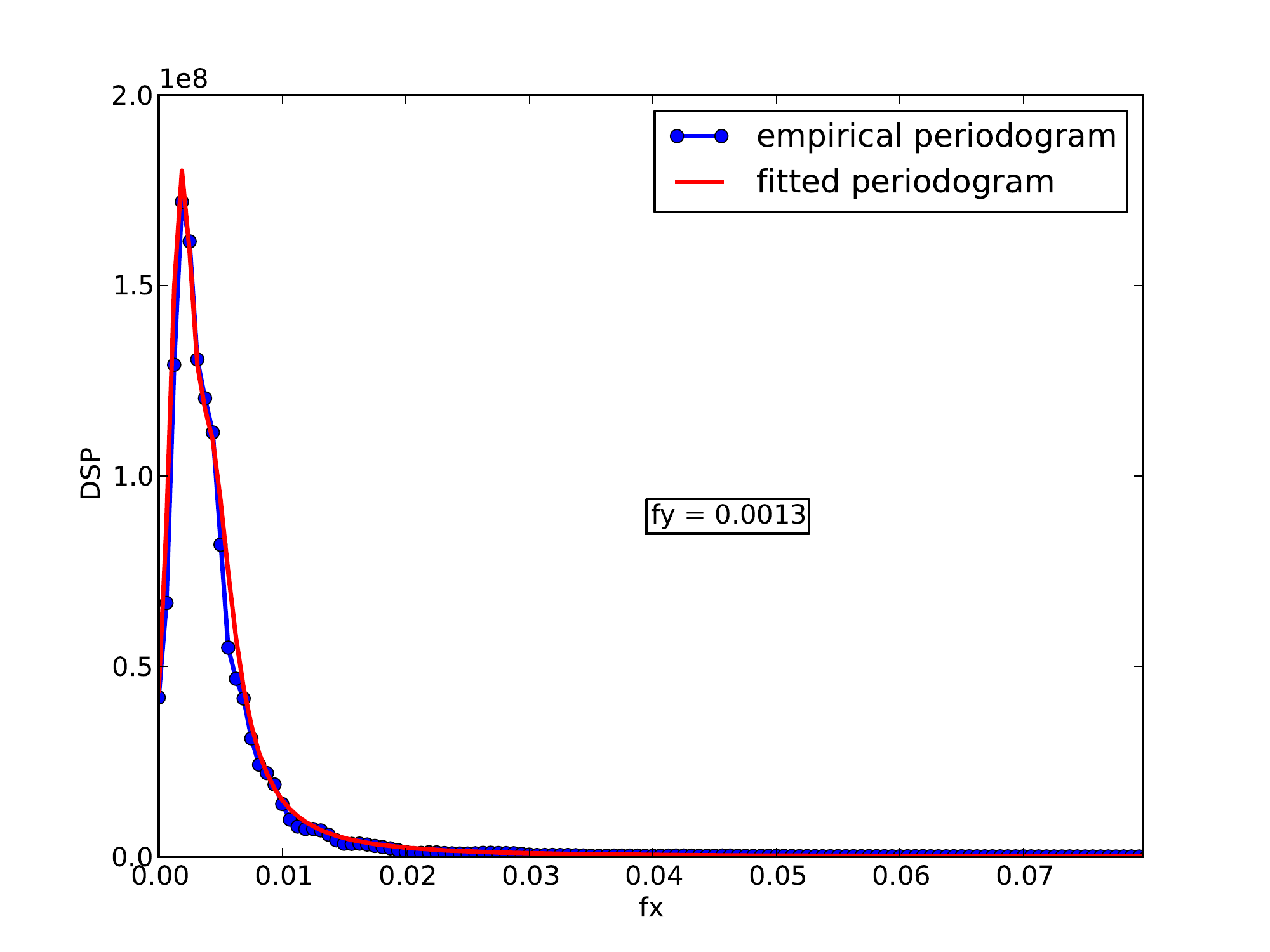}
      \caption{Case \#2: Cut of the periodograms in the $X$ direction}\label{fig-IV.23}
    \end{center}
\end{figure}

The average empirical periodogram obtained from $L=35$ realizations of
the maximal principal stress field $\sigma_I$ at 3.5\% of macroscopic
strain is plotted in Figure~\ref{fig-IV.23}-a. Three types of
theoretical periodograms have been fitted as in the previous section,
which lead to the conclusion that the mixed model that combines a
Gaussian and an exponential component is best suited. The fitted
parameters are gathered in Table~\ref{tab:06} where the parameters fitted
for Case \#1 are also recalled for the sake of comparison.

\begin{figure}[!ht]
    \begin{center}
      \includegraphics[width=0.48\textwidth]{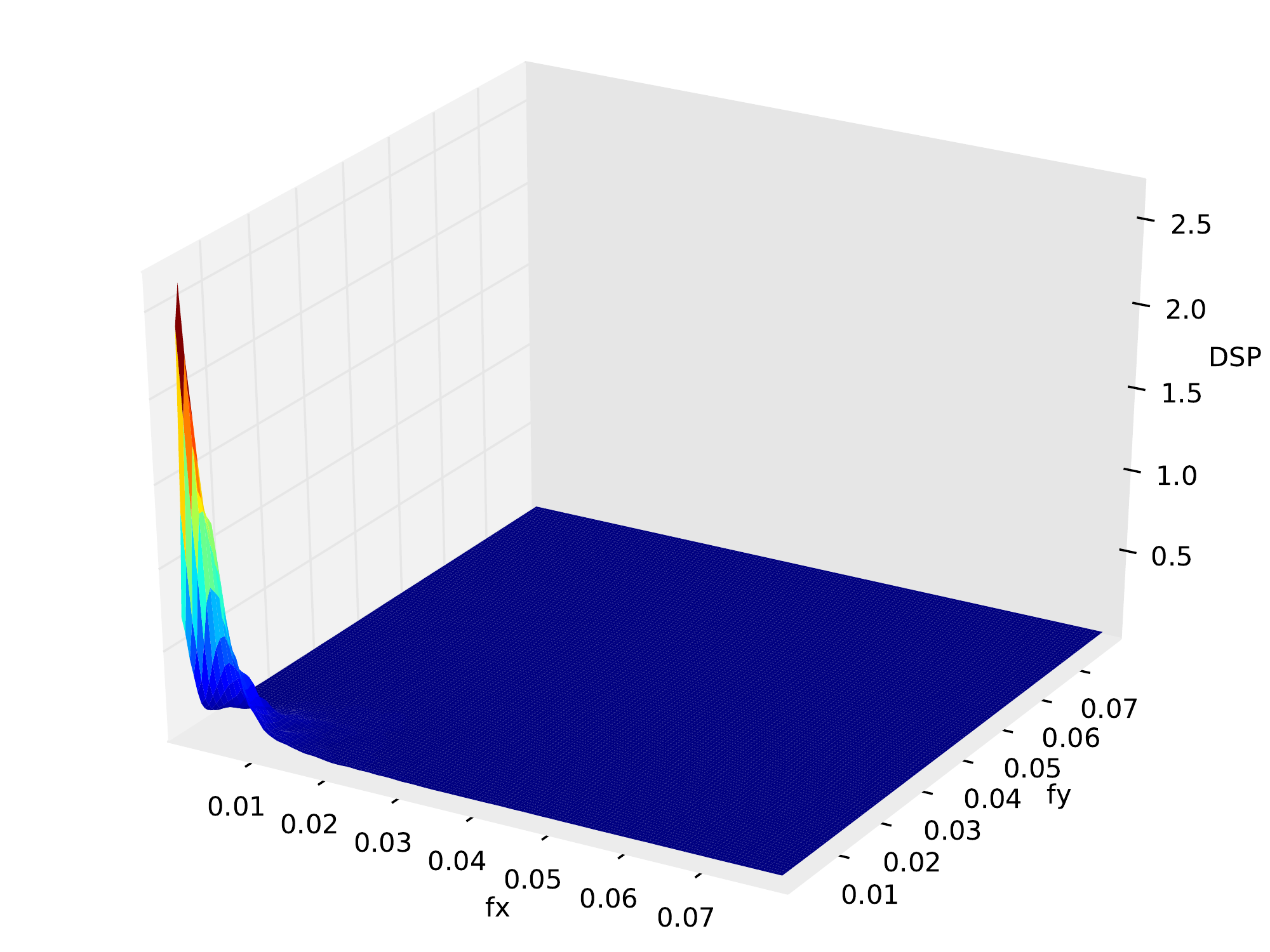}
      \includegraphics[width=0.48\textwidth]{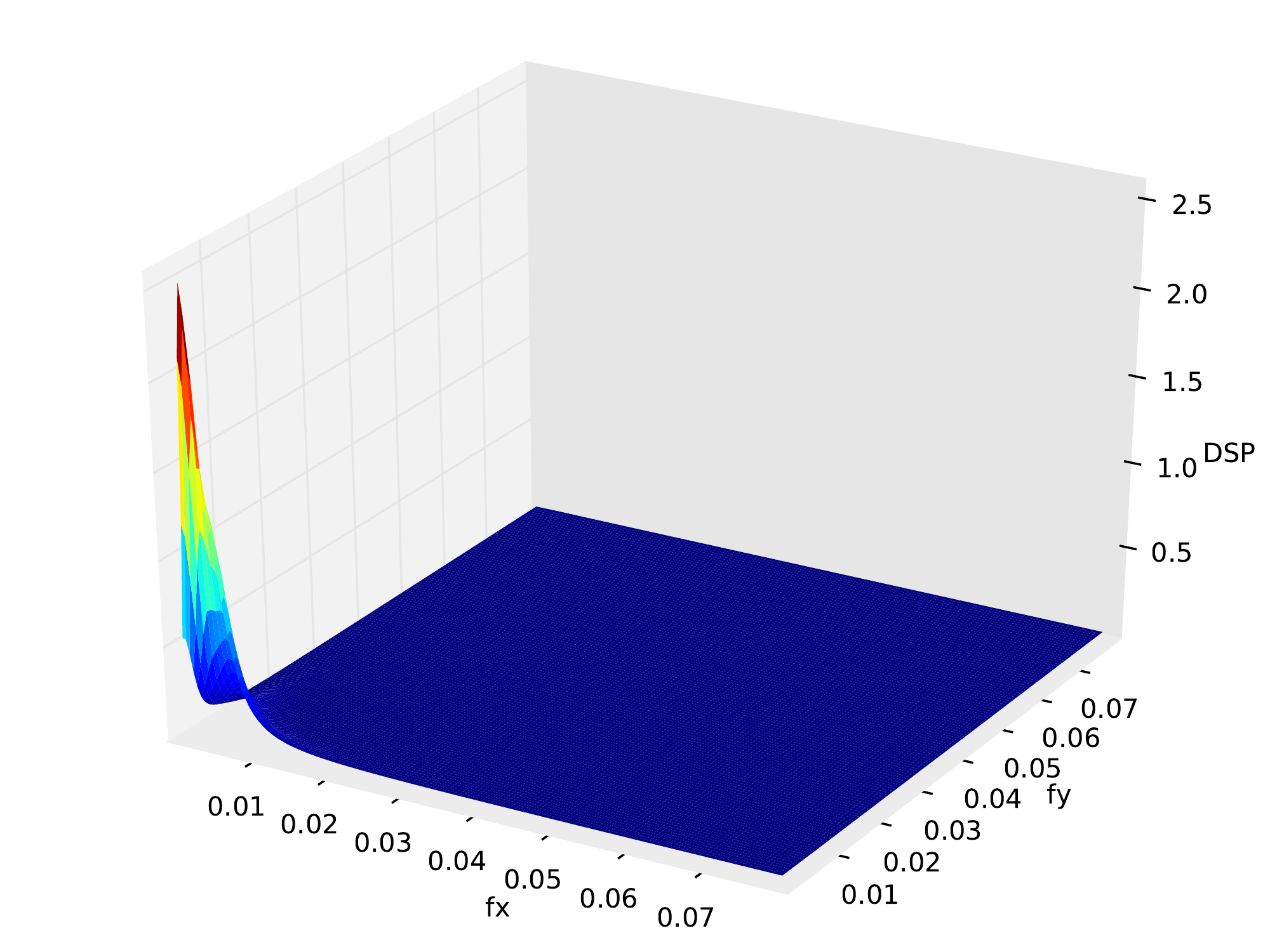}
      \caption{ Case \#2: (a) Average empirical periodogram of the
        stress field at $3.5\%$ macroscopic strain -- (b) best fitted
        periodogram}\label{fig-IV.22b}
    \end{center}
\end{figure}

\begin{figure}[!ht]
    \begin{center}
      \includegraphics[width=0.48\textwidth]{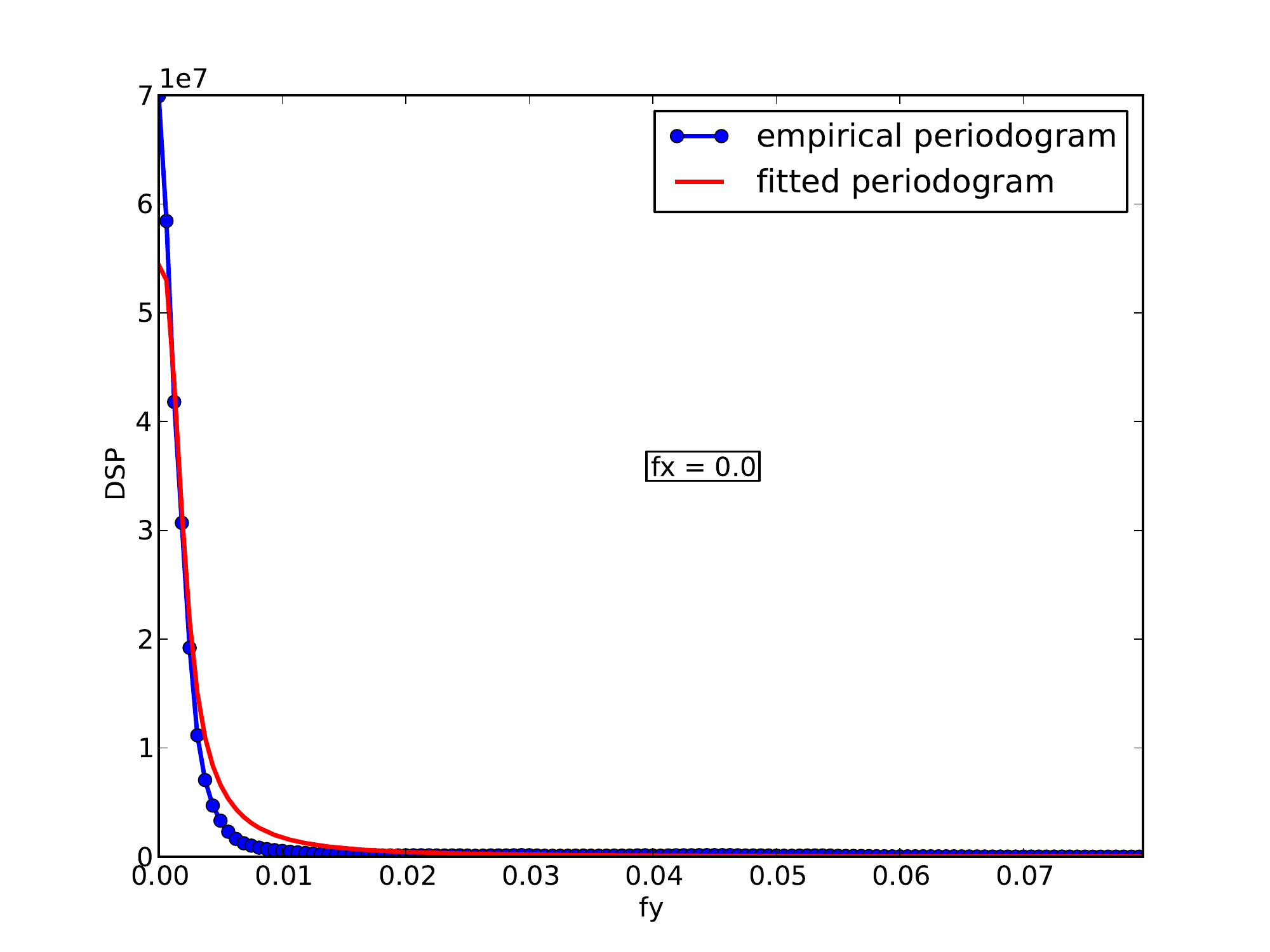}
      \includegraphics[width=0.48\textwidth]{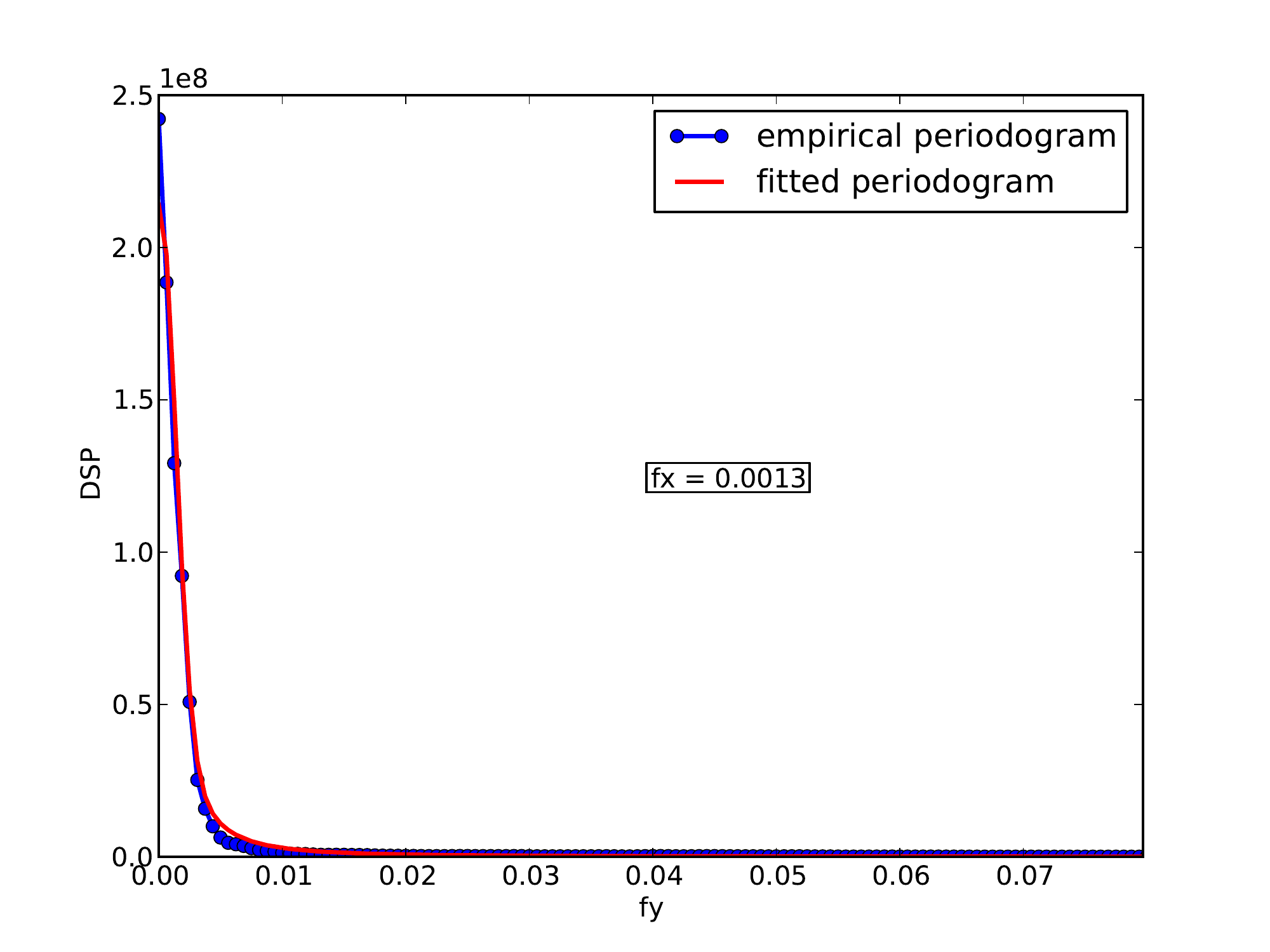}
      \caption{Case \#2: Cut of the periodograms in the $Y$ direction}
      \label{fig-IV.24}
    \end{center}
\end{figure}

\begin{figure}[!ht]
    \begin{center}
      \includegraphics[width=0.48\textwidth]{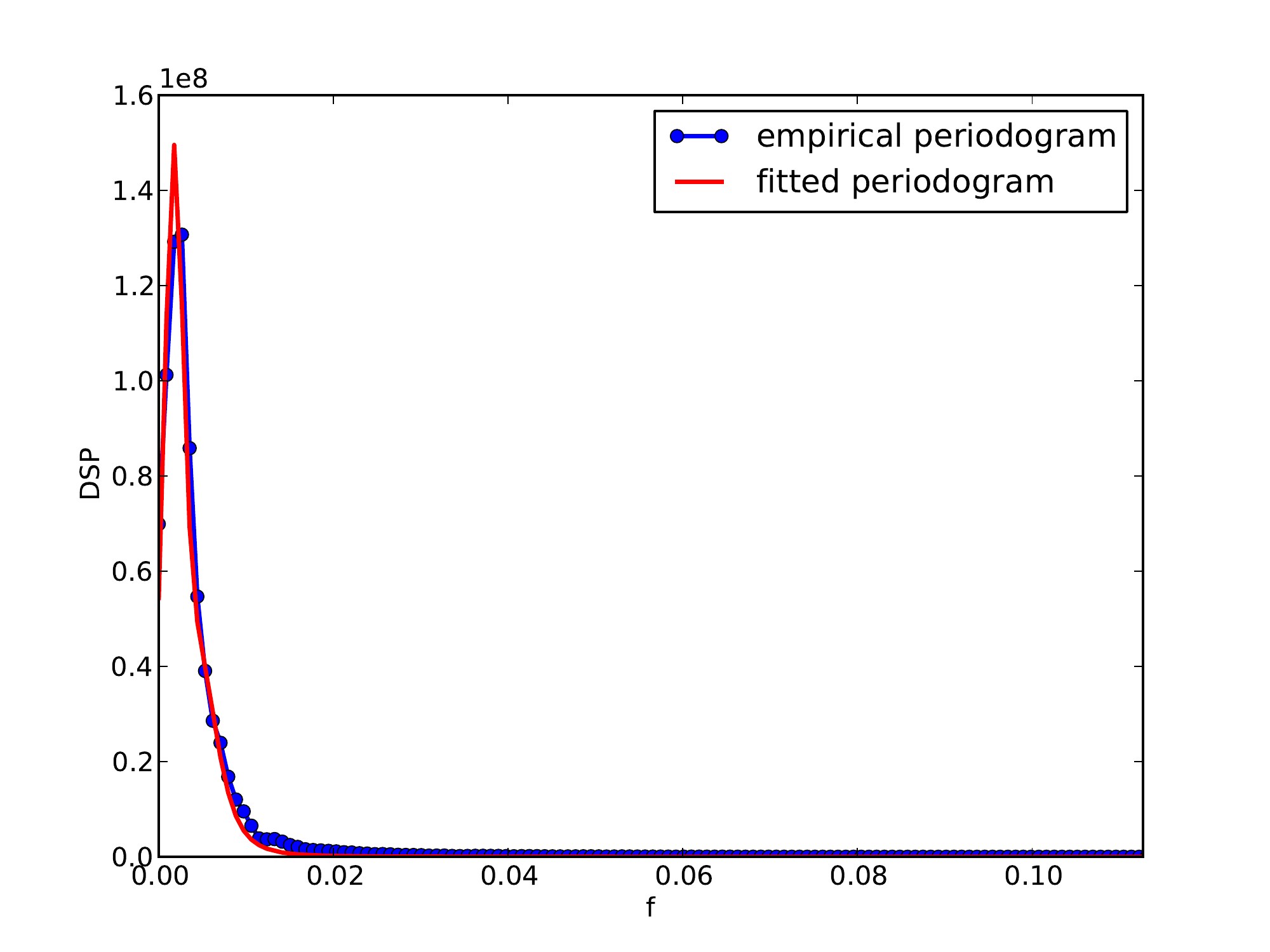}
      \caption{ Case \#2: Cut of the periodograms along
        the diagonal $f_x = f_y$} \label{fig-IV.25}
    \end{center}
\end{figure}

In order to check the accuracy of the fitting, two-dimensional cuts of
the empirical (resp. fitted periodogram) are plotted in
Figure~\ref{fig-IV.23} (cut along the $X$ direction),
Figure~\ref{fig-IV.24} (cut along the $Y$ direction),
Figure~\ref{fig-IV.25} (cut along the diagonal $f_x=f_y$ ). Again the
fitting is remarkably accurate, meaning that the fitted model of
periodogram accurately represents the spatial variability of the maximal
principal stress field.

It can be observed from the figures in Table~\ref{tab:06} that the
fitting is of equal quality in both cases (relative error less than $2
\, 10^{-3}$). As far as the contribution of each component of the
periodogram is concerned, the symmetry reported in Case \#1 is not
existing anymore since the standard deviation of the exponential
contribution ($\sigma_2 = 81.6$) is much greater than that of the
Gaussian part ($\sigma_1 = 35.8$). The total variance of the field is
7940~MPA${}^2$, corresponding to a standard deviation of 89.1~MPa and a
coefficient of variation of 12\%. Thus there is a little more scattering
in the random stress field obtained in Case~\#2 when considering both
the random grain geometry and orientations.

\begin{table}[!ht] 
  \caption{Fitted parameters and error estimates for the  mixed
    ``Gaussian + exponential'' periodogram} \label{tab:06} 
  \begin{tabular}{cccccccccccc}
    \hline\noalign{\smallskip}
    \multirow{2}{*}{Case} & $\epsilon$ & \multicolumn{5}{c}{Gaussian} & \multicolumn{5}{c}{Exponential} \\
    \cline{3-12}
    & (Eq.(\ref{eq-IV.33})) & $\sigma_1$ & $l_{x1}$ & $l_{y1}$ & $f^{(1)}_{x0}$ & $f^{(1)}_{y0}$ & $\sigma_2$ & $l_{x2}$ & $l_{y2}$ &
    $f^{(2)}_{x0}$  &   $f^{(2)}_{y0}$ \\
    \hline
    Case \#1 & $0.0017$ & $54.7$ & $138.4$ & $159.1$ & $0.00244$ & $0$ & $57.6$ & $57.5$ & $63.5$& $0.00562$ & $0.0028$\\
    Case \#2 & $0.0018$ & $35.8$ & $269.5$ & $174.5$ & $0.00172$ & $0$ & $81.6$ & $67.2$ & $70.4$& $0.004$ & $0$\\
    \noalign{\smallskip}\hline
  \end{tabular}
\end{table}

The correlation lengths associated with the exponential part do not
differ much in Case~\#2 compared to Case~\#1 (corresponding here to 1.5
to 2.4~$D_g$). In contrast the correlation lengths related to the
Gaussian part are increased, which tends to produce less rapidly varying
realizations. This may be explained by the fact that the grain
boundaries are ``averaged'' in Case~\#2 whereas they were fixed in
Case~\#1. The stress concentrations that are usually observed at the
grain boundaries are thus smoothed in Case~\#2 compared to Case~\#1.

\subsection{Influence of the number of realizations}
\label{sec:6-3}

\begin{figure}[!ht]
    \begin{center}
      \includegraphics[clip, trim = 50mm 152mm 60mm 130 mm,width = 0.48\textwidth]{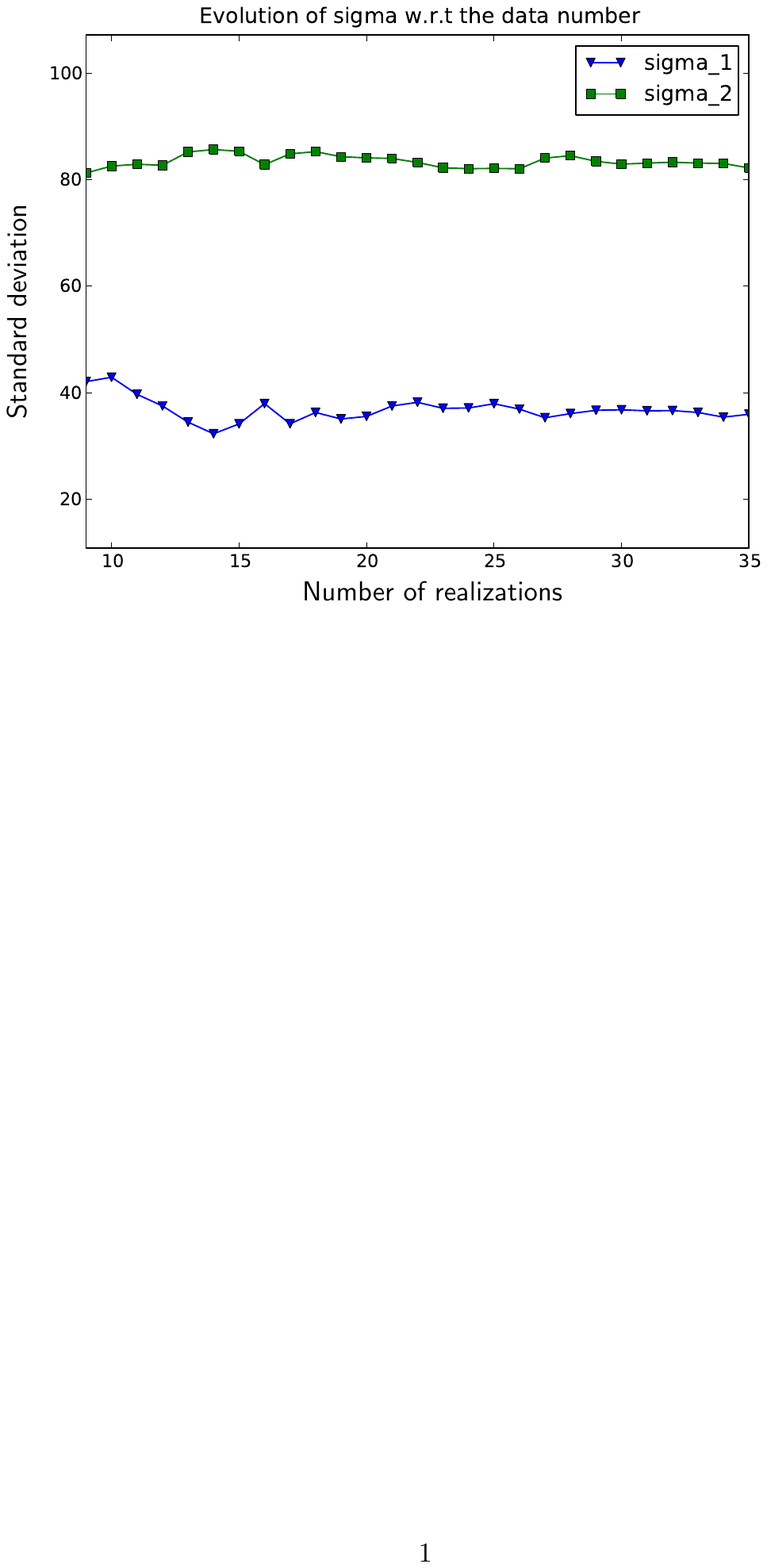}
    \includegraphics[clip, trim = 50mm 152mm 60mm 130 mm,width = 0.48\textwidth]{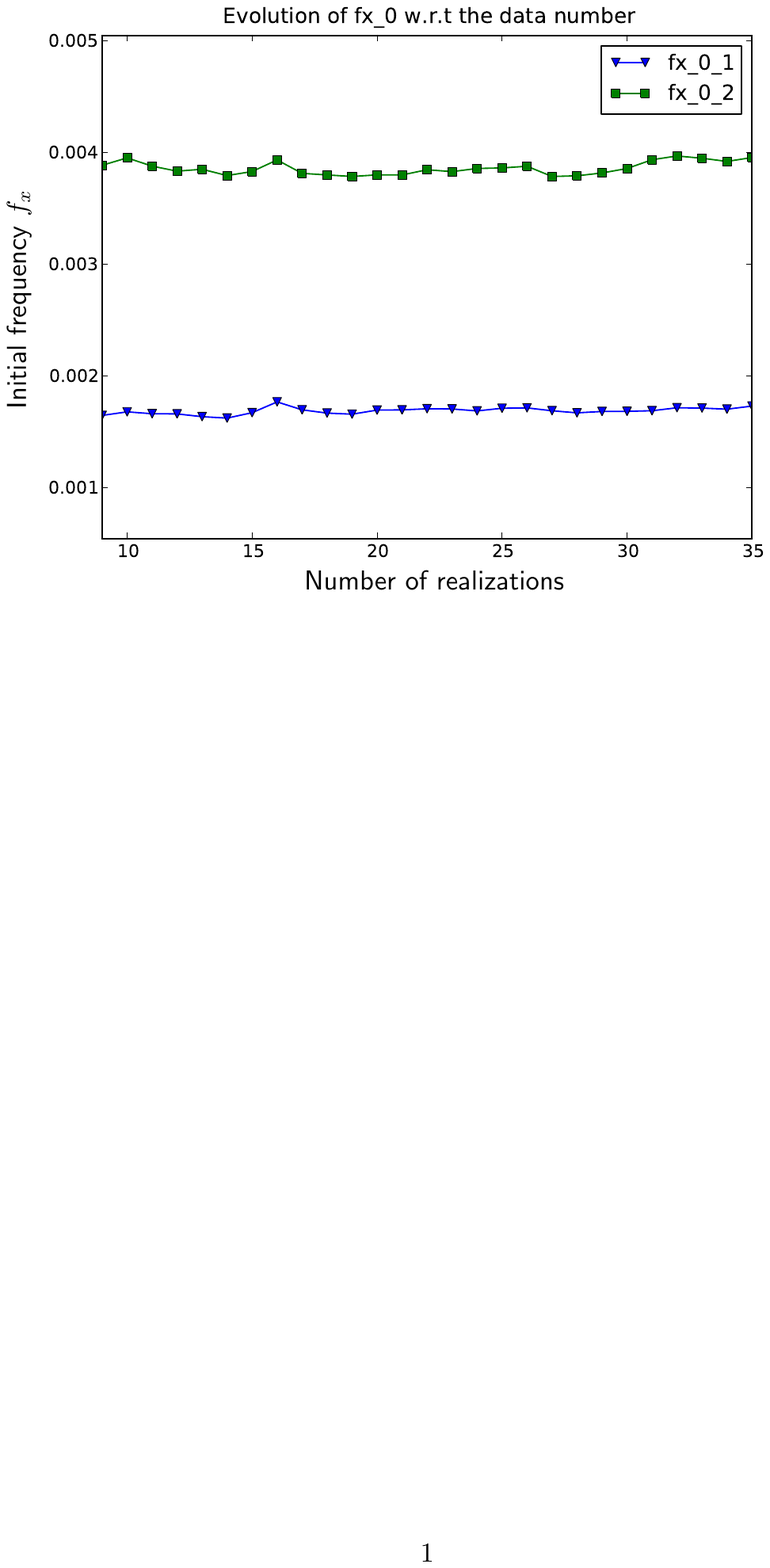}
    \caption{Case \#2: Evolution of the fitted standard deviations and
      the initial frequencies $f_{(x)_0}^{(1,2)}$ with respect to the
      number of realizations $K=8 \enu 35$}
      \label{fig-IV.26}
    \end{center}
\end{figure}

\begin{figure}[!ht]
    \begin{center}
      \includegraphics[clip, trim = 50mm 152mm 60mm 130 mm,width = 0.48\textwidth]{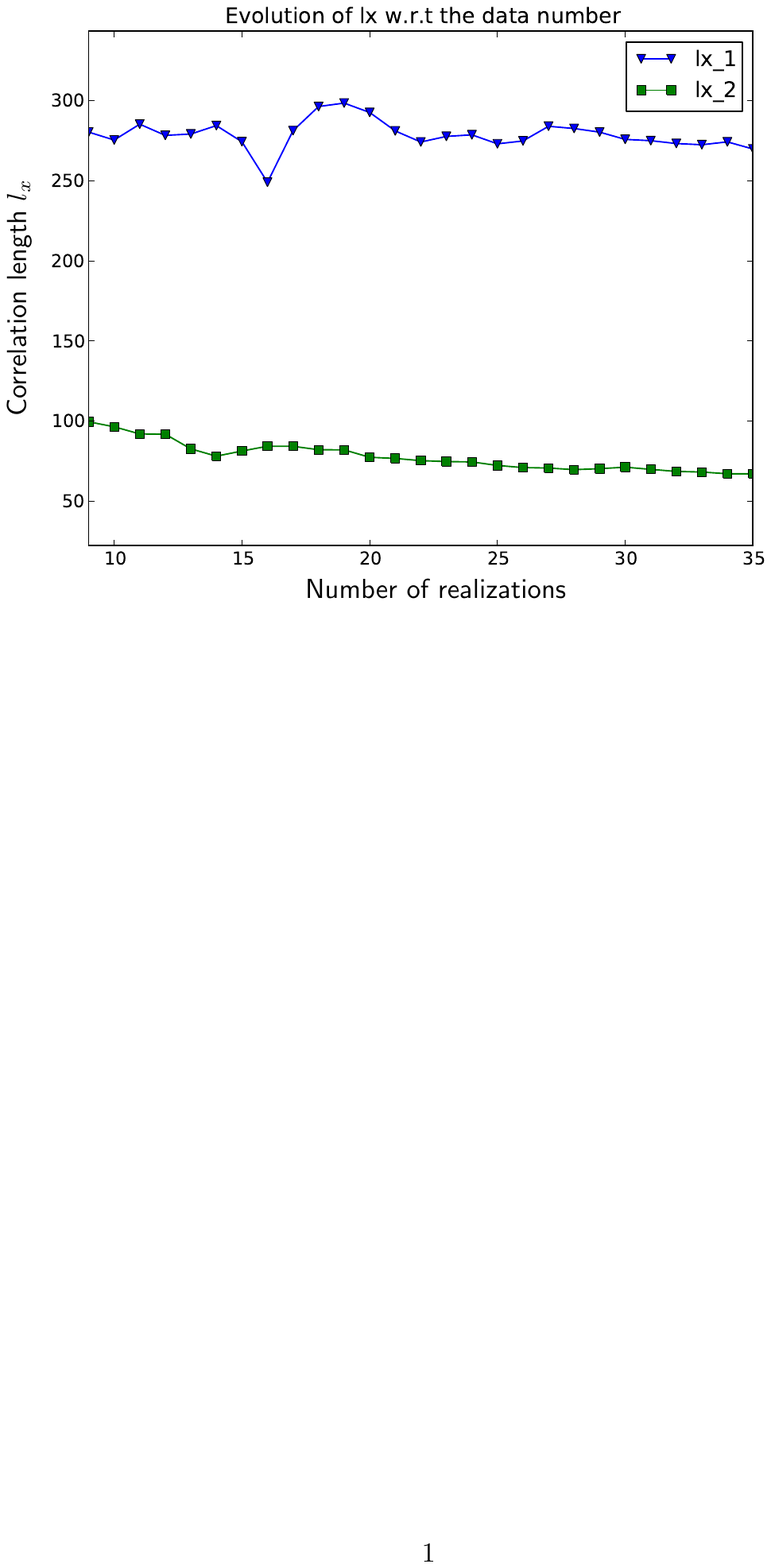}
      \includegraphics[clip, trim = 50mm 152mm 60mm 130 mm,width = 0.48\textwidth]{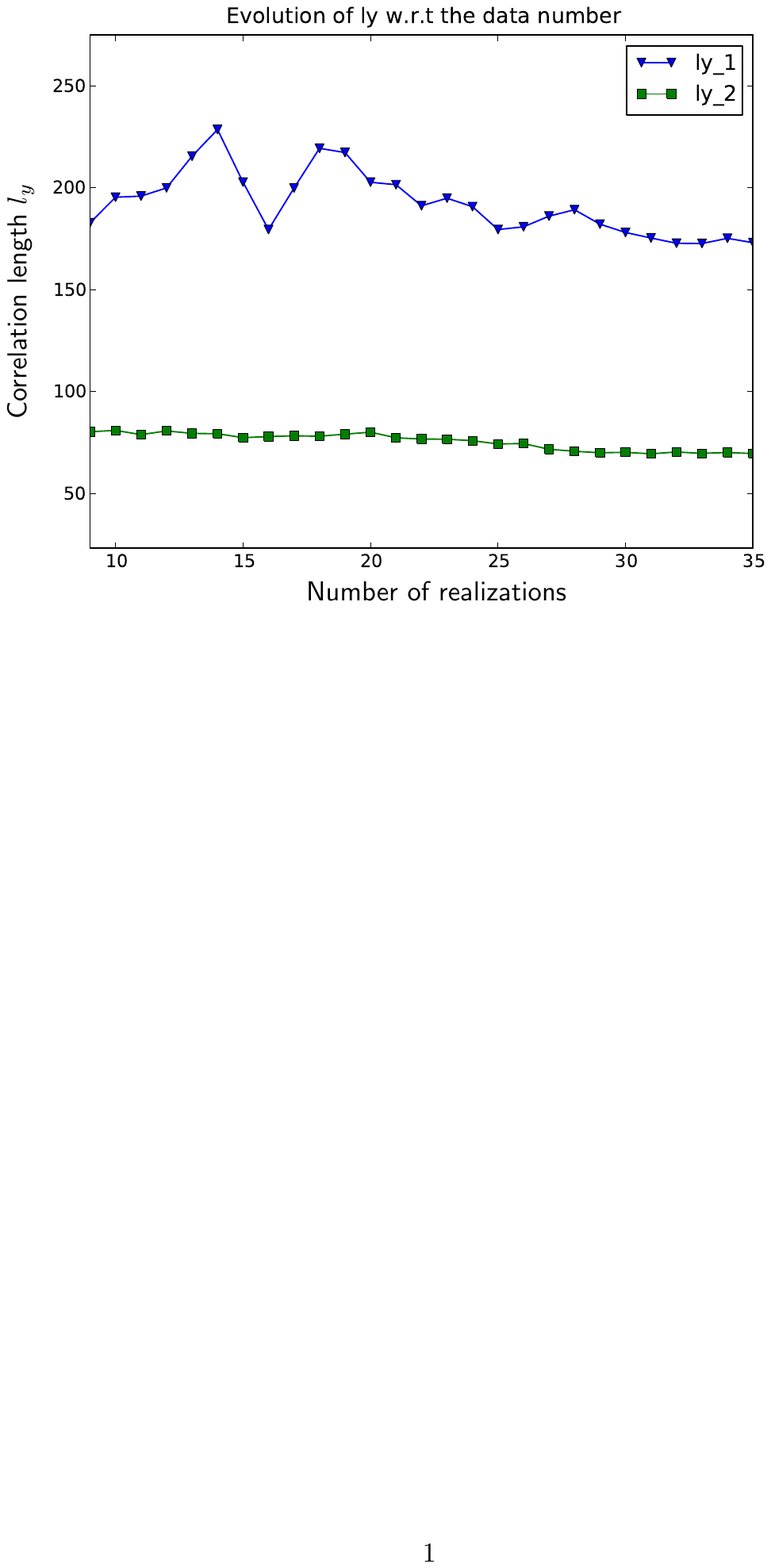}
      \caption{Case \#2: Evolution of the fitted correlation lengths in
        the $X$ (left) (resp. $Y$ (right)) directions with respect to
        the number of realizations $K=8 \enu 35$}
      \label{fig-IV.27}  
    \end{center}
\end{figure}

In this section one considers the stability of the fitted parameters as
a function of the number of available realizations $K$ used in the
average periodogram method. The procedure applied in the previous
paragraph is run using $K=8,9 \enu 35$ realizations of the stress field.
The evolution of the standard deviations $(\sigma_1, \sigma_2)$ and the
initial frequencies $f_{(x)_0}^{(1,2)}$ is shown in
Figure~\ref{fig-IV.26} (note that $f_{(y)_0}^{(1,2)} =0$ in the present
case).  The evolution of the correlation lengths $l_{(x,y)(1,2)}$ is
shown in Figure~\ref{fig-IV.27}.

From these figures it clearly appears that the fitted parameters are
almost constant when the number of realizations of the stress field used
in their estimation increases. The minimal number of $K=8$ could be used
here without significant errors although it is recommended to keep a
value of $K=20$ as in Case \#1 for robustness.

\subsection{Influence of the macroscopic strain level}
\label{sec:6-4}
Finally the evolution of the parameters of the fitted periodograms as a
function of the macroscopic strain $E_{YY}$ is investigated.  For this
purpose the identification method is applied using the realizations of
the maximal principal stress fields corresponding to various levels of
the loading curve, \ie various values of the equivalent macroscopic
strain $E_{YY} = 0. \enu 3.5\%$.

\begin{figure}[!ht]
  \begin{center}
    \includegraphics[clip, trim = 50mm 152mm 60mm 130 mm, width = 0.48\textwidth]{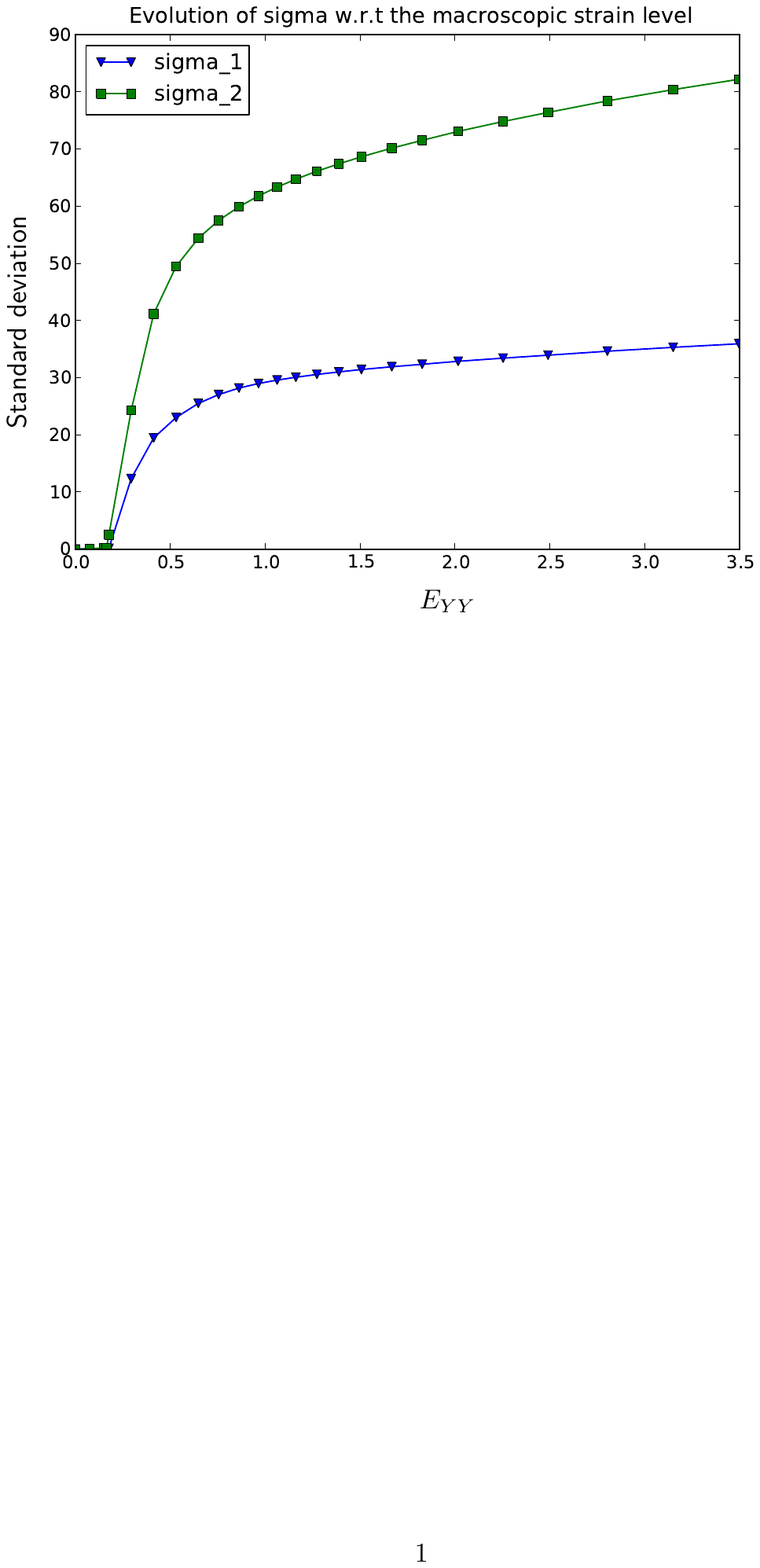}
    \includegraphics[clip, trim = 50mm 152mm 60mm 130 mm ,width = 0.48\textwidth]{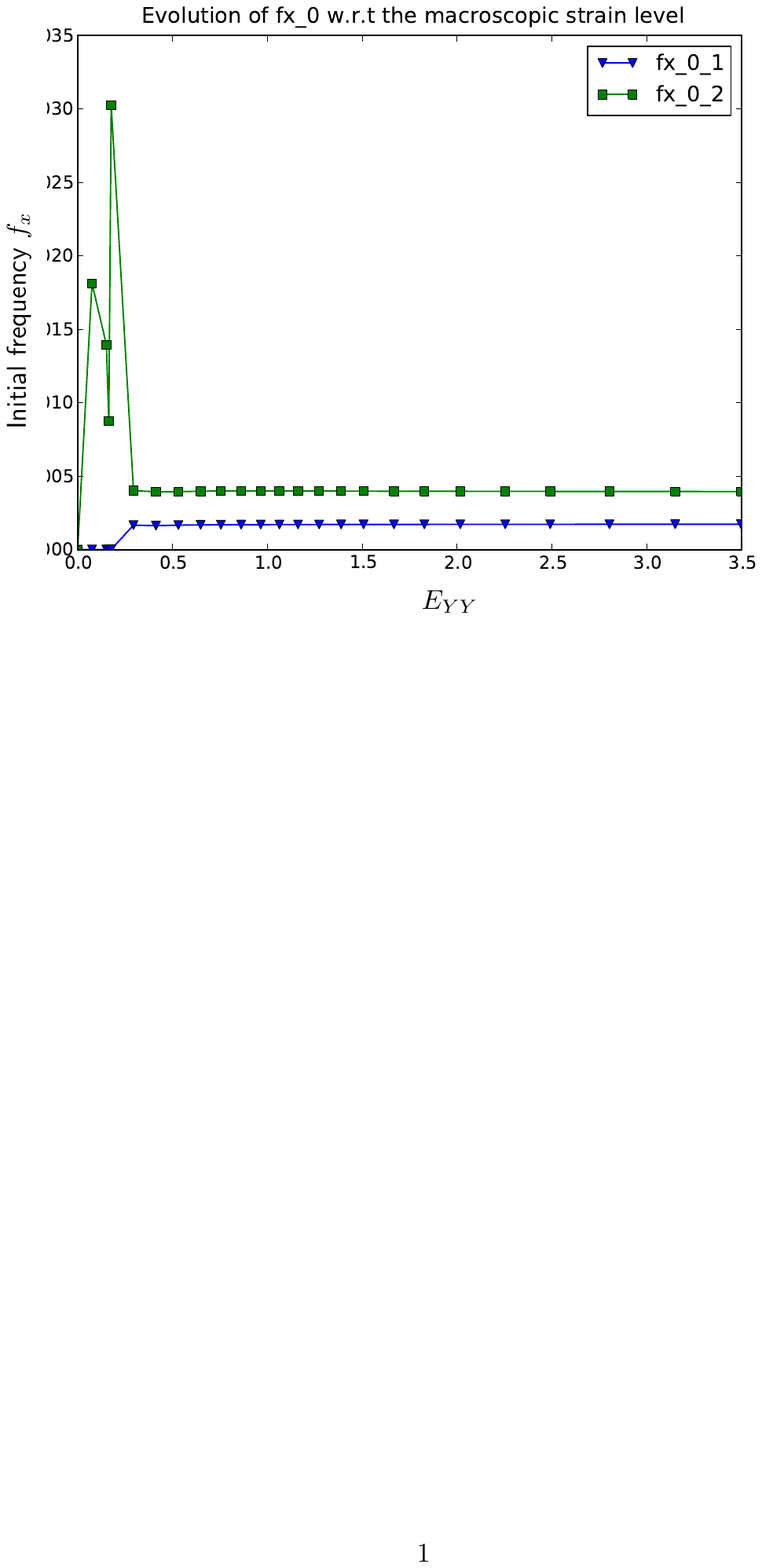}
    \caption{Case \#2: Evolution of the fitted standard deviations
      (left) with respect to the load level (macroscopic strain $E_{YY}
      = 0. \enu 3.5\%$) (resp. the initial frequencies $f_{(x,y)_0}^{(1)}$ (right))}
      \label{fig-IV.28}
    \end{center}
\end{figure}

\begin{figure}[!ht]
    \begin{center}
      \includegraphics[clip, trim = 50mm 152mm 60mm 130 mm,width = 0.48\textwidth]{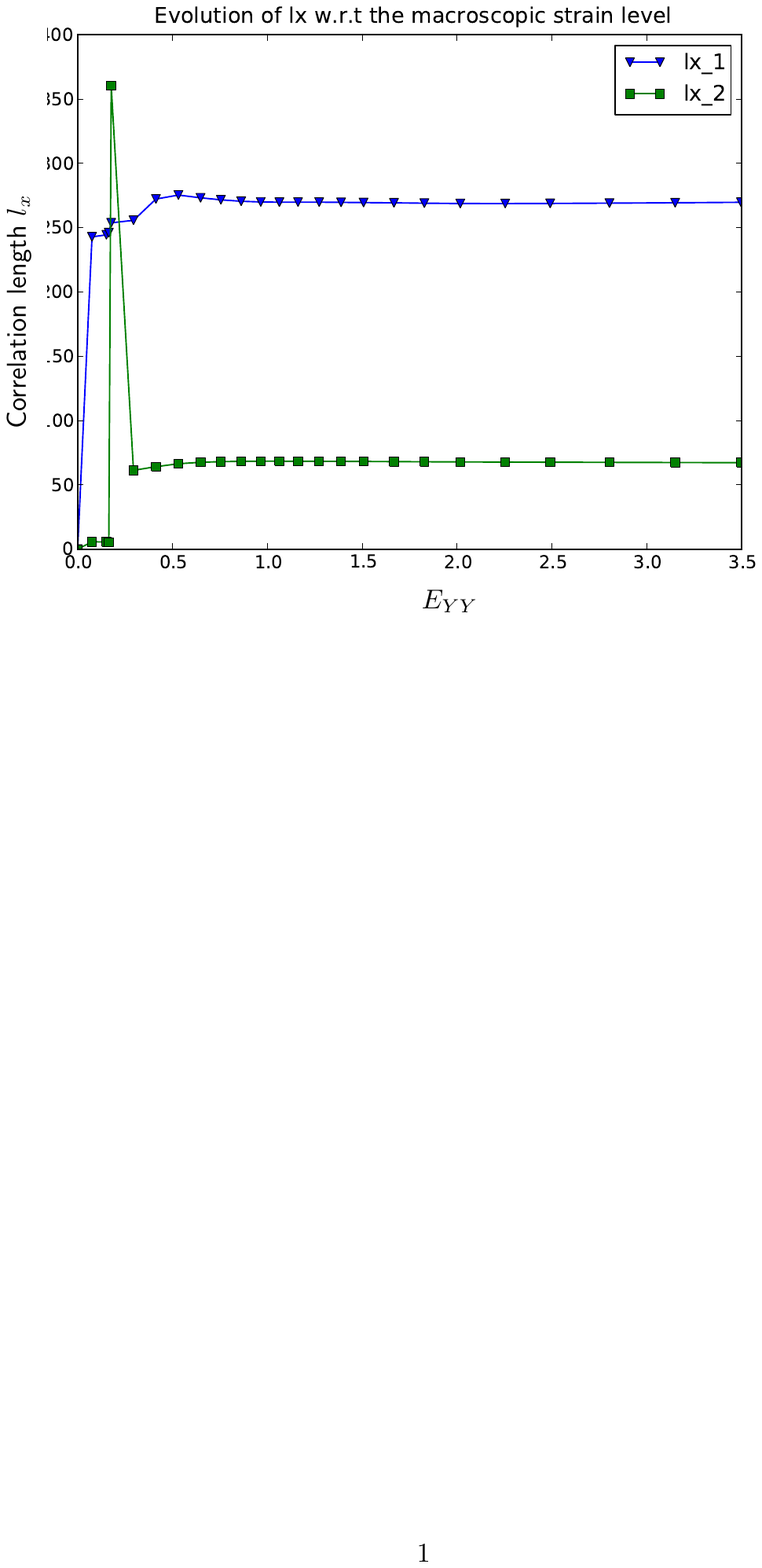}
      \includegraphics[clip, trim = 50mm 152mm 60mm 130 mm,width = 0.48\textwidth]{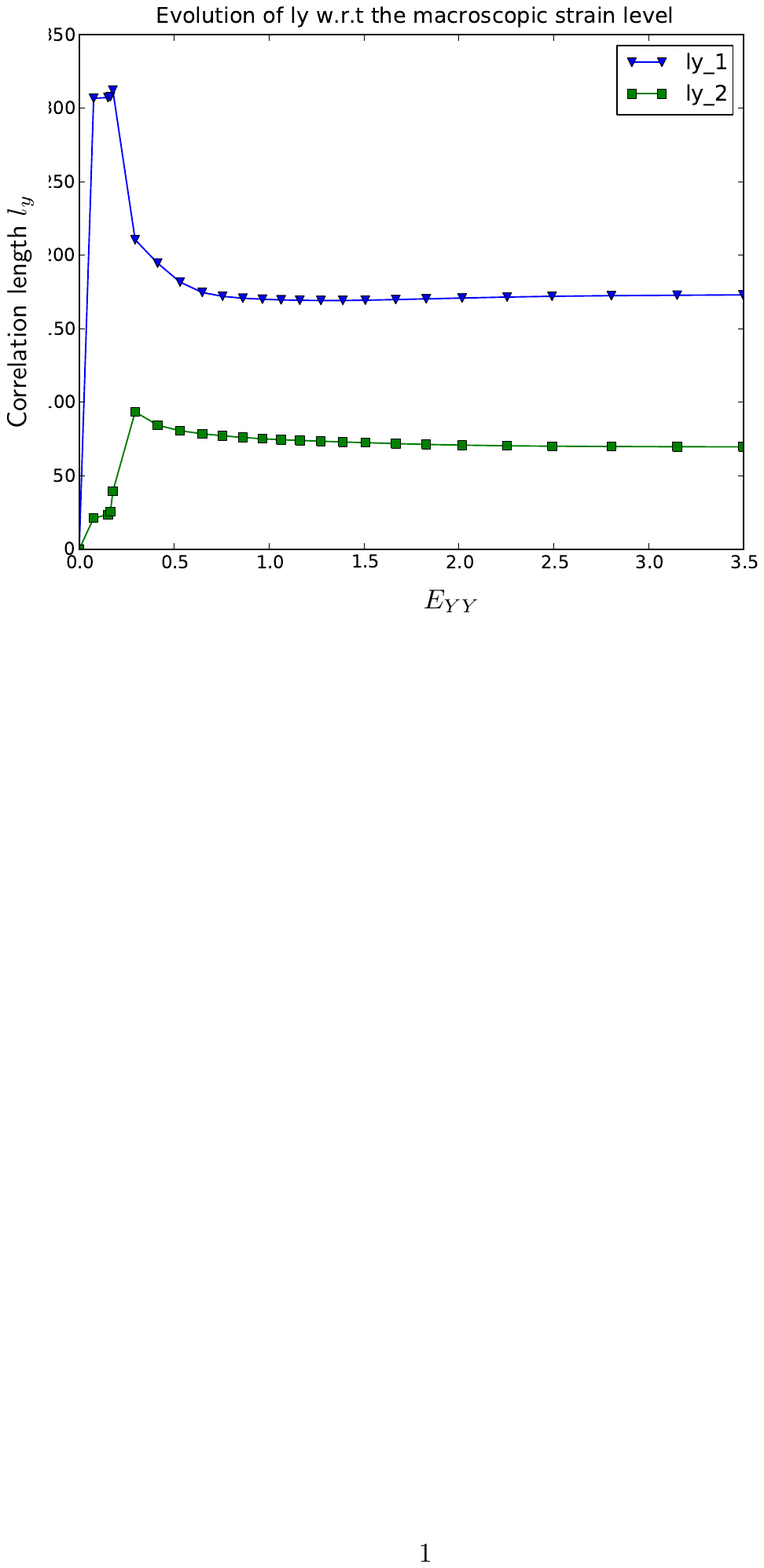}
      \caption{Case \#2: Evolution of the fitted correlation lengths in
        the $X$ (left) (resp. $Y$ (right)) directions with respect to
        the load level (macroscopic strain $E_{YY} = 0. \enu 3.5\%$)}
      \label{fig-IV.29}  
    \end{center}
\end{figure}

The evolution of the two standard deviations look similar to the results
obtained in Case~\#1 (Figure~\ref{fig-IV.28}). It is observed that the
ratio $\sigma_2 / \sigma_1$ is almost constant all along the loading
path up to 3.5\% strain. As far as the initial frequencies are
concerned, there is a complete independance with the load level as soon
as $E_{YY}$ is greater than $\sim 0.5\%$, \ie when plasticity has
settled in the aggregate. The same conclusion can be drawn for the
various correlation lengths.

\clearpage

\section{Conclusions} 
\label{conclusion}

The distribution of stresses in a material at a microscopic scale (where
heterogeneities such as grain structures are taken into account) has
been given much attention in the context of computational homogenization
methods. However the current methods usually stick to a deterministic
formulation. Starting from the premise that any representative volume
element (such as a polycrystalline aggregate) is a single specific
realization of a random quantity, the present paper aims at using
methods of computational stochastic mechanics for representing the
(random) stress field.

After recalling the basic mathematics of Gaussian random fields, the
paper presents a {\em periodogram method} for estimating the parameters
describing the spatial fluctuation of a random field from a collection
of realizations of this field. This method is adapted in two dimensions
from well-known techniques originating from signal processing. 

The material under consideration, namely the 16MND5 steel used in
nuclear pressure vessels is then presented together with a local
modelling by polycrystalline finite element calculations. From a
collection of 35 realizations of the (maximal principal) stress field,
the spatial correlation structure of the latter is identified. By
fitting various theoretical periodograms, a mixed model combining a
Gaussian and an exponential-type contribution is retained. These two
contributions may be empirically interpreted as follows: The Gaussian
part corresponds to the fluctuation from grain to grain ; the (less
smooth) exponential component corresponds to the sharp grain boundaries
stress concentrations.

Two cases are considered, namely a ``fixed-geometry'' case in which only
the crystallographic orientations changes within the 35 realizations
(fixed grain boundaries), and a ``variable geometry'' in which the grain
geometry is randomly sampled for each realization. In both cases, a good
convergence of the procedure is observed when the number of realizations
increases. A set of 20~realizations is recommended, although good
results are already obtained for $\sim$8~realizations in Case \#2.

Moreover it is shown that the correlation lengths are of the same order
of magnitude as the grain size. The initial frequencies that are
required for a best fitting of the periodogram and that translate into
some kind of spatial periodicity in the covariogram could be explained
by spurious edge effects due to the limited size of the aggregate. This
should be investigated more in details in further analysis. 

Another important result is drawn from the comparison of the fitted
parameters at various load levels. Once plasticity is settled within the
aggregate, the parameters describing the spatial fluctuations of the
field are almost constant. Moreover the variance of the field (sum of
the variance of each component of the periodogram) increases
proportionally to the mean strain/stress curve, meaning that the
coefficient of variation of the stress field is almost constant (around
11\% for the fixed geometry and 12\% for the variable geometry).

The results presented in this paper should be confirmed by additional
investigations under different types of loading (\eg biaxial
loading). The tools that are presented here may be applicable to
three-dimensional aggregates and stress fields at a much larger
computational cost though. This work is currently in progress.

The identified stress fields may eventually be re-simulated: new
realizations of the stress fields are straightforwardly obtained at a
low computational cost by random field simulation techniques such as
the spectral approach or the circulant embedding method
\citep{Preumont1990,SudretCFRAC2011,SudretDangIcasp2011}. This allow us
to apply local approach to fracture analysis (such as that presented in
\citet{Mathieu:2006b}) for the assessment of the brittle fracture of
metallic materials, as shown in \citet{SudretDang2013}.

\textbf{Acknowledgements}\\
  The second author is funded by a CIFRE grant at Phimeca Engineering
  S.A. subsidized by the French {\em Agence Nationale de la Recherche et
    de la Technologie} (convention number 027/2010). The research
  project is supported by EDF R\&D under contract \#8610-AAP5910056413.
  These supports are gratefully acknowledged.


\bibliographystyle{spr-chicago}      

\end{document}